\documentclass[sw,table]{iosart2x}

\usepackage{graphicx}
\usepackage{xcolor}
\usepackage{footnote}
\usepackage[utf8]{inputenc}
\usepackage[ruled,vlined]{algorithm2e}
\usepackage{listings}
\usepackage{multirow}
\usepackage{amsthm}
\usepackage{subcaption}
\usepackage{appendix}

\theoremstyle{definition}
\newtheorem{definition}{Definition}

\pubyear{0000}
\volume{0}
\firstpage{1}
\lastpage{1}

\begin{document}
\makeatletter
\let\put@numberlines@box\relax
\makeatother
\begin{frontmatter}

\title{Enhancing Virtual Ontology Based Access over Tabular Data with Morph-CSV}
\runningtitle{Enhancing Virtual OBDA over Tabular Data with Morph-CSV}


\author[A]{\inits{D.}\fnms{David} \snm{Chaves-Fraga}\ead[label=e1]{dchaves@fi.upm.es}%
\thanks{Corresponding author. \printead{e1}.}},
\author[A]{\inits{E.}\fnms{Edna} \snm{Ruckhaus}\ead[label=e2]{eruckhaus@fi.upm.es}},
\author[A]{\inits{F.}\fnms{Freddy} \snm{Priyatna}\ead[label=e3]{fpriyatna@fi.upm.es}},
\author[B]{\inits{M.}\fnms{Maria-Esther} \snm{Vidal}\ead[label=e4]{maria.vidal@tib.eu}}
and
\author[A]{\inits{O.}\fnms{Oscar} \snm{Corcho}\ead[label=e5]{ocorcho@fi.upm.es}}

\runningauthor{Chaves-Fraga et al.}
\address[A]{Ontology Engineering Group, \orgname{Universidad Polit\'ecnica de Madrid}, \cny{Spain}\printead[presep={\\}]{e1,e2,e3,e5}}
\address[B]{\orgname{TIB - Leibniz Information Centre for Science and Technology and L3S Leibniz University of Hannover},
\cny{Germany}\printead[presep={\\}]{e4}}


\begin{abstract}
Ontology-Based Data Access (OBDA) has traditionally focused on providing a unified view of heterogeneous datasets (e.g., relational databases, CSV and JSON files), either by materializing integrated data into RDF or by performing on-the-fly querying via SPARQL query translation. In the specific case of tabular datasets represented as several CSV or Excel files, query translation approaches have been applied by considering each source as a single table that can be loaded into a relational database management system (RDBMS). Nevertheless, constraints over these tables are not represented (e.g., referential integrity among sources, datatypes, or data integrity); thus, neither consistency among attributes nor indexes over tables are enforced. As a consequence, efficiency of the SPARQL-to-SQL translation process may be affected, as well as the completeness of the answers produced during the evaluation of the generated SQL query. Our work is focused on applying implicit constraints on the OBDA query translation process over tabular data. We propose Morph-CSV, a framework for querying tabular data that exploits information from typical OBDA inputs (e.g., mappings, queries) to enforce constraints that can be used together with any SPARQL-to-SQL OBDA engine. Morph-CSV relies on both a constraint component and a set of constraint operators. For a given set of constraints, the operators are applied to each type of constraint with the aim of enhancing query completeness and performance. We evaluate Morph-CSV in several domains: e-commerce with the BSBM benchmark; transportation with a benchmark using the GTFS dataset from the Madrid subway; and biology with a use case extracted from the Bio2RDF project. We compare and report the performance of two SPARQL-to-SQL OBDA engines, without and with the incorporation of Morph-CSV. The observed results suggest that Morph-CSV is able to speed up the total query execution time by up to two orders of magnitude, while it is able to produce all the query answers.

\end{abstract}


\end{frontmatter}

\section{Introduction}
\label{sec:intro}
Guided by the Open Data principles, governments and private organizations are regularly publishing vast amounts of public data in open data portals. For example, almost a million datasets are available in the European Open Data Portal (EODP)\footnote{\url{https://www.europeandataportal.eu}}, and many of them are available in tabular formats (e.g., CSV, Excel), as observed in Table \ref{tab:odp}. Both the simplicity of a tabular representation and the variety of tools to manage a table (e.g., Excel, Calc) have influenced the popularity of tabular formats to represent open data.  

Albeit extensively utilized, tabular representations imposed various data management challenges to advanced users (e.g., developers, data scientists). The lack of a unified way to query tabular data, something available in other formats (e.g., RDB, JSON, XML), hinders the integration of sources, especially those having datatype inconsistencies. Moreover, data may not be normalized, and information about relationships or column names are not always descriptive or homogeneous. Hence, data consumers are usually forced to apply ad-hoc or manual data wrangling processes to consume data via open data portals. 

Following Linked Data~\cite{bizer2011linked} and FAIR initiatives~\cite{wilkinson2016fair}~\footnote{\url{https://www.go-fair.org/fair-principles/}}, data providers are encouraged to make data available in an RDF-based representation following the 5-star linked data principles\footnote{\url{https://5stardata.info/en/}}. The Ontology-Based Data Access (OBDA)~\cite{poggi2008linking} paradigm facilitates the transformation of heterogeneous data into RDF.
An OBDA corresponds to a data integration system (DIS)~\cite{Lenzerini02} over heterogeneous data sources. A DIS unified schema is defined in terms of ontologies, while mapping rules establish a correspondence between the unified schema concepts and the DIS data sources. An OBDA can be materialized or virtual. In a materialized OBDA, the integration of the DIS data sources is physically represented in RDF~\cite{poggi2008linking}. Contrary, in a virtual OBDA, data integration is performed on the fly during query processing; DIS mapping rules are used to translate SPARQL queries into queries against the DIS data sources~\cite{calvanese2017ontop,priyatna2014formalisation}.
Features like functions in mappings~\cite{de2017declarative,junior2016funul} and metadata~\cite{tennison2015model}, (i.e., annotations) are usually used in materialized OBDAs to overcome the aforementioned challenges of tabular data.

Traditional virtual OBDA approaches, usually, rely on loading tabular data into SQL-based systems\footnote{\url{https://github.com/oeg-upm/morph-rdb/wiki/Usage\#csv-files}}$^,$\footnote{\url{https://ontop-vkg.org/tutorial/mapping/primary-keys.html}}(e.g., MySQL, Apache Drill, Spark SQL, Presto) to perform query translation techniques. However, the correctness and optimization of these techniques are supported by the main assumption about the existence of constraints over the source data (i.e., a good physical design of the relational database instance). Their absence during a virtual OBDA process over tabular data directly impacts  completeness and performance of these techniques. Completeness is affected because of heterogeneity issues in data sources (e.g., datatype CSV columns are simply treated as string-type SQL columns). Furthermore, performance is impacted because indexes are not created based on basic relational constraints, i.e., primary and foreign key constraints are not defined in the schema. Consequently, query translation optimization techniques that commonly exploit indexes (e.g., ~\cite{rodriguez2015efficient,priyatna2014formalisation}) may not produce the expected results whenever the constraints have not been applied, or the indexes have not been created.
\begin{table}[t]
\centering
\caption{Most commonly used formats and percentage over the total number of datasets to expose data in mature EU open data portals in October 2019. Each dataset may be shared in different formats.}
\label{tab:odp}
\begin{tabular}{c|c|c|c}
\hline
\rowcolor{orange!20} 
\textbf{Data Portal} & \textbf{1st Format}  & \textbf{2nd Format} & \textbf{3rd Format} \\ \hline
Spain                & \textbf{CSV (50\%)}  & \textbf{XLS (35\%)}  & JSON (33\%)          \\ 
Norway               & \textbf{CSV (77\%)}    & GEOJSON (17\%)         & JSON (14\%)            \\ 
Italy               & \textbf{CSV (76\%)}  & JSON (35\%)          & XML (25\%)           \\ 
Croatia              & \textbf{XLS (63\%)}    & \textbf{CSV (40\%)}   & HTML (33\%)           \\ \hline
\end{tabular}
\end{table}

OBDA annotations such as the W3C recommendation to annotate tabular data, CSVW~\cite{tennison2015model} and some extensions of standard mapping rules (e.g., RML+FnO~\cite{de2017declarative}) are commonly used to describe constraints over an OBDA tabular dataset. For example, we can standardize a column indicating its format, define integrity constraints, or declare data types. The majority of OBDA query translation engines~\cite{priyatna2014formalisation,endris2019ontario} do not include this information. Those engines that have partially included the constraints (e.g., Squerall~\cite{mami2019squerall} parses RML+FnO mapping rules) are not fully documented; i.e., there is no explanation of how these constraints are taken into account. The definition of a workflow that includes the exploitation of these tabular annotations during a virtual OBDA process will ensure correct and optimized SPARQL-to-SQL translations.

\noindent\textbf{Problem Statement:} 
We address the limitations of current OBDA query translation techniques over tabular data, which enforce and demand lots of unreproducible and hard manual work for the application of constraints to ensure efficient query processing and query completeness\footnote{\url{https://github.com/oeg-upm/morph-rdb/wiki/Usage\#csv-files}}$^,$\footnote{\url{https://ontop-vkg.org/tutorial/mapping/primary-keys.html}}$^,$\footnote{\url{https://ontop-vkg.org/tutorial/mapping/foreign-keys.html}}. Our goals are to (i) define a framework that includes the application of a set of constraints over tabular data, and (ii) define a set of efficient operators that apply each type of constraint to improve query completeness and performance (e.g., removal of duplicates, normalization of input sources or application of transformation functions). 

\noindent\textbf{Proposed Solution:} 
We propose a set of new steps to be aligned with the current OBDA workflow. Further, we implement Morph-CSV, and evaluate its behavior embedded on top of two well known open source SPARQL-to-SQL engines, in comparison with previous approaches.
\begin{figure*}[th]
    \centering
    \includegraphics[width=0.6\linewidth]{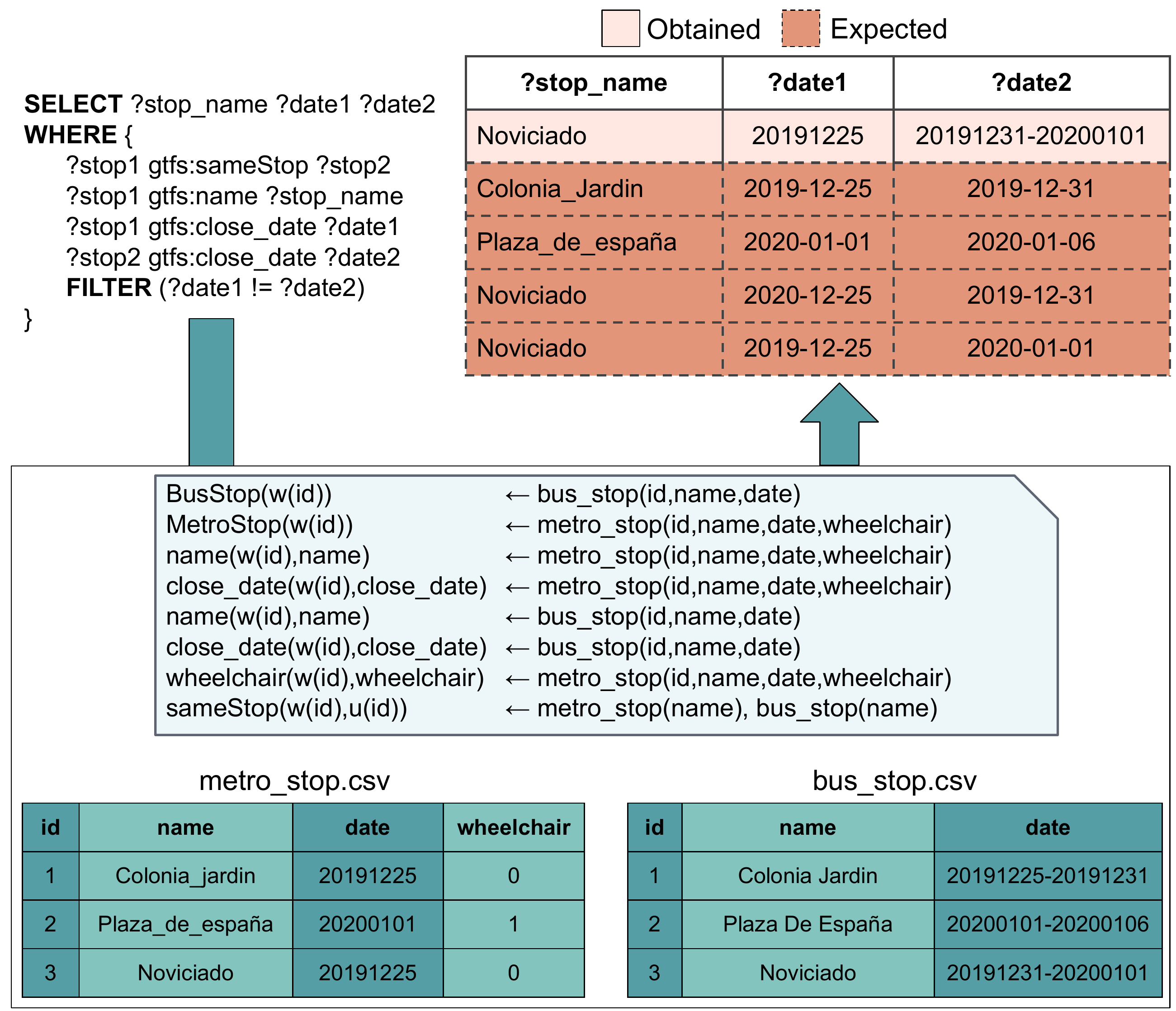}
    \caption{\textbf{Motivating Example.} SPARQL query evaluation over two tabular data files in the transport domain through a common OBDA approach. It loads the files as single tables in an SQL-based system and uses the mapping rules for query translation. The number of results differs with respect to the expected results due to the heterogeneity of the raw data. Additionally, query performance may be affected by the join condition between the two tables, the absence of indexes and the loading of columns that are not needed to answer the input query (wheelchair).}
    \label{fig:example}
\end{figure*}

\noindent\textbf{Contributions:} Our main contributions are as follows:
\begin{enumerate}
\item Definition of the concept of Virtual Tabular Dataset (VTD) composed by a tabular dataset and its corresponding OBDA annotations, as well as its alignment with the current definition and assumptions of the OBDA framework~\cite{xiao2018obdasurvey}.
\item Morph-CSV, a framework that implements a constraint-based OBDA workflow for tabular datasets; it receives a VTD and a SPARQL query as inputs and outputs an OBDA instance. Morph-CSV performs the following steps: (i) generation of the constraints based on information on the VTD; (ii) selection of sources and attributes needed to answer the query; (iii) pre-processing of the selected sources applying some of the constraints; and (iv) physical implementation of the corresponding RDB instance and associated schema, ensuring effectiveness of the SPARQL-to-SQL translations and optimizations. Morph-CSV is engine agnostic, i.e., it can be embedded on top of any SPARQL-to-SQL engine.
\item Evaluation of Morph-CSV re-using in the backend two well-known open source SPARQL-to-SQL engines: Morph-RDB~\cite{priyatna2014formalisation} and Ontop~\cite{calvanese2017ontop}; two benchmarks (BSBM~\cite{bizer2009berlin} and GTFS-Madrid-Bench~\cite{chaves2020gtfs}), and a real-world testbed from the Bio2RDF project~\cite{belleau2008bio2rdf} are used in the study.
\end{enumerate}

The rest of the paper is structured as follows: Section~\ref{sec:example} motivates the problem of OBDA query translation over tabular data with an example in the transport domain. Section~\ref{sec:challenges} describes the identified challenges for querying and integrating tabular data, and current proposals of OBDA annotations for tabular data that address these challenges. Section~\ref{sec:approach} presents Morph-CSV, an approach for enhancing OBDA query translation over tabular data through the application on the fly of a set of constraints. Section~\ref{sec:eval} reports the results of our empirical study together with a general discussion in Section~\ref{sec:discussion}. We present the related work in Section~\ref{sec:related} and our conclusions and future work in Section~\ref{sec:conclusions}.

\section{Motivating Example}
\label{sec:example}
Since May 2017, the publication of a new directive by the EU Commission on discoverability and access to public transport data across Europe\footnote{\url{https://ec.europa.eu/transport/themes/its/road/action_plan/nap}} has motivated the development of solutions for multi-modal travel information services. This document states that transport data should be available through national access points (NAP), e.g., databases, data warehouses, and repositories. Consider the \emph{de-facto standard} for publishing open data in the transport domain, GTFS\footnote{\url{https://developers.google.com/transit/gtfs/reference/}}. This model enables the representation of transport-related concepts such as \textit{schedules}, \textit{stops,} and \textit{routes}, using 15 different inter-related CSV files called GTFS feed. Following best modeling practices recommended in this model specification, each feed comprises entities of one type of transportation mode (e.g., metro, train, and tram). Linking these feeds based on their stops enables route planners to offer multi-modal routes, a route that can be created using various transportation types.
Albeit straightforward and simple to use, GTFS feeds do not allow for the definition of integrity constraints such as primary or foreign keys. As a consequence, data integrity cannot always be guaranteed. 

Consider the GTFS feeds from the metro and buses of Madrid's city; they have several stops and stations in common. Different transport authorities create them, and the names of their stops are defined in various manners. Although these types of entities can be represented, the unique identification and relationships among them cannot be explicitly expressed. Figure~\ref{fig:example} depicts a portion of these two GTFS feeds. As it is usual in open datasets, stop names do not follow a standard structure (e.g., ``Colonia Jardin'' in \textit{bus\_stops.csv} and ``Colonia\_jardin'' in \textit{metro\_stops.csv}). A similar issue is present in closing dates, where there are multi-valued cells, and their format is not the standard one (e.g., yyyy-MM-dd). Suppose a user is interested in collecting information about bus and metro stops with the same name and information related to their closing dates during holidays; Figure~\ref{fig:example} presents the SPARQL query describing this request. Following the approach commonly employed by typical OBDA engines, the two files would be loaded into an SQL-system and treated as single tables. The obtained result set only contains one answer where the stop names in the two data sources are identical (``Noviciado''). However, the expected result set should include more answers by joining among the bus and metro's stop names through the normalization of multi-valued date columns. 

Query's performance may also be affected whenever a join condition is executed between the stop names of both files. Furthermore, the absence of possible indexes in these joining columns makes ineffective the typical optimizations applied in a SPARQL-to-SQL process. Nonetheless, to effectively exploit the indexes to scale-up the execution of the translated queries, the satisfaction of the unique and foreign integrity constraints should be ensured. The manual and ad-hoc definition of the relational schema representing these tables and the corresponding integrity constraints will overcome this problem. Nevertheless, this task is time-consuming, and reproducibility is not ensured. In this paper, we propose Morph-CSV, a constraint-based OBDA framework capable of exploiting standard tabular data annotations (e.g., RML or CSVW) to generate the required constraints ensure the integrity of the tabular schema in terms of unique identifiers and foreign keys. Moreover, Morph-CSV applies metadata annotation from CSVW to generate domain-specific constraints. As a result, Morph-CSV enhances query completeness and performance of SPARQL-to-SQL techniques, in compliance with OBDA assumptions.

\section{Ontology Based Data Access over Tabular Data}
\label{sec:challenges}
This section describes a set of challenges demanded be addressed whenever tabular data is queried in a virtual OBDA framework. Further, we describe relevant OBDA proposals for annotating tabular datasets and their alignment with the identified challenges.

\subsection{Querying challenges under virtual OBDA for tabular data}
There are specific challenges on querying tabular datasets using an OBDA approach that have not been tackled by existing techniques. We will describe those challenges and explain how they may have a negative effect in terms of completeness and performance of query-translation approaches:
\begin{itemize}
    \item \textbf{Updated results:} Existing frameworks load all of the tabular input files that are specified as sources in the OBDA mapping rules into a SQL database before executing the query-translation process. This step has to be repeated whenever a SPARQL query is evaluated to ensure up-to-date results, resulting in unnecessary longer loading time, affecting, thus, OBDA performance.
    \item \textbf{Normalization:} Tabular data formats do not provide restrictions on how to structure data. As a result, cells may contain multiple values, and one file may represent multiple entities. Having non-normalized tables may affect the completeness of the query. When a tabular source with multiple-valued cells is loaded into an RDB table, the cell's value is interpreted by the RDBMS as an atomic value, reducing, thus, completeness for queries that filter or ``join'' on the corresponding column. Representing several entities in a single file may lead to duplicate answers, and in turn, decrease query answering performance.
    \item\textbf{Heterogeneity:} Tabular data normally contain values that need to be transformed before query evaluation (e.g., column default values or normalization of date formats). Since there may be different formats for the same datatype or default values that may have not been included in the dataset, query completeness can be affected.
    \item \textbf{Lightweight Schema:} Most of the tabular data only provide minimal information about their underlying schema in the form of column names in the header, if at all present. Also, although there is implicit information on keys and relationships among sources, there is no way to specify primary key or foreign key constraints. The same can be said on indexes and datatypes. The existence of this type of information is assumed~\cite{xiao2018obdasurvey} in an OBDA approach for performing optimizations in query evaluation techniques. Therefore, the lack of this information affects the performance of OBDA engines.
\end{itemize}

Although some of the aforementioned challenges are not only specific to tabular datasets and are proposed in several data integration approaches~\cite{golshan2017data,halevy2006data,doan2012principles} there are two main reasons why it is important to address these problems in this context: first, as we reflect in Section \ref{sec:intro}, the number of tabular datasets available in the web of data is enormous and still growing and these challenges were not taken into account in previous OBDA proposals; second, although there are declarative proposals to handle these issues in the state of the art like CSV on the Web~\cite{tennison2015model} for metadata annotations, or mapping languages that include transformation functions to deal with heterogeneity (e.g., RML+FnO~\cite{de2017declarative} or R2RML-F~\cite{debruyne2016r2rml}), there is not yet a proposal that exploits the information from these inputs including their application in the form of constraints into a common OBDA workflow.

\begin{table*}[t]
\centering
\caption{Properties of CSVW and RML+FnO that can be used to address the challenges of dealing with tabular data in a virtual OBDA approach}
\label{tab:features}
\resizebox{0.9\textwidth}{!}{%
\begin{tabular}{c|l|l}
\hline
\rowcolor{orange!50} 
\textbf{General Challenge} & \multicolumn{1}{c|}{\textbf{Detailed Challenges}} & \multicolumn{1}{c}{\textbf{Relevant Properties}} \\ \hline
 Updated results & Select relevant sources and columns &  SPARQL + RML+FnO\\ \hline
\multirow{4}{*}{\begin{tabular}[c]{@{}c@{}}Lightweight\\ Schema\end{tabular}} & Describe the corresponding concept & rr:class \\ \cline{2-3} 
 & Describe the corresponding property & rr:predicateMap \\ \cline{2-3} 
 & Specify NOT NULL constraint & csvw:required \\ \cline{2-3} 
 & Column datatype & csvw:datatype \\ \hline
\multirow{7}{*}{Heterogeneity} & Domain values & csvw:minimum, csvw:maximum \\ \cline{2-3} 
 & Specify the format of a column & csvw:format \\ \cline{2-3} 
 & Transform value & fnml:functionValue \\ \cline{2-3} 
 & Default for missing values & csvw:default \\ \cline{2-3} 
 & Specify NULL values & csvw:null \\ \cline{2-3} 
 & Add header to a CSV file & csvw:rowTitles \\ \hline
\multirow{4}{*}{Normalization} & Primary Key & csvw:primaryKey \\ \cline{2-3} 
 & Foreign Key & csvw:foreignKey \\ \cline{2-3} 
 & Relationships between columns & rr:parentTriplesMap + rr:joinCondition \\ \cline{2-3} 
 & Mutiple entities in one source & rr:TriplesMap + rml:logicalSource \\ \cline{2-3} 
 & Support for multiple values in one cell & csvw:separator \\ \cline{2-3} 
 \hline
\end{tabular}%
}
\end{table*}

\subsection{OBDA annotations for tabular data}
R2RML~\cite{das2012r2rml} is a W3C Recommendation for describing transformation rules from RDB to RDF and a widely used mapping language in virtual OBDA approaches. RML~\cite{dimou2014rml} extends R2RML; it provides support to a variety of data formats, e.g., XML, CSV, and JSON. Both languages provide basic transformation functions to concatenate strings, which are especially useful for generating URIs from columns/fields of the dataset. Recently, RML has been integrated with the Function Ontology (FnO)~\cite{de2016ontology} to support other types of transformations. Additionally, for tabular data, CSVW metadata~\cite{tennison2015model} is a W3C Recommendation to describe tabular datasets. Although there are other proposals in the state of the art to deal with some of the aforementioned challenges~\cite{junior2016funul,debruyne2016r2rml}, Morph-CSV relies on these two proposals because they cover the identified challenges. Additionally, this election is supported by the fact that CSVW is a recommendation from the W3C and RML+FnO (besides being a extended version of a W3C recommendation) has been previously applied in other projects~\cite{de2017declarative,mami2019squerall} and is widely used by several materialization engines, e.g.,  RMLMapper\footnote{https://github.com/RMLio/rmlmapper-java}, CARML\footnote{https://github.com/carml/carml/} and RocketRML~\cite{csimcsek2019rocketrml}. Finally, relevant benefits of these annotations are that both of them are defined in a declarative manner. Thus, the maintainability, the readability, and the understanding of the virtual OBDA approach is improved and independent from any specific programming language.

In Table \ref{tab:features}, we summarize the relevant properties from RML+FnO and CSVW that can be used to address the  challenges identified in the previous section. Additionally, we provide a detailed description of these properties:
\begin{itemize}
    \item \textbf{Metadata.} The property \texttt{csvw:rowTitles} can be used to specify column names in case the first row is not used to specify them.
    
    \item \textbf{Transformation functions.} String concatenation functions are supported by both CSVW (\texttt{csvw:aboutUrl}, \texttt{csvw:valueUrl}) and the RML property (\texttt{rr:template}). In addition, more complex functions can be declaratively specified using RML+FnO, specifically, with the \texttt{fnml:functionValue} property. Finally, two special cases of transformation functions in the context of OBDA are related to how default values and NULL representations have to be generated in the RDB instance. These two cases can be handled by CSVW properties: \texttt{csvw:defaultValue} and \texttt{cvwv:null}.
    
    \item \textbf{Domain Constraints.} CSVW allows for the specification of the datatype (\texttt{csvw:datatype} property) and format (\texttt{csvw:format} property) of tabular columns. CSVW also provides a couple of properties (e.g., \texttt{csvw:mininum} or \texttt{csvw:maximum}) to specify the range of numerical columns and a property \texttt{csvw:required} to specify the NOT NULL constraint over the column of a table.
    
    \item \textbf{Integrity Constraints.} In CSVW the property \texttt{csvw:primaryKey} can be used to declare explicitly the primary key of a table. As for the foreign key, the use of RML's properties \texttt{rr:parentTriplesMap} together with the property \texttt{rr:joinCondition} can be seen as an indication that the parent column used over this rule could be a foreign key, or at least that a relation exists. CSVW provides an explicit way to declare whether a column is a foreign key, using the \texttt{csvw:foreignKeys} property. 
    
    \item \textbf{Normalization.} The property \texttt{csvw:separator} from CSVW indicates the character used to separate multiple values in the cells of a CSV column, which is relevant when a CSV file is in 1NF. Multiple RML TriplesMap using the same data source can be used as an indication that the source contains multiple concepts (2NF).
\end{itemize}

\section{The Morph-CSV Framework}
\label{sec:approach}

The formal framework presented in~\cite{xiao2018obdasurvey} defines an OBDA specification as a tuple $P$ = $\langle O,S,M\rangle$ where $O$ is an ontology, $S$ is the source schema, and $M$ a set of mappings. Additionally, an OBDA instance is defined as a tuple $PI$ = $\langle P,D\rangle$ where P is an OBDA specification and $D$ is a data instance conforming to $S$. In a virtual OBDA framework, queries are posed over a conceptual layer and then translated to queries over the data layer using information in the mappings. There is a set of assumptions over the framework that support the possibility of doing query translation and  ensuring  semantic preservation in the process, together with the application of  optimization techniques proposed in the state of the art. To motivate our proposal, we have to establish what are the main assumptions made in previous proposals and their impact when data is represented in tabular form.
\begin{figure*}[t]

\begin{subfigure}{.48\textwidth}
  \centering
  \includegraphics[width=0.9\linewidth]{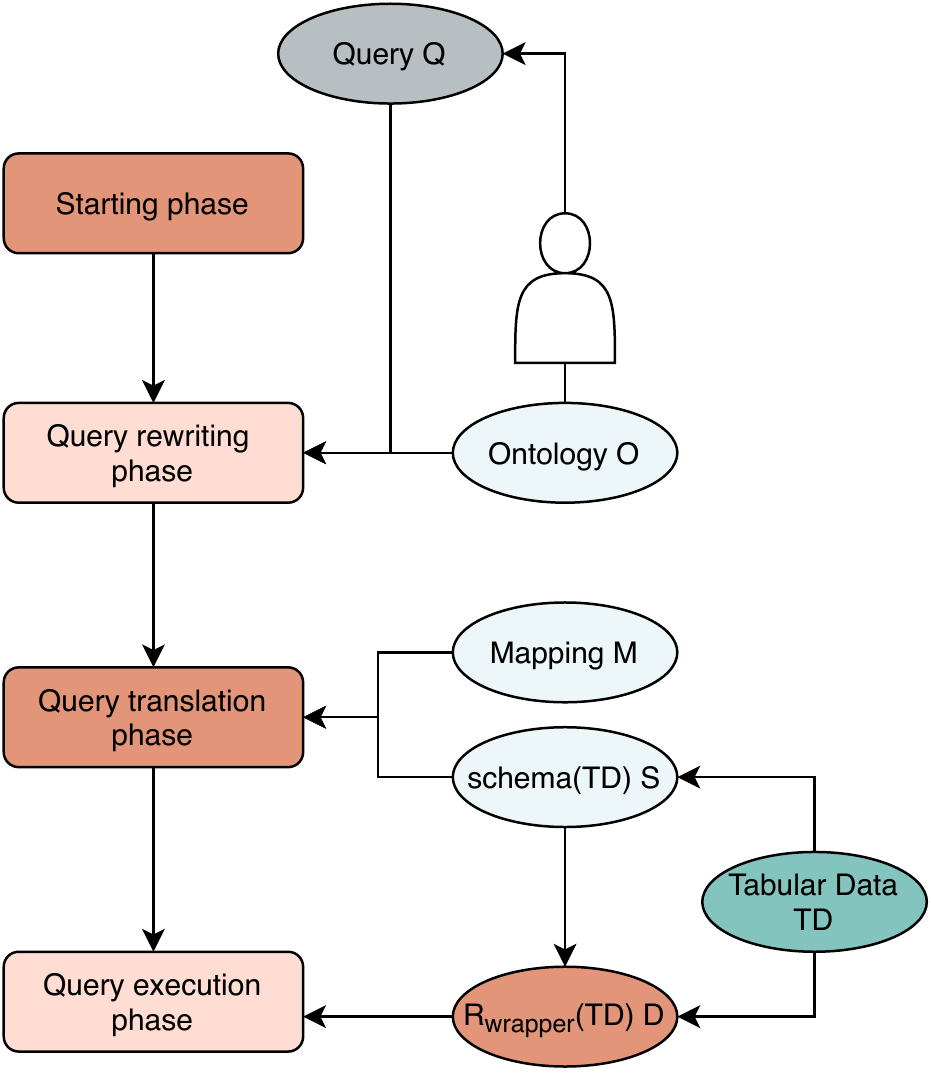}  
  \caption{Baseline approach.}
  \label{fig:naive}
\end{subfigure}
\begin{subfigure}{.48\textwidth}
  \centering
  \includegraphics[width=1\linewidth]{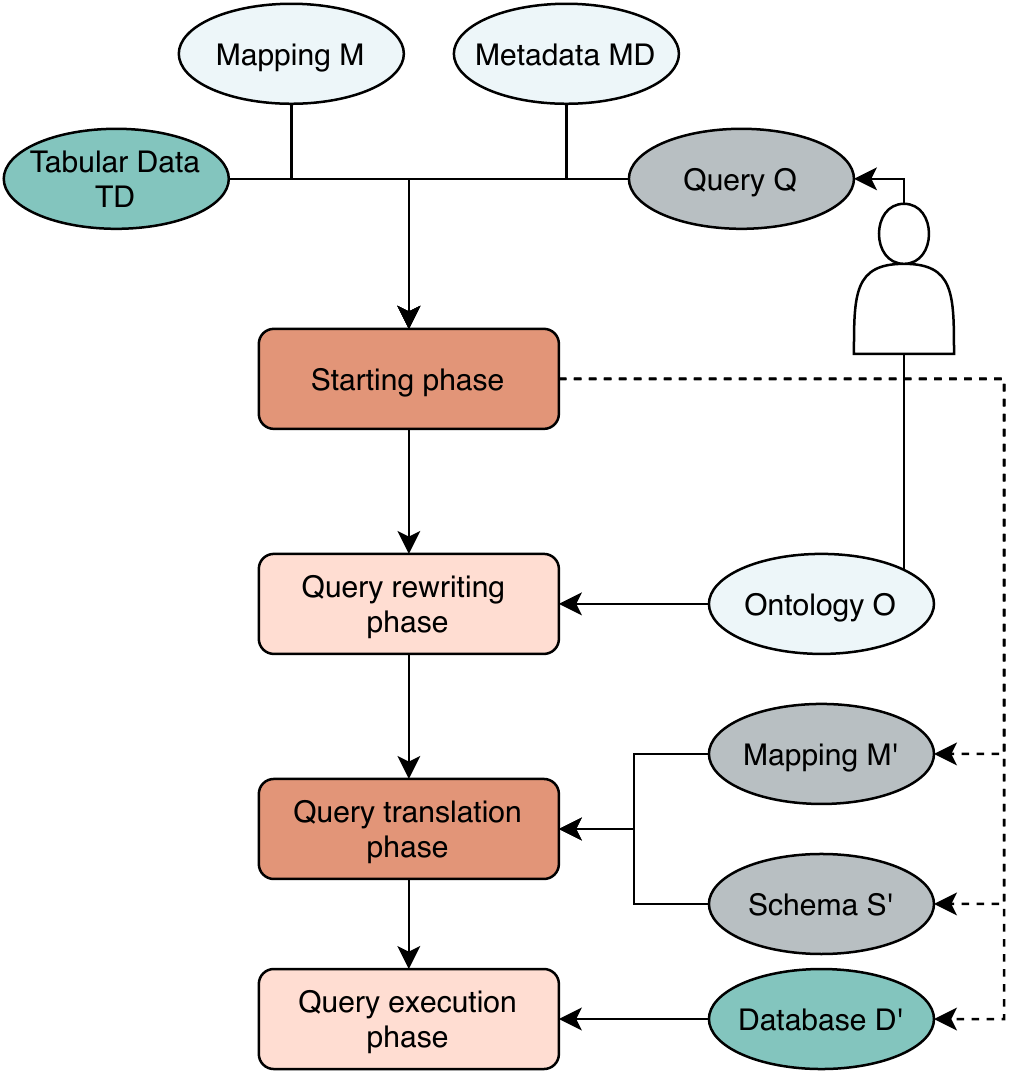} 
  \caption{Enhanced virtual OBDA workflow.}
  \label{fig:vtd}
\end{subfigure}
\caption{\textbf{Virtual OBDA for tabular data approaches.} The baseline approach creates the schema and relational database instance extracting file and columns names from the tabular dataset. The proposed workflow exploits the information from the mapping rules and metadata to extracted a set of constraints and applying them over the tabular data to generate the schema and the relational database instance.}
\label{fig:obda}
\end{figure*}

\subsection{OBDA assumptions}
Analyzing the definition of OBDA in~\cite{xiao2018obdasurvey} and its extension for NoSQL databases defined in~\cite{botoeva2019ontology} we identified a set of assumptions made over the framework and their impact when the dataset is tabular:
\begin{itemize}
    \item There is a native query language $QL$ for $D$. For a tabular dataset, there is no native query language for querying this format, which generates an important difference with other common formats for exposing raw data on the web such as JSON and XML as they include methods to query them (JSONPath, XPath). This is the main issue that needs to be solved in order to query tabular datasets in a virtual OBDA context and has a direct impact on the rest of the assumptions.
    \item $S$ typically includes a set of domain and integrity constraints. In the case of querying a tabular dataset $D_{tabular}$, $S$ is defined using column names extracted from $D_{tabular}$ and it does not include any  constraint types (neither domain nor integrity constraints). This has a negative impact not only in terms of query execution time but also over query result completeness as there will be queries that cannot be executed due to the lack of explicit domain constraints.
    \item $D$ is an RDB instance or  a NoSQL database instance, 
    that includes an RDB wrapper able to provide a relational view over $S$ and $D$. In the context of a tabular dataset $D_{tabular}$, $D$=$R_{wrapper}(D_{tabular})$ where $R_{wrapper}$ is a relational database wrapper that satisfies $S$.
\end{itemize}

\begin{figure*}[t]
    \centering
    \includegraphics[width=1\linewidth]{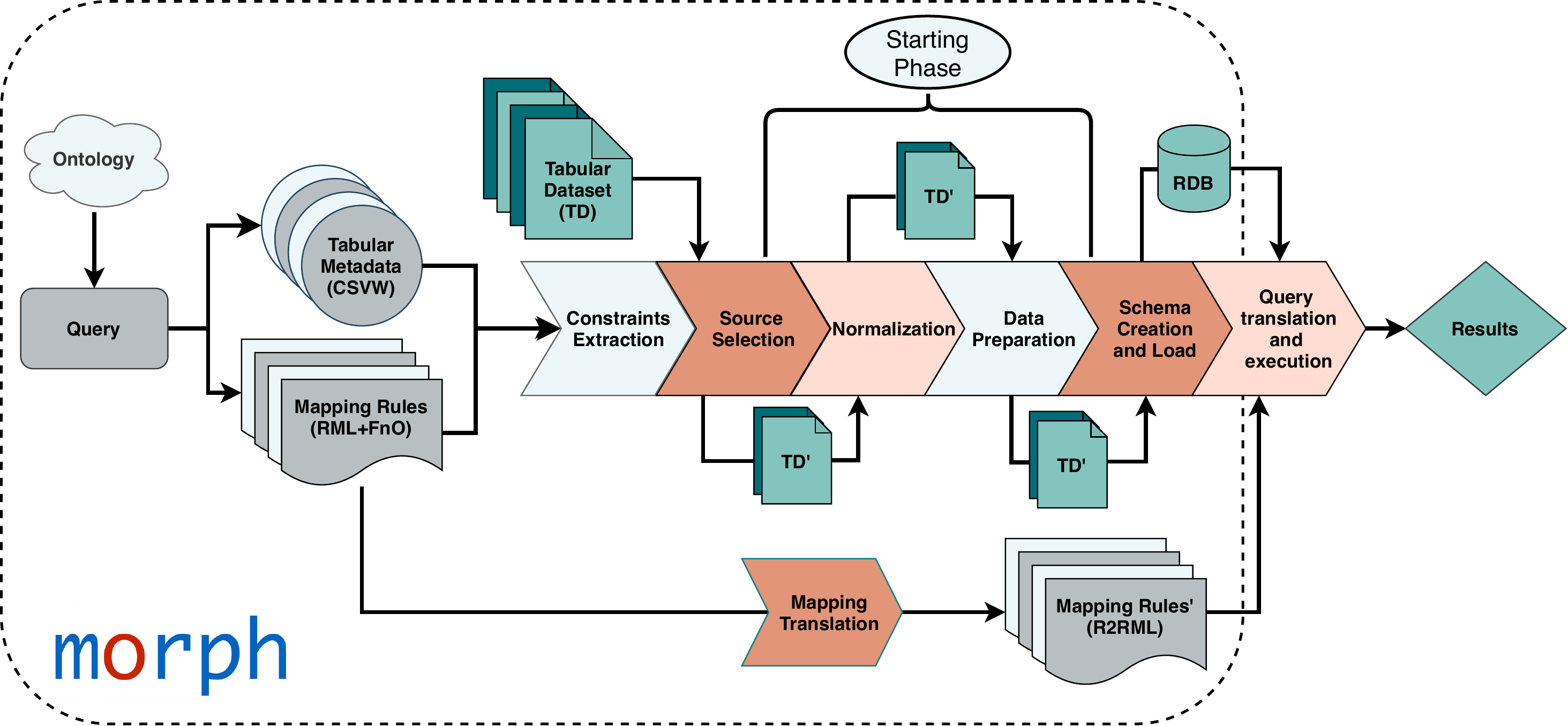}
    \caption{\textbf{The Morph-CSV Framework.} Morph-CSV extends the starting phase of a typical OBDA system including a set of steps for dealing with the identified tabular data querying challenges. The framework first, extracts the constraints from mappings and tabular metadata and then, implements them in a set of operators that are run before executing the query translation and query execution phases, which can be delegated to any SPARQL-to-SQL engine. The mapping rules are translated accordingly to the modified tabular dataset to allow its access by the underlying OBDA engine.}
    \label{fig:workflow}
\end{figure*}
\subsection{From a virtual tabular dataset to an OBDA instance}
Based on the previous OBDA assumptions, we define the concepts and functions to address the problem of querying a tabular dataset in OBDA.
\begin{definition}
\label{def:vtd}
A virtual tabular dataset is defined as a tuple $VTD$=$\langle D_{tabular},O,M,MD\rangle$ where $D_{tabular}$ is a tabular dataset that is composed of a set of data sources, defined as $\mathcal{D}_{tabular}$ = $\{s_1,\ldots, s_n\}$ and where each $s_i$ is a tabular relation defined over the domains of the attributes $Att(s_i)=\{A_{i1},\ldots,A_{im}\}$\footnote{A relation is defined as the subset of the Cartesian product of the domains of the attributes.}, where $m$ is the number of attributes of $s_i$. $O$ is an ontology, and $M$ is a set of global as view mappings between $O$ and $schema(D_{tabular})$\footnote{The set of the attributes of each tabular relation in $D_{tabular}$, i.e., $schema(D_{tabular})=\{Att(s_i),\ldots,Att(s_n)\}$}. $MD$ is a set of metadata tabular (domain) annotations, where for each $s_i$ there exists a set $\{(A_{i1},Type(A_{i1})),\ldots,(A_{im},Type(A_{im}))\}$ in $MD$. 
\end{definition}

\textit{Example 1.} The virtual tabular dataset of the GTFS of Madrid's metro system can be defined as $VGTFS^{metro}_{madrid}$ where the dataset is composed by of 10 different tabular sources in CSV format $GTFS_{tabular}$, $LinkedGTFS$\footnote{\url{https://lov.linkeddata.es/dataset/lov/vocabs/gtfs}} is the ontology, the mappings $RML+FnO_{GTFS}$, following the RML+FnO~\cite{de2017declarative} specification, define the relation between the input sources and the ontology and, finally, the metadata $CSVW_{GTFS}$ is defined according to the W3C recommendation, CSVW~\cite{tennison2015model}, specifying a set of constraints extracted from the GTFS reference data model\footnote{\url{https://developers.google.com/transit/gtfs/reference}}.

Given a $VTD$, we define the function $\theta(VTD)=PI$ where $PI$ is an OBDA instance $PI=\langle P,D\rangle$ where $D$=$R_{wrapper}(D_{tabular})$ and $P=\langle O,S,M\rangle$ is an OBDA definition where $S$ does not contain any type of constraint.
We extend the function $\theta(VTD)$ with the aim of enhancing the virtual OBDA baseline approach over tabular data. We define $\theta^{++}(VTD)$=$PI$ as a function that extracts a set of constraints from $M$ and $MD$ and then applies them over $D_{tabular}$ to obtain $PI$. More in detail, the function can be expressed as $\theta^{++}(VTD)$=$\gamma(D_{tabular},O,M,\psi(M,MD))$ where the function $\psi(M,MD)= C$ extracts a set of constraints from OBDA annotations for tabular data. Then, $\gamma(D_{tabular},O,M,C)$ applies the constraints $C$ over $D_{tabular}$ to create a relational database schema $S^{'}$ and its corresponding instance $D^{'}$. In summary, the final output is an OBDA instance $PI^{'}=\langle P^{'},D^{'}\rangle$, where $D^{'}$ is a relational database instance that is compliant with the main assumptions of the OBDA framework and $P^{'}=\langle O,S^{'},M^{'}\rangle$ where $S^{'}$ contains a set of domain and integrity constraints and $M^{'}$ are the mapping rules that define the relations between $O$ and $S^{'}$. Following the proposed workflow in Figure \ref{fig:vtd}, the user first defines the query based on the concepts defined in the ontology, and then, during the starting phase, the $\theta^{++}(VTD)$ is performed. During the execution of the function, first, the constraints from mappings and annotations ($\psi(M,MD)$) are extracted, and then the OBDA instance $PI^{'}$ is generated where the constraints are applied to efficiently create the schema $S^{'}$ and the relational database instance $D^{'}$. Mapping rules are also translated, from $M$ to $M{'}$ to be aligned with the new created schema.
\begin{table*}[t]
\centering
\caption{Summary of constraints, corresponding functions and OBDA annotations applied by Morph-CSV}
\label{tab:summary}
\resizebox{0.9\textwidth}{!}{%
\begin{tabular}{l|l|l|l|l}
\hline
\rowcolor{orange!50}
\multicolumn{1}{c|}{\textbf{Step}} & \multicolumn{1}{c|}{\textbf{Constraint/Improvement}} & \multicolumn{1}{c|}{\textbf{Rule/Annotation}} & \multicolumn{1}{c|}{\textbf{Function}} & \multicolumn{1}{c}{\textbf{Challenge}} \\ \hline
\multirow{2}{*}{Extraction} & \multirow{2}{*}{Reduce search space} & SSG from Query & select\_annotations & \multirow{2}{*}{Selection} \\ \cline{3-4}
 &  & Mapping Rules & select\_sources &  \\ \hline
\multirow{2}{*}{\begin{tabular}[c]{@{}l@{}}Data \\ Normalization\end{tabular}} & 2NF & csvw:separator & split & \multicolumn{1}{c}{\multirow{2}{*}{Normalization}} \\ \cline{2-4}
 & 3NF & \begin{tabular}[c]{@{}l@{}}TriplesMap with\\ same source\end{tabular} & cut & \multicolumn{1}{c}{} \\ \hline
\multirow{3}{*}{\begin{tabular}[c]{@{}l@{}}Data \\ Preparation\end{tabular}} & \multirow{2}{*}{Standarization} & \begin{tabular}[c]{@{}l@{}}csvw:null, csvw:default\\ csvw:format, etc.\end{tabular} & sub & \multicolumn{1}{c}{\multirow{3}{*}{Heterogeneity}} \\ \cline{3-4}
 &  & fnml:functionValue & create & \multicolumn{1}{c}{} \\ \cline{2-4}
 & Duplicates & - & duplicates & \multicolumn{1}{c}{} \\ \hline
\multirow{4}{*}{\begin{tabular}[c]{@{}l@{}}Schema \\ Creation and\\ Load\end{tabular}} & Primary Key & csvw:primaryKey & primaryKey & \multirow{4}{*}{\begin{tabular}[c]{@{}l@{}}Lightweight \\ Schema\end{tabular}} \\ \cline{2-4}
 & Foreign Key & csvw:foreignKey & foreignKey &  \\ \cline{2-4}
 & DataType & csvw:datatype & datatype &  \\ \cline{2-4}
 & Index & \begin{tabular}[c]{@{}l@{}}selectivity on mapping \\ join conditions\end{tabular} & index &  \\ \hline
\end{tabular}%
}
\end{table*}
\textit{Example 2.} The process of applying the function $\theta^{++}(VGTFS^{metro}_{madrid})$ generates the OBDA instance $PIGTFS^{metro}_{madrid}$. The features of this output are a relational database schema $GTFS_{schema}$, a relational database instance $GTFS_{SQL}$ compliant with the defined schema, and a set of mapping rules following the R2RML W3C recommendation, $R2RML_{GTFS}$, that represent the relations between $GTFS_{schema}$ and the $LinkedGTFS$ ontology.

Constraints are conjunctive rules specified for tabular data that restrict the valid data in one or more tables. $C$ is a set of constraints, where each constraint $c$ is a logical statement that expresses the condition that needs to be satisfied by the data in order to be valid. Each constraint is applied through a function.

\textit{Example 3.} CSVW allows expressing a primary key constraint for a table. The function $\psi(M,MD)=C$ generates the corresponding constraints in the form of a function $primaryKey(t,a)$ that applies this constraint to a source $t$ and a set of columns $a$, and generates a primary key in the output schema. 

Given an OBDA instance $PI$=$\langle\mathcal{P,D}\rangle$, we define the function $eval(Q,PI)$, that retrieves a SPARQL answer set that is the result of the translation of $Q$ from SPARQL to SQL using the mapping rules $M$ defined in $P$, and then evaluating the query directly over $D$.

\subsection{Problem statement and solution}
Based on the preliminaries and assumptions on the OBDA framework, we now define the problem that we address in this paper and Morph-CSV, our proposed solution.

\paragraph{}\noindent\textbf{Problem statement:} Given a $VTD$, the problem of OBDA query translation over tabular data is defined as the problem of explicitly enforcing implicit constraints $C$ extracted from mapping rules $M$ and metadata $MD$ on a tabular dataset $D_{tabular}$, such that: 
\begin{itemize}
    \item The number of results obtained in the evaluation of the SPARQL query $Q$ over the function $eval(Q,\theta^{++}(VTD))$ is equal or greater than the number of results in the evaluation of the same query $Q$ over the function $eval(Q,\theta(VTD))$, i.e., $\#answers(eval(Q,\theta^{++}(VTD))) \geq \\ \#answers(eval(Q,\theta(VTD)))$.
    \item The total execution time of evaluating a SPARQL query $Q$ over $eval(Q,\theta^{++}(VTD))$ is 
    less than or equal than the total execution time of the same SPARQL query $Q$ over the function $eval(Q,\theta(VTD))$, i.e.,\\ $time(eval(Q,\theta^{++}(VTD))) \leq \\
    time(eval(Q,\theta(VTD)))$. 
\end{itemize}

\noindent\textbf{Proposed solution:} We propose Morph-CSV, an alternative to the traditional OBDA workflow for query translation when the input is a tabular dataset (see Figure \ref{fig:vtd} and Appendix \ref{apppendix:algorithm}). Morph-CSV relies on the function $eval(Q,\theta^{++}(VTD,\psi(M,MD)))$, to apply the tabular dataset constrains. Thus, Morph-CSV extends a typical OBDA workflow by including a set of steps for a maintainable extraction and efficient application of constraints. The workflow proposal is as follows:
\begin{itemize}
    \item \textbf{Constraint Extraction}: the evaluation of the function $\psi(M,MD)$ produces as output the set of constraints $C$; it exploits the information defined in the annotations of $M$ and $MD$, i.e., the set of metadata tabular annotations and mapping rules, respectively. At implementation level they are expressed as CSVW specifications and RML+FnO mapping rules. 
    \item \textbf{Source Selection}: in this step the sources required to evaluate the SPARQL query $Q$ are selected. The required data sources correspond to the set of sources in the result of unfolding~\cite{poggi2008linking} $Q$ according to the mapping rules in $M$. 
    \item \textbf{Normalization}: metadata and mapping rules are used to extract functional dependencies between the attributes of the data sources. The algorithm by Beeri et al. \cite{Beeri1978ASI} is followed to transform tabular data sources into tabular relations that meet third normal form (3NF).  
    \item \textbf{Data Preparation}: application of the transformation functions based on the extracted domain constraints and on a set of optimization techniques that adapt the ideas proposed in~\cite{jozashoori2019mapsdi,iglesias2020sdm,jozashoori2020funmap} to a virtual OBDA environment. 
    \item \textbf{Schema Creation and Load}: creation of the schema and loading the data into the database instance applying a set of rules for index creation. 
    \item \textbf{Query Translation and Execution}:the evaluation of the query $Q$ is delegated to any OBDA SPARQL-to-SQL engine.
\end{itemize}
 We show the workflow of Morph-CSV in Figure \ref{fig:workflow} with the inputs and outputs of each step.

\begin{figure*}[t!]
\begin{subfigure}{0.45\textwidth}
  \centering
  \includegraphics[width=1\linewidth]{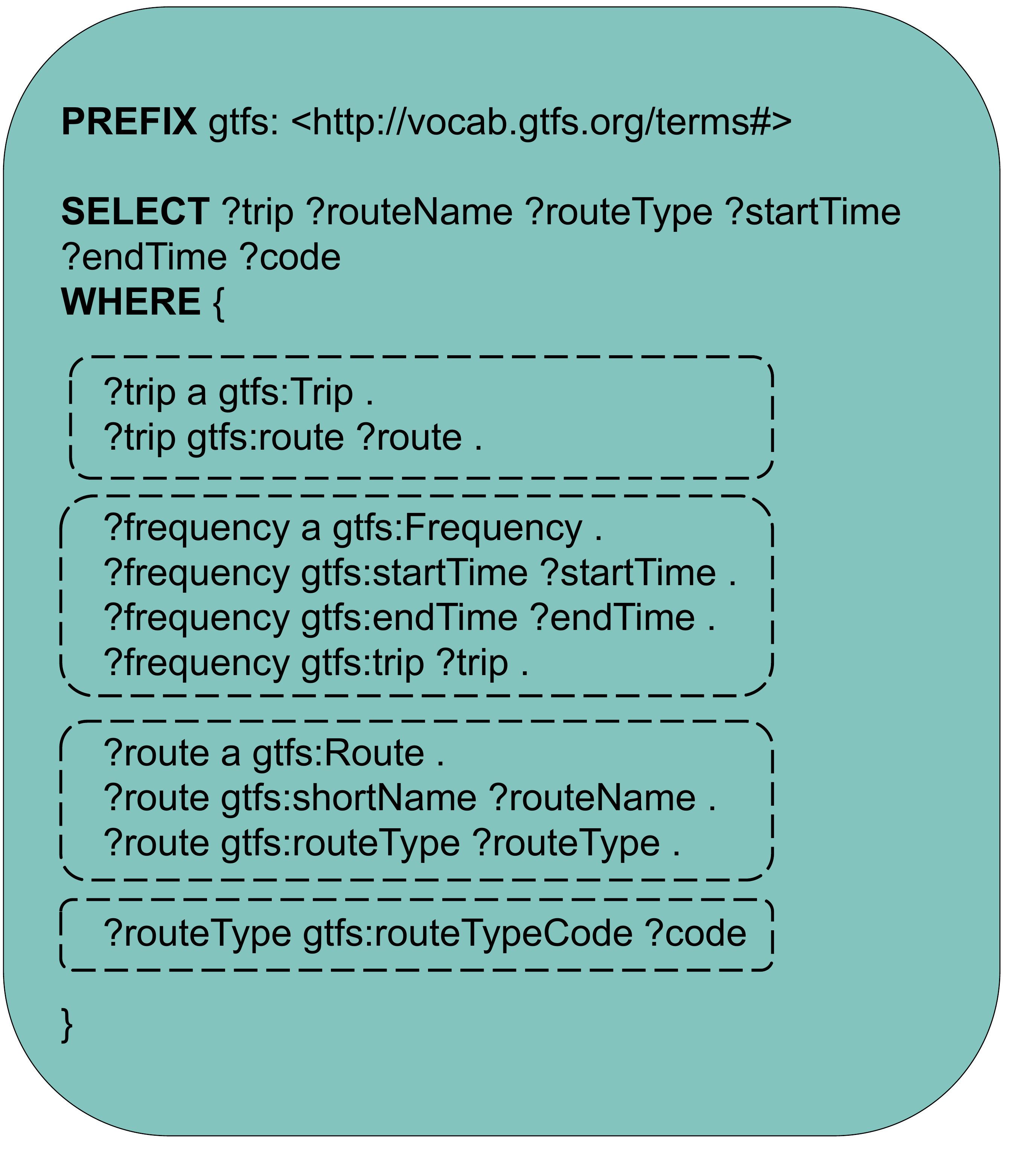}  
  \caption{Input SPARQL query.}
  \label{fig:selectionq}
\end{subfigure}
\begin{subfigure}{0.54\textwidth}
  \centering
  \includegraphics[width=1\linewidth]{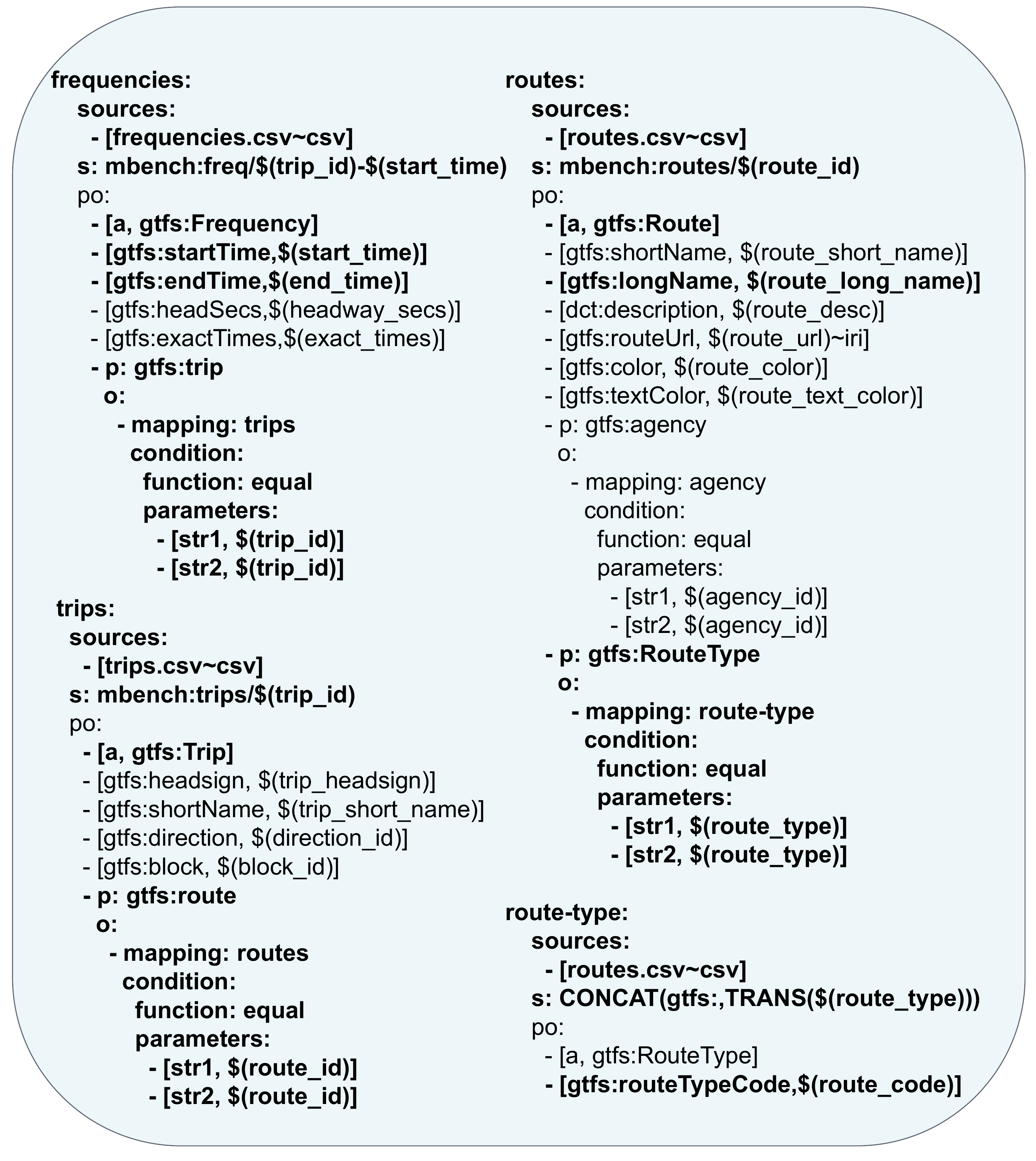}  
  \caption{Mapping rules selection.}
  \label{fig:selectionm}
\end{subfigure}
\caption{\textbf{Selection of Mapping Rules.} Based on the SPARQL query relevant rules are selected (in bold), the rest are discarded. These rules are serialized in YARRML \cite{Heyvaert2018yarrrml}}
\label{fig:selection}
\end{figure*}

\subsection{Steps performed in the Morph-CSV framework}
We describe in detail the steps proposed in Morph-CSV together with an example extracted from the benchmark for virtual knowledge graph access, Madrid-GTFS-Bench, using the query shown in Figure \ref{fig:selectionq}, the GTFS feed from the  Madrid metro as source data, and the corresponding RML+FnO mapping rules and CSVW annotations\footnote{Resources at: \url{https://github.com/oeg-upm/gtfs-bench}}.

\subsubsection*{Constraint Extraction}
The first step performed by Morph-CSV is the extraction of the constraints that are applied to improve query execution and completeness. Morph-CSV benefits from having declarative and standard approaches to generalize this step: CSVW~\cite{tennison2015model} for the metadata; and RML+FnO~\cite{de2017declarative} for mapping rules and specific transformation functions. Thus, maintainability, understandability and readability of this process are improved in comparison with ad-hoc pre-processing approaches. 

Most of the constraints such as PK-FK relations, datatypes or NULL values are explicitly declared in the metadata of the sources. However, there are a set of implicit constraints such as the conditions for the normalization of sources and the creation of indexes, that require complex rules to extract them and that are explained in detail in the corresponding steps. The summary of the constraints, associated functions, and properties used from OBDA annotations to extract them, are shown in Table~\ref{tab:summary}.

\subsubsection*{Source selection}
The second step is to select the relevant sources to answer the input query. The baseline approach delegates this step to the RDBMS: it loads all the sources of the dataset in the RDB instance because it does not have information about which sources are going to be queried. This has a negative impact in the total execution time of a query. Taking the input mapping rules, Morph-CSV performs query unfolding, and pushes down source selection by executing the function $select(Q,M)$, divided into two main steps. First, Morph-CSV performs an operation to select only the relevant annotations for answering the input query, $select\_annotations(Q,M)$. It first creates the set of star shaped groups SSG$_1\ldots $SSG$_n$ of the query~\cite{vidal2010efficiently} (triple patterns with the same subject)\footnote{As usual in these approaches, we assume bounded predicates in the triple patterns}. Then, for each SSG$_i$ and \texttt{rr:TriplesMap} $TM_j$ defined in $M$, the engine selects the $TM_j$ where the predicates in SSG$_i$ are contained in the set of \texttt{rr:PredicateObjectMap} (POMs) defined in $TM_j$. Finally, for each selected \texttt{rr:TriplesMap} $TM_j$, Morph-CSV only selects the POMs according to the predicates defined in the SSG$_i$, hence, removing from each $TM_j$ irrelevant rules for the input query. Using these mapping rules $M^{'}$, only relevant metadata annotations are also selected, $MD^{'}$. The obtained mapping rules in this step, $M^{'}$ and annotations $MD^{'}$, substitute the original ones in $VTD$. An example of this step is shown in Figure \ref{fig:selection}, where the input query asks for trips, their route type, routes names and corresponding time frequencies. Morph-CSV first creates the SSGs, 3 in this case, and using the predicates of each SSG, the \texttt{rr:TriplesMap} are selected from the general GTFS mapping document, discarding the rest of the rules. Then, it only selects the necessary POMs for evaluating the query such as \texttt{gtfs:startTime}, \texttt{gtfs:shortName} and \texttt{gtfs:routeType} (Figure \ref{fig:selectionm}).

\begin{figure}[h]
\begin{subfigure}{0.5\textwidth}
  \centering
  \includegraphics[width=1\linewidth]{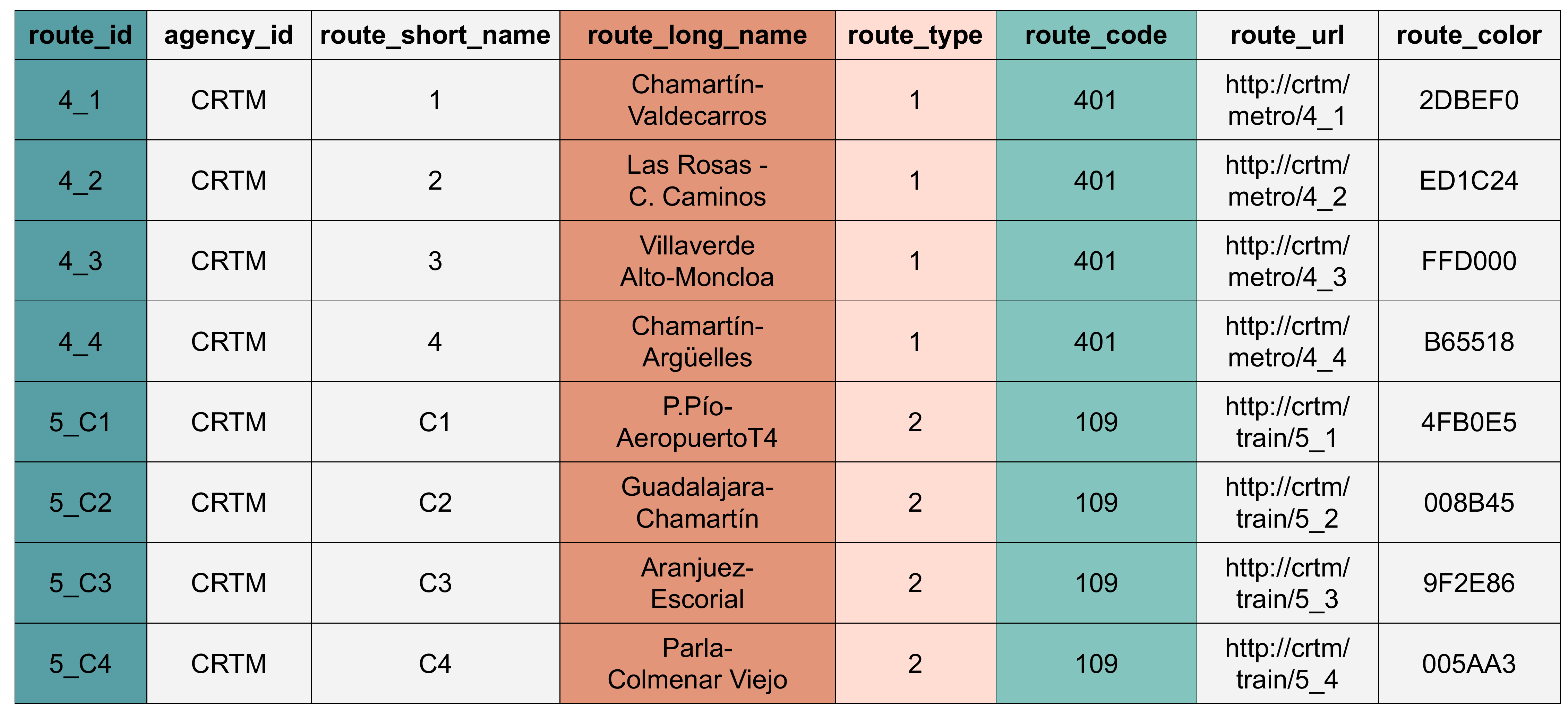}  
  \caption{Original routes.csv input source.}
  \label{fig:selection2i}
\end{subfigure}
\begin{subfigure}{0.5\textwidth}
  \centering
  \includegraphics[width=0.8\linewidth]{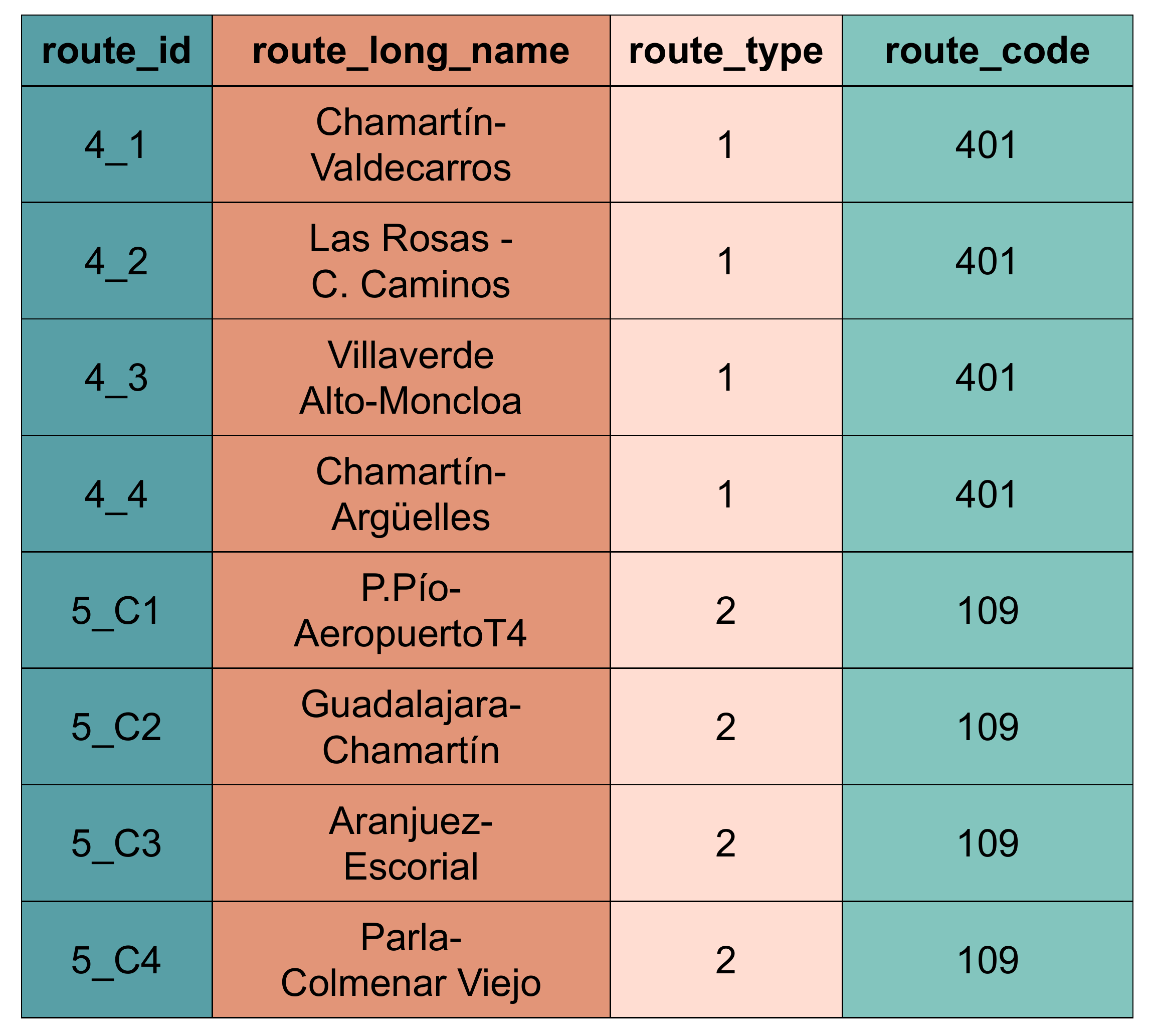}  
  \caption{Output of routes.csv source.}
  \label{fig:selection2r}
\end{subfigure}
\caption{\textbf{Source Selection.} Based on the selection of the rules, only the route\_id and trip\_id columns are selected, discarding the rest of the fields.}
\label{fig:selection2}
\end{figure}

Second, Morph-CSV runs $select\_sources(M)$, where it projects, from the input $D_{tabular}$, the sources and columns that are referenced in $M$, hence, relevant sources for the input query. The output of this function generates a set of new tabular sources $s_i\ldots s_n$ that substitute the original $D_{tabular}$ in $VTD$. Following the previous example, Figure \ref{fig:selection2} shows the selection of the relevant columns of  source \textit{routes.csv}, where Morph-CSV has the original source as  input (Figure \ref{fig:selection2i}), and discards the unnecessary columns of the source based on the mapping rules, obtaining as output the source with the relevant columns for evaluating the input query (Figure \ref{fig:selection2r}). Note that in this step, unnecessary sources from the input GTFS feed such as \textit{agency.csv} and \textit{stops.csv} are also discarded.

\subsubsection*{Normalization}
There are two functions for performing data normalization. The first one is the treatment of multi-values in a column. In this case, Morph-CSV performs the function $split(A_{ij},sep)$ where $A_{ij}$ is the multi-valued column of source $s_{j}$ and $sep$ is the character defined in the CSVW metadata using the \texttt{csvw:separator} property. The output is a modified $VTD$ with a new source $s_t$ containing the separated values in one column with a common identifier $ID_{ij}$ in another column and an $s_{j}^{'}$ source where the values of $A_{ij}$ are substituted by the identifier defined in $s_t$, $ID_{ij}$. Additionally, this function modifies the mapping document $M$ with a new \texttt{rr:TriplesMap} $TM_t$ generated for the new source $s_t$ and a \texttt{rr:joinCondition} between the \texttt{rr:TriplesMap} of $s_j$, $TM_j$ and $TM_t$. The application of this function is known as the normalization step for second normal form (2NF)~\cite{codd1979extending}. The problems of not performing this step are already mentioned in Section \ref{sec:example}, where the multi-valued columns affect  the query completeness.
 
\begin{figure}[t]
\begin{subfigure}{0.5\textwidth}
  \centering
  \includegraphics[width=0.8\linewidth]{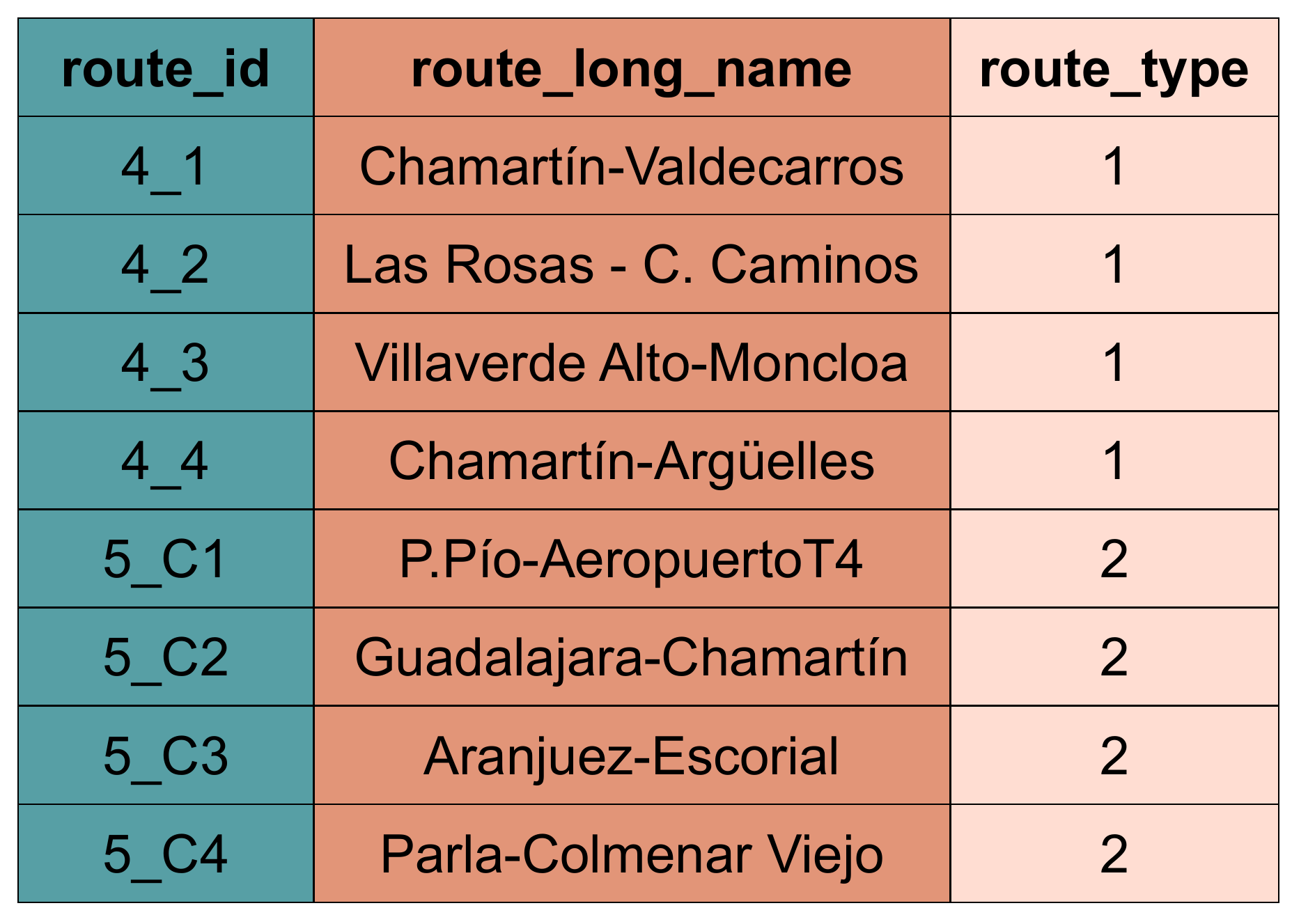}  
  \caption{Routes.csv after the 3NF normalization step.}
  \label{fig:norm1}
\end{subfigure}
\begin{subfigure}{0.5\textwidth}
  \centering
  \includegraphics[width=0.8\linewidth]{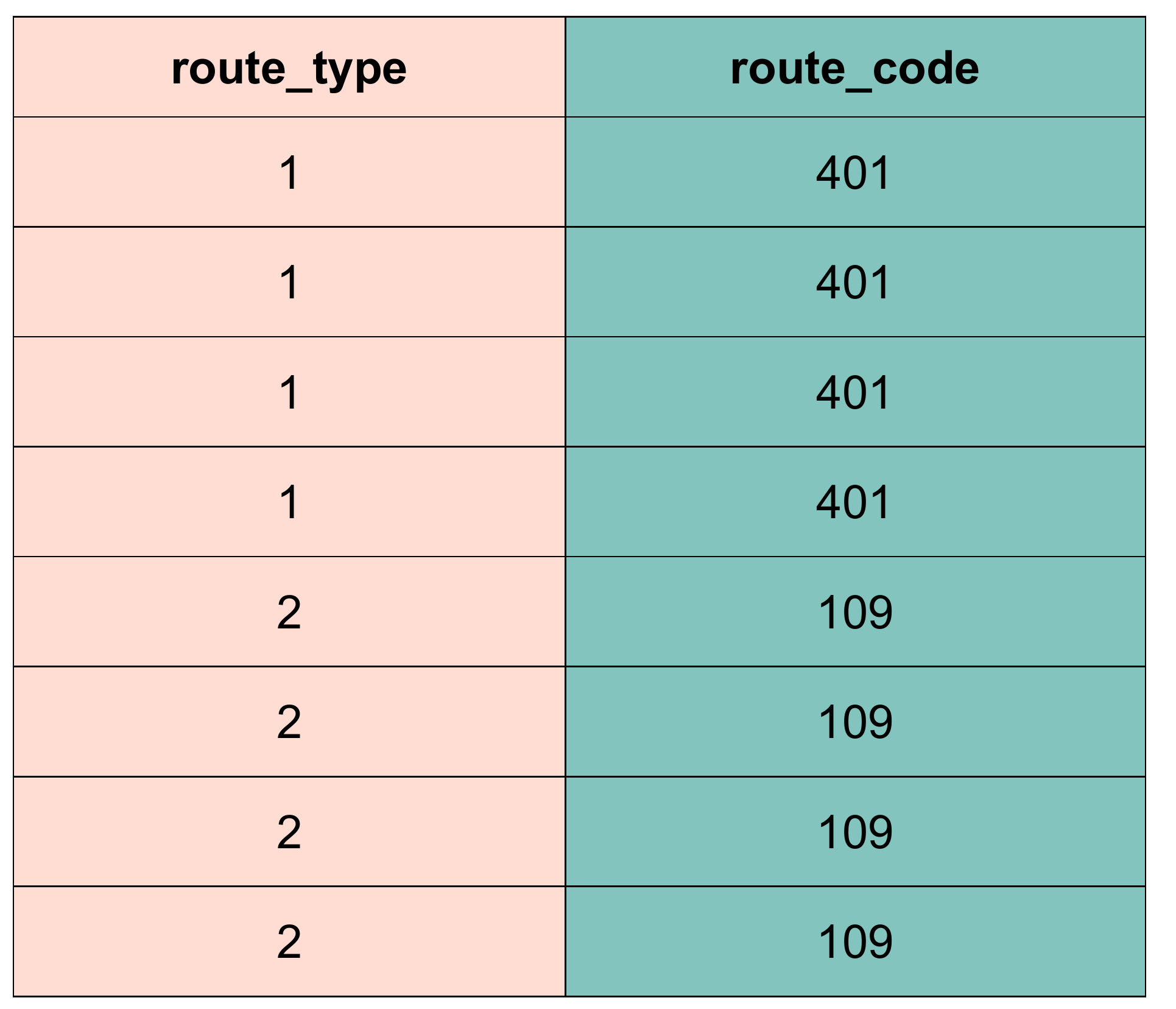}  
  \caption{Route\_type.csv file generated with Morph-CSV.}
  \label{fig:norm}
\end{subfigure}
\caption{\textbf{Normalization.} 3NF Normalization step over the \textit{routes.csv} file generating other file with the data for \texttt{gtfs:RouteType} class.}
\label{fig:normalization}
\end{figure}

The second function is the treatment of multiple entities in the same source. Morph-CSV takes the mapping rules and executes the function $cut(\mathcal{M},\mathcal{D}_{tabular})$. This function analyzes the mapping rules $\mathcal{M}$, and performs a 3NF~\cite{codd1979extending} normalization step over ${D}_{tabular}$ when there are two sets of mapping rules ($TM_j$ and $TM_i$) that have the same source, and the intersection of their columns in the rules only contains the join condition references. Following a similar approach as in 2NF, the output is a modified $VTD$ with a set of new sources $s_i\ldots s_n$, each one with the corresponding columns of each entity. For example, in Figure \ref{fig:normalization} we show the 3NF normalization of the \textit{routes.csv} file, that generates an auxiliary source for the \texttt{rr:TriplesMap} with the \textit{gtfs:RouteType} entity data (Figure \ref{fig:normalization}), removing that information from \textit{routes.csv}. In several data integration approaches, normalization steps are not taken into account in order to improve query execution (reducing the number of joins among sources). However, in the case of RDF, where each entity of a class has a unique URI (subject), joins cannot be reduced (see input mapping in Figure \ref{fig:selectionm}). This means that taking into account normalization steps in an OBDA context not only helps to improve query completeness, but also helps to improve performance. Additionally, normalization is also essential for allowing Morph-CSV to efficiently run data preparation steps, as we show in the next step.

\subsubsection*{Data preparation}
\begin{figure}[t]
    \centering
    \includegraphics[width=1\linewidth]{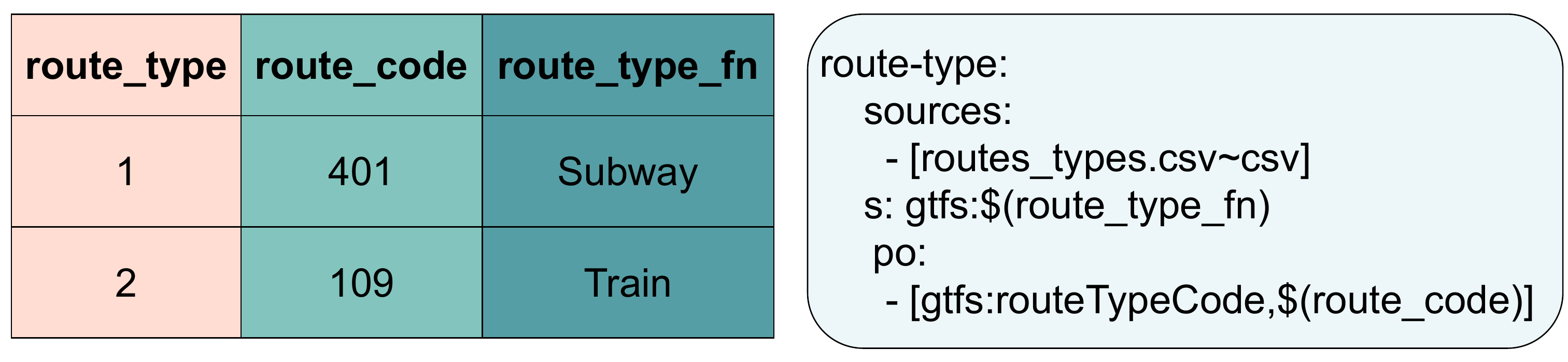}
    \caption{Data preparation of \textit{route-types.csv} file.}
    \label{fig:preparation}
\end{figure}

In this step, Morph-CSV addresses the challenge of \textit{Heterogeneity} and executes three different functions: $duplicates$, $sub$ and $create$. First, Morph-CSV removes all duplicates in the raw data, not only the original ones, but also other duplicates that can appear during the normalization step (see Figure \ref{fig:norm}). It applies the ideas described in~\cite{jozashoori2019mapsdi}, performing $duplicates(s_{j})$ where $s_{j}$ is a source in $D_{tabular}$. As it has already been demonstrated in~\cite{jozashoori2019mapsdi,iglesias2020sdm,jozashoori2020funmap}, this step not only has a high impact on the behavior of these engines, but in this case, it also reduces the number of operations performed by Morph-CSV $sub$ and $create$, as they are defined as deterministic functions. The first one is defined as $sub(exp(A_{ij}),val)$ where $exp(A_{ij})$ is a boolean function over column $A_{ij}$ of source $s_{j}$ that when true, the value of $A_{ij}$ is substituted by $val$. There are multiple substitution functions that Morph-CSV executes such as default values, null values and date formats. This function is one of the most important for enhancing the completeness of the query (e.g., enforcing the default values of a column). The second function creates a new column in a specific source $s_{j}$. It is defined as $create(c(A_{nj},\ldots,A_{mj}))$, where $c(A_{nj},\ldots,A_{mj})$ is the application of a set of transformation functions over the columns $A_{nj},\ldots,A_{mj}$ in source $s_{j}$. This function is used to push down the application of ad-hoc transformation functions, usually defined inside the mapping rules~\cite{junior2016funul,de2017declarative}, thus, avoiding the incorporation of them inside the SQL translated query. In Figure \ref{fig:preparation} we show the \textit{route\_type.csv} file after the execution of this step. First, Morph-CSV removes the duplicates of the file obtaining as output a file with only two rows. Then, it executes the transformation function defined in the mapping rules and creates a new column in the file, generating the desired value for the subject of the class according to the LinkedGTFS ontology, ``Subway''. Additionally, the engine substitutes the definition of the transformation functions in the mapping rules by a reference to the created column. In this manner, Morph-CSV efficiently performs the $sub$ and $create$ functions directly over the raw data and together with the normalization step. Thus, the number of joins in the input query is reduced.

\subsubsection*{Schema creation and load}
\begin{figure}[t]
    \centering
    \includegraphics[width=1\linewidth]{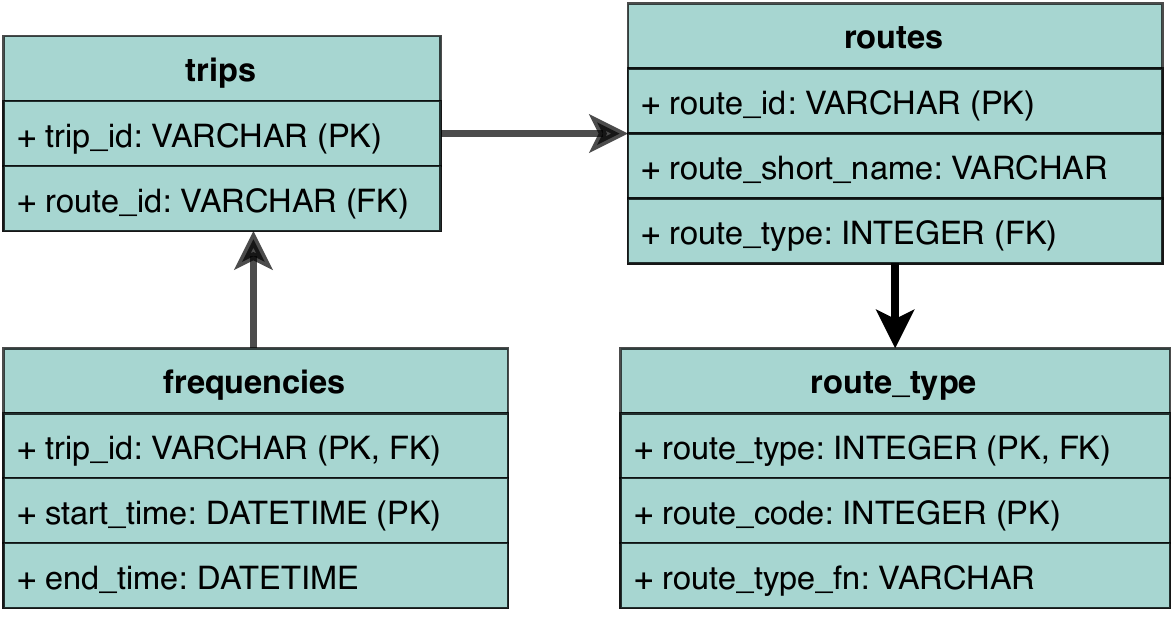}
    \caption{\textbf{Generated schema.} The schema generated by Morph-CSV extracting domain and integrity constraints from the annotations and based on the identified sources selected from the input query.}
    \label{fig:rdb}
\end{figure}

The final step before translating and executing the query is the creation of an SQL schema applying the rest of the identified constraints, and loading the selected tabular data sources. Besides the typical integrity constraints that can be extracted from CSVW annotations (PK/FK), Morph-CSV implements a rule for creating indexes in the RDB instance in order to optimize the execution of query joins. In tabular datasets, it is common that the join conditions defined in the mapping rules are based on columns that are not part of PK-FK relations; thus, they are not indexed and OBDA optimizations do not have the desired effect. To address this problem, Morph-CSV gets the \texttt{rr:child} and \texttt{rr:parent} references of the mapping rules and calculates their selectivity on the fly. Then, taking this selectivity into account Morph-CSV decides to create, or not, an index over these columns. Additionally, the mapping document is translated so that it is aligned with the RDB schema that has been created. Figure \ref{fig:rdb} shows the RDB schema generated by Morph-CSV for the input query in Figure \ref{fig:selectionq}, with the applied domain and integrity constraints. 

\paragraph{}
There are two main points that make the contributions of Morph-CSV relevant: (i) it incorporates the steps to the standard OBDA workflow without modifying the rest of the steps, hence, it can also benefit from optimizations in other steps of the workflow like query rewriting (reasoning)~\cite{mora2014kyrie2} or query translation (SPARQL-to-SQL)~\cite{priyatna2014formalisation}, and (ii) the reliance of the approach on declarative and standard annotations for OBDA allows the generalization of the proposed steps, usually solved in an ad-hoc manner, not only automatizing the process but also improving its maintainability, understandability and readability.

\section{Evaluation}
\label{sec:eval}

This section reports on the results of the empirical evaluation conducted to test the effect of respecting constraints, on the fly, during OBDA query translation over tabular data. The hypotheses we want to validate in our work are:
\begin{itemize}
\item \textit{H1)} The application of a set of domain and integrity constraints over tabular data sources to create an RDB instance ensures the effectiveness of SPARQL-to-SQL optimizations proposed in the state of the art.
\item \textit{H2)} Extending a common OBDA workflow with a set of additional steps to deal with the challenges for querying tabular data,  does not  impact negatively on the total query execution time.
\item \textit{H3)} The exploitation of declarative and standard annotations in the process of querying tabular data in an OBDA environment, guarantees the independence of the solution and its application over different domains.
\end{itemize}
Aligned with the defined hypothesis, our aim is to answer the following research questions: \textbf{RQ1)} What is the effect of combining different types of constraints over a tabular dataset? \textbf{RQ2)} What is the impact of the constraints when the tabular dataset size increases? \textbf{RQ3)} What is the effect of different kinds of SPARQL query shapes in the extraction and application of constraints?. To answer these questions, we have performed three evaluations in different domains: e-commerce, transportation, and biology. Our first evaluation is in the e-commerce domain, in which we used the Berlin SPARQL Benchmark (BSBM)~\cite{bizer2009berlin}. Our second evaluation is in the transportation domain in which we used the GTFS-Madrid-Bench~\cite{chaves2020gtfs}. This benchmark focuses on measuring the performance of ontology based data access for heterogeneous data sources, based on the publicly-released public transportation data in GTFS format. One of the resources provided by GTFS-Madrid-Bench is a tabular dataset together with its corresponding mappings and annotations together with a set of representative SPARQL queries. Finally, our third evaluation is in the domain of biological data, in which we extend one of our previous proposals~\cite{iglesias2019enhancing} for the generation of an OBDA layer over Bio2RDF tabular datasets. Appendix \ref{apppendix:queries} presents the features of the queries together with the constraints and number of sources used by Morph-CSV. In all of the evaluations the common configurations are:

\noindent\textbf{Engines.} The baselines of our study are two open source SPARQL-to-SQL OBDA engines: Ontop\footnote{\url{https://github.com/ontop/ontop}}$^,$\footnote{We modified the default configuration of Ontop extending the maximum used memory from 512Mg to 8Gb} v3.0.1 and Morph-RDB v3.9.15\footnote{\url{https://github.com/oeg-upm/morph-rdb}}. We select these two engines as they are open source engines (others such as Ultrawrap~\cite{sequeda2013ultrawrap} are not openly available) and also the ones that incorporate the set of most relevant optimizations in the SPARQL-to-SQL query translation process~\cite{priyatna2014formalisation,rodriguez2015efficient}.  To evaluate the baseline approach, we manually generate the relational database schemes of each benchmark without any kind of constraints, and measure the load and query execution times. In order to measure the impact of the additional steps proposed by Morph-CSV\footnote{\url{https://doi.org/10.5281/zenodo.3731941}}$^,$\footnote{\url{https://github.com/oeg-upm/morph-csv}}, we integrate our solution on top of the two OBDA engines in two different configuration: Morph-CSV$^-$ that does not include the source selection step, hence, it loads and applies all the constraints over the input data source each time a query has to be answered, and Morph-CSV that implements the full proposed workflow\footnote{We name the combined engines as follows: a) Morph-CSV: Morph-CSV+Morph-RDB, and Morph-CSV+Ontop; b) Morph-CSV$^-$: Morph-CSV$^-$+Morph-RDB, and Morph-CSV$^-$+Ontop}. To ensure the reproducibility of the experiments, we also provide all of the resources in a docker image.

\noindent\textbf{Metrics.} We measure the loading time of each query and the total query execution time (including the steps proposed by Morph-CSV or baseline when appropriate), and the number of answers obtained (see Appendix \ref{appendix:completeness}). Additionally, we detail the times of each proposed step of our workflow in the execution of each query using Morph-CSV in both configurations (see Appendix \ref{appendix:loadingtime}) following the recommendations proposed in the GTFS-Madrid-Bench~\cite{chaves2020gtfs}. Each query was executed 5 times with a timeout of 1 hour in cold mode, that means that the corresponding database is generated each time a query is going to be evaluated in order to ensure up to date number of answers. Regarding the completeness of the queries, both BSBM benchmark and GTFS-Madrid-Bench provide an RDF materialized version of the input sources that has been loaded in a triplestore (Virtuoso in the case) and used as gold standard. To analyze the completeness of each query, we compare the cardinality of the result set of each configuration against the gold standard assuming its correctness. In the case of the Bio2RDF use case, we cannot compare our results with any gold standard as the last dump version of the project~\cite{dumontier2014bio2rdf} is not comparable with the current status of the input sources, as we declare in one of our previous works~\cite{iglesias2019enhancing}. The experiments were run in an Intel(R) Xeon(R) equipped with a CPU E5-2603 v3 @ 1.60GHz 20 cores, 64GB memory and with the O.S. Ubuntu 16.04LTS.

\subsection{BSBM}
The Berlin SPARQL Benchmark~\cite{bizer2009berlin} is one the most popular benchmarks in the Semantic Web field that not only tests the performance of RDF triple stores, but also tests approaches that perform SPARQL-to-SQL query translations providing an RDB instance. It is the chosen benchmark to test the capabilities of many state-of-the-art OBDA engines~\cite{priyatna2014formalisation,calvanese2017ontop,mami2019squerall}. 
 
\begin{figure*}[th]

\begin{subfigure}{.48\textwidth}
  \centering
  \includegraphics[width=1\linewidth]{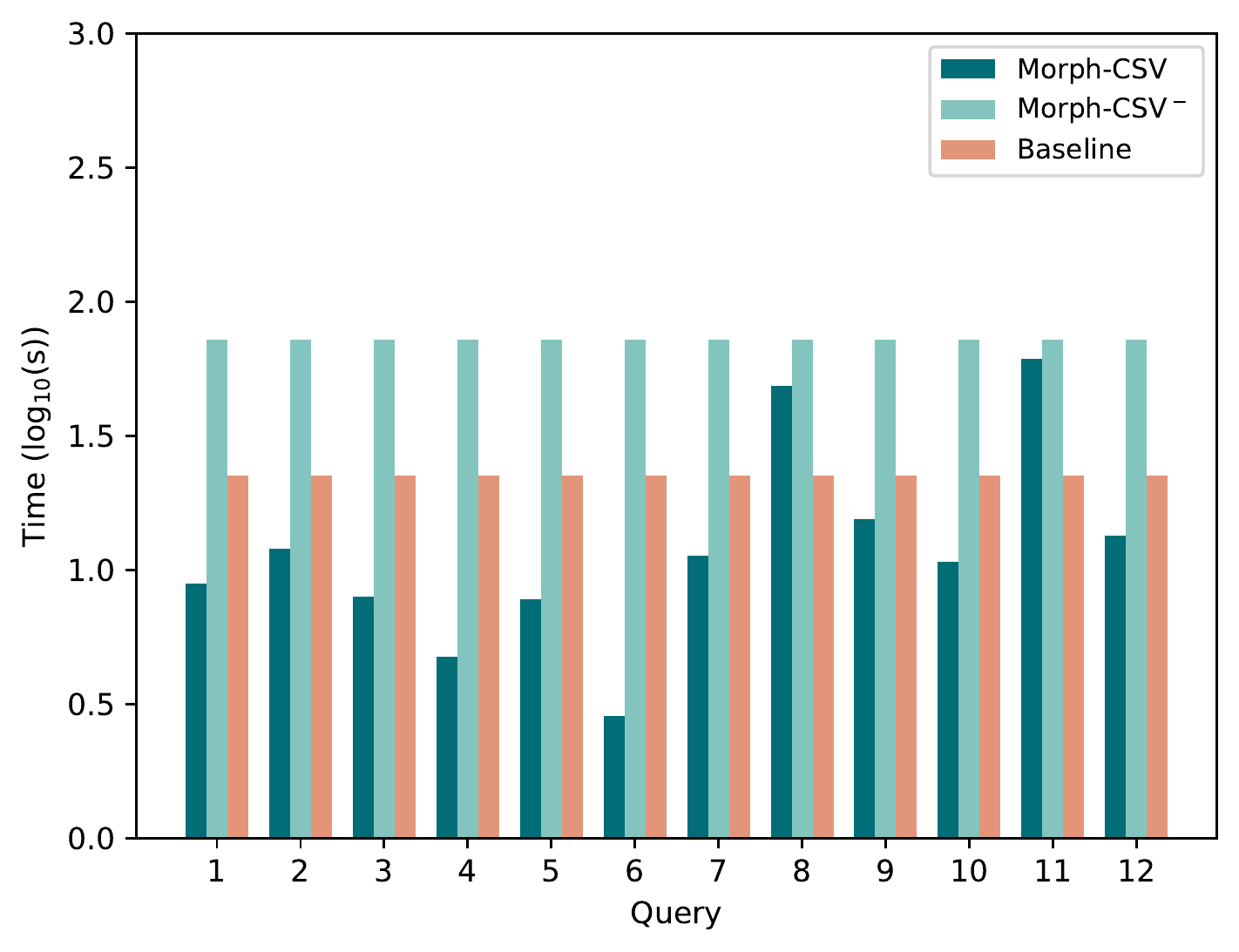}  
  \caption{Loading time for BSBM 45K.}
  \label{fig:bsbmload45}
\end{subfigure}
\begin{subfigure}{.48\textwidth}
  \centering
  \includegraphics[width=1\linewidth]{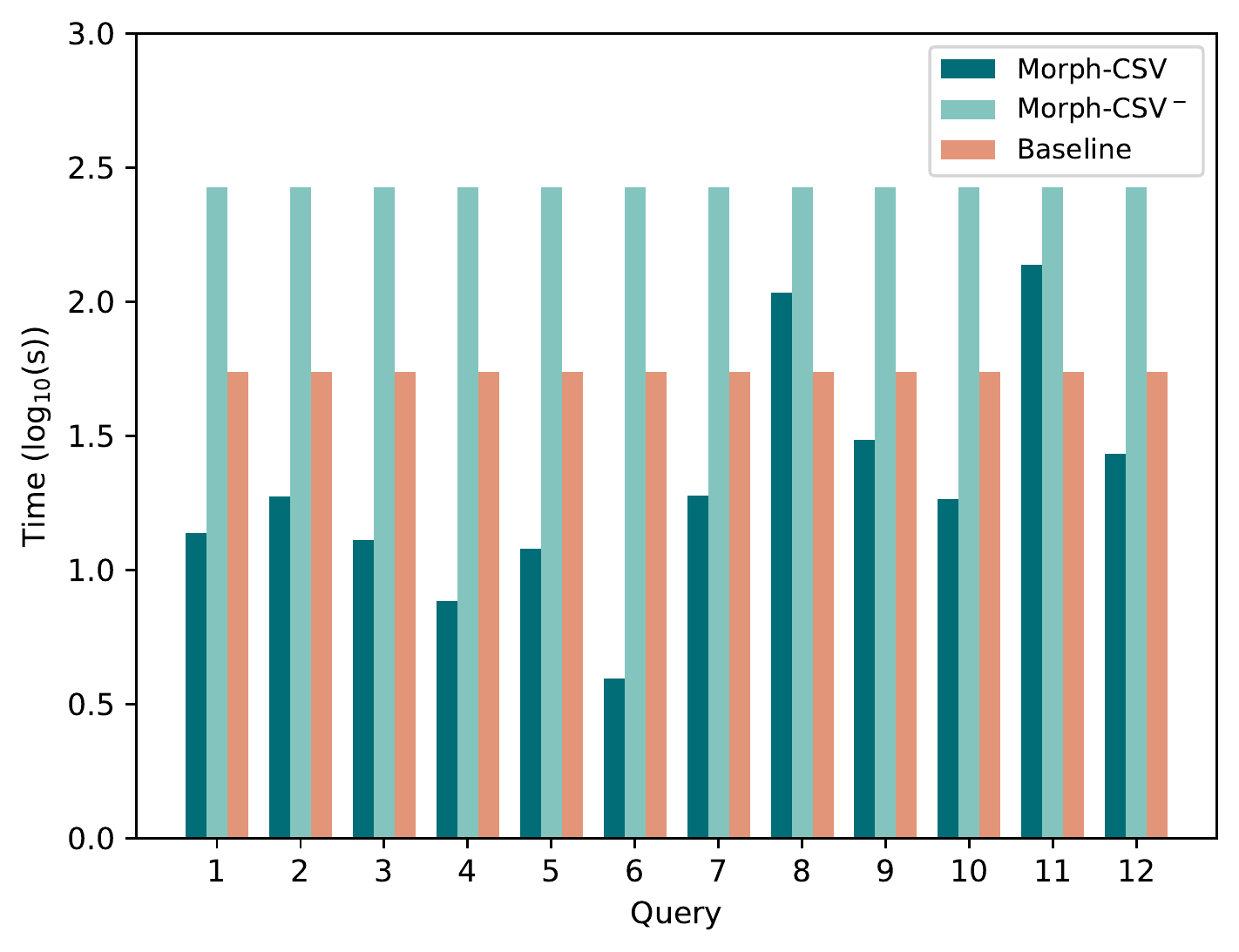}  
  \caption{Loading time for BSBM 90K.}
  \label{fig:bsbmload90}
\end{subfigure}
\begin{subfigure}{.48\textwidth}
  \centering
  \includegraphics[width=1\linewidth]{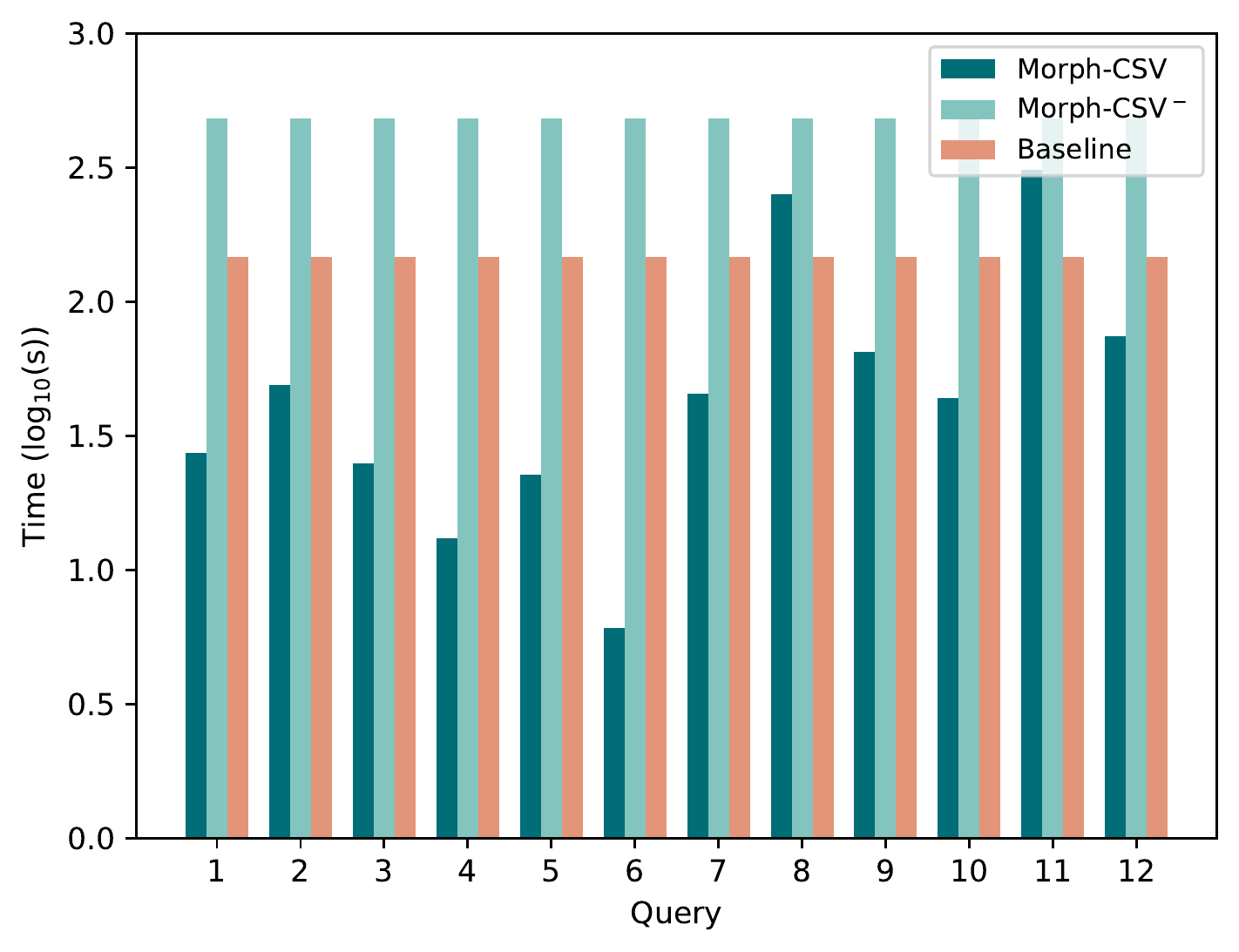}  
  \caption{Loading time for BSBM 180K.}
  \label{fig:bsbmload180}
\end{subfigure}
\begin{subfigure}{.48\textwidth}
  \centering
  \includegraphics[width=1\linewidth]{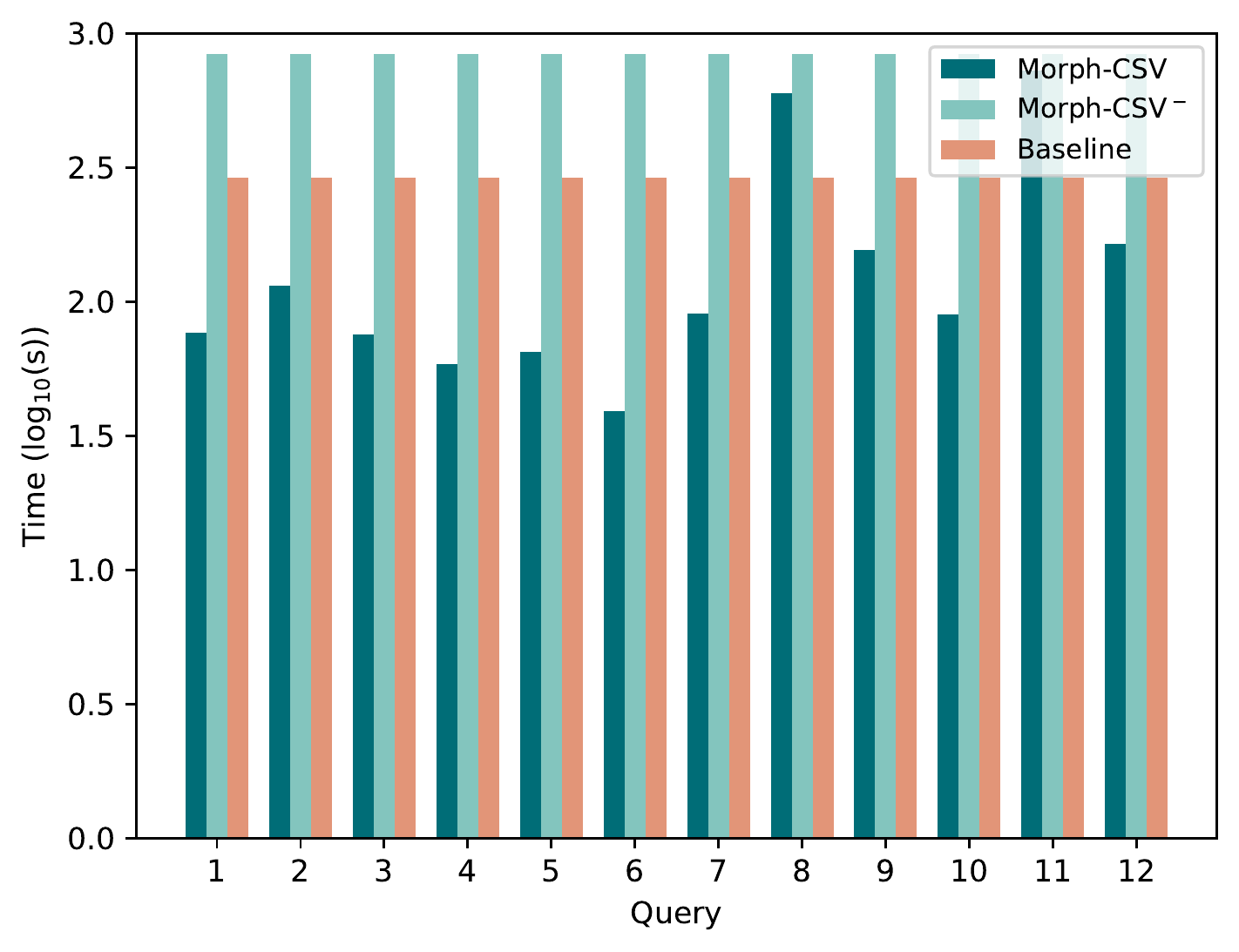}
  \caption{Loading time for BSBM 360K.}
  \label{fig:bsbmload360}
\end{subfigure}
\caption{\textbf{Loading Time of Tabular Datasets in BSBM.} Loading time in seconds of the tabular datasets from the BSBM benchmark with number of products 45K, 90K, 180K and 360K. The baseline approach (red columns) and Morph-CSV$^-$ (light green) are constant for each dataset and query, while Morph-CSV (dark green) depends on the query and number of constraints to be applied over the selected sources.}
\label{fig:bsbmload}
\end{figure*}

\begin{figure*}[th]

\begin{subfigure}{.48\textwidth}
  \centering
  \includegraphics[width=1\linewidth]{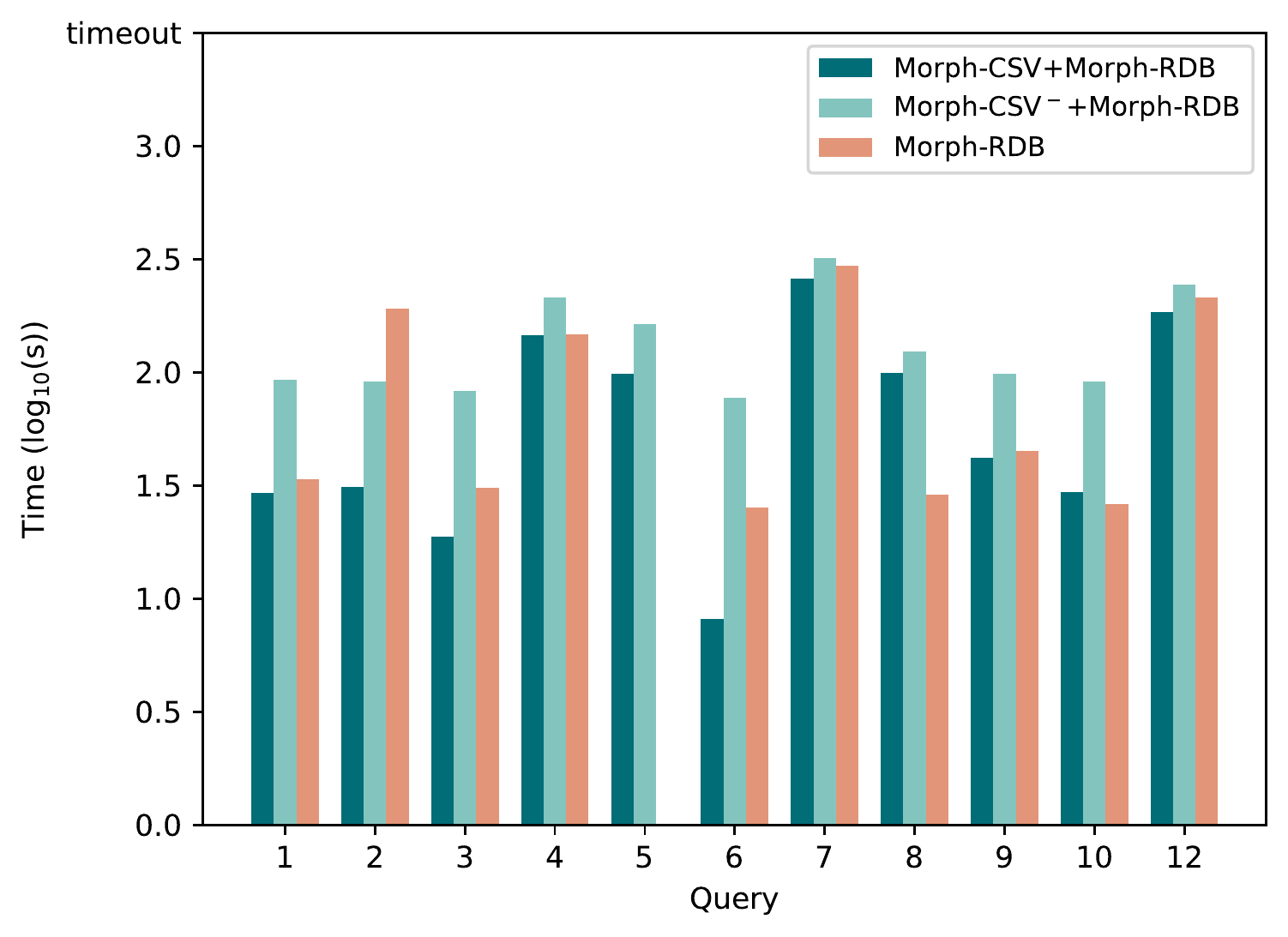}  
  \caption{Query execution time for BSBM-45 with Morph-RDB.}
  \label{fig:morphbsbm45}
\end{subfigure}
\begin{subfigure}{.48\textwidth}
  \centering
  \includegraphics[width=1\linewidth]{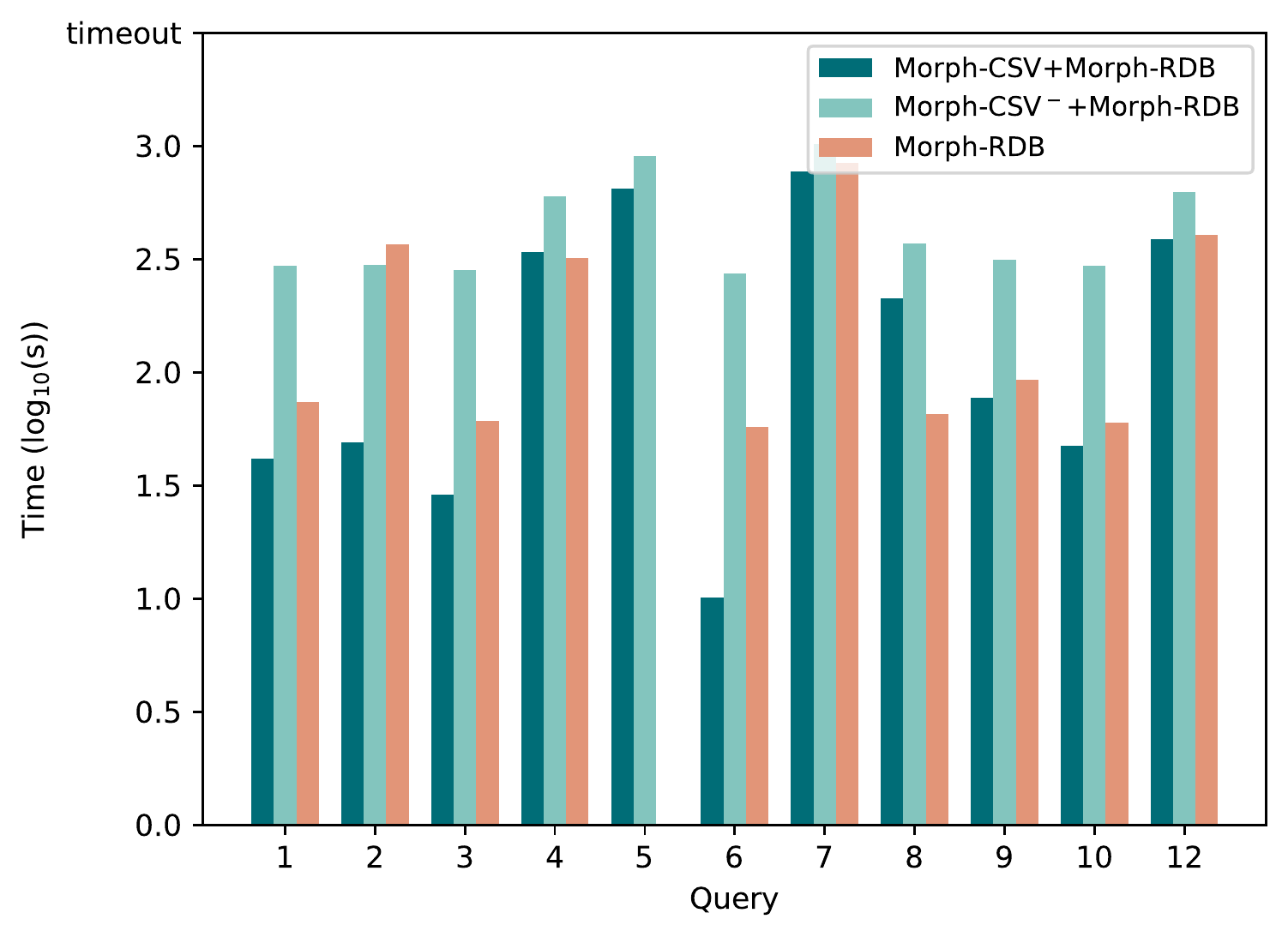}  
  \caption{Query execution time for BSBM-90 with Morph-RDB.}
  \label{fig:morphbsbm90}
\end{subfigure}
\begin{subfigure}{.48\textwidth}
  \centering
\includegraphics[width=1\linewidth]{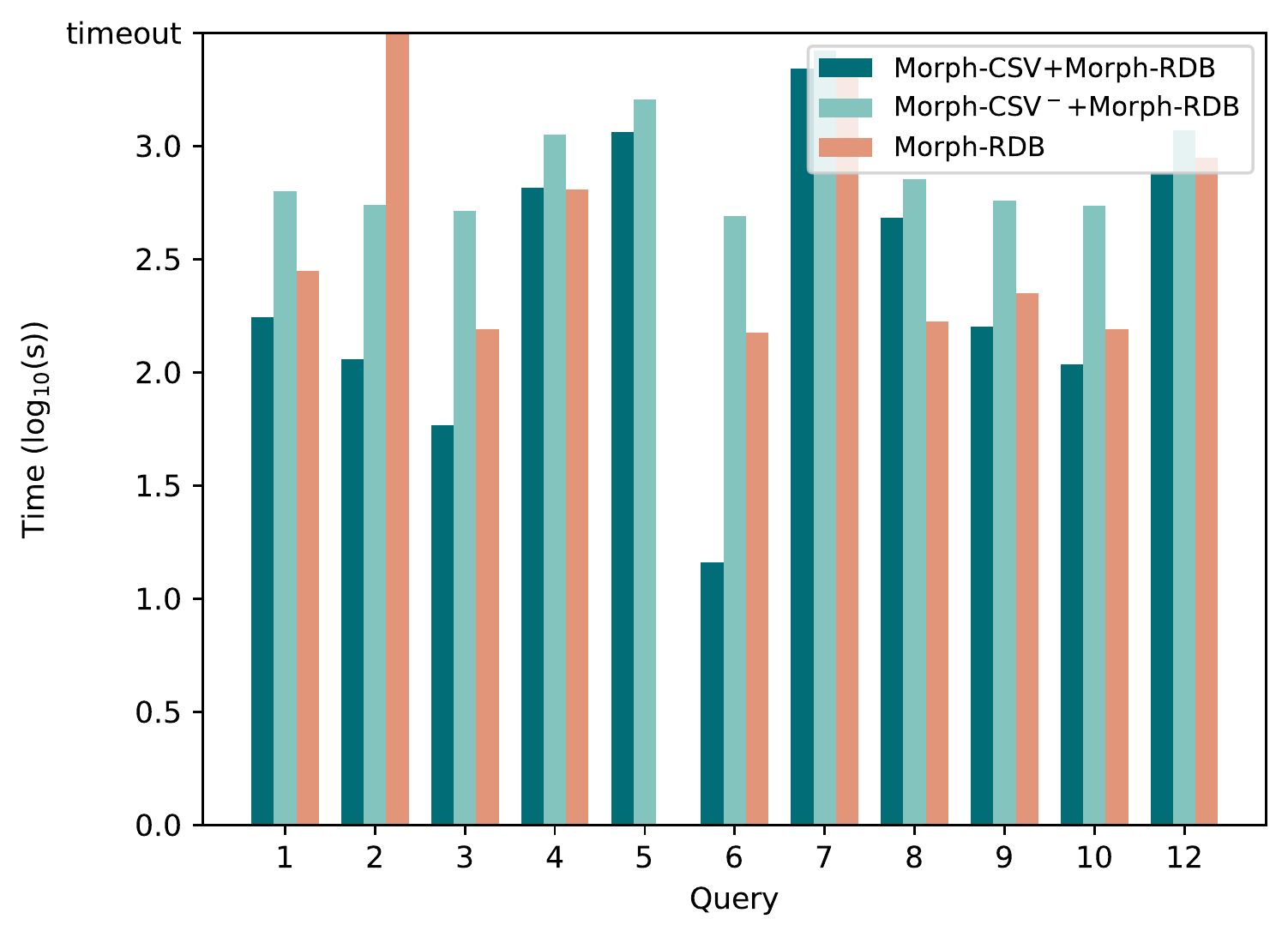}  
  \caption{Query execution time for BSBM-180 with Morph-RDB.}
  \label{fig:morphbsbm180}
\end{subfigure}
\begin{subfigure}{.48\textwidth}
  \centering
  \includegraphics[width=1\linewidth]{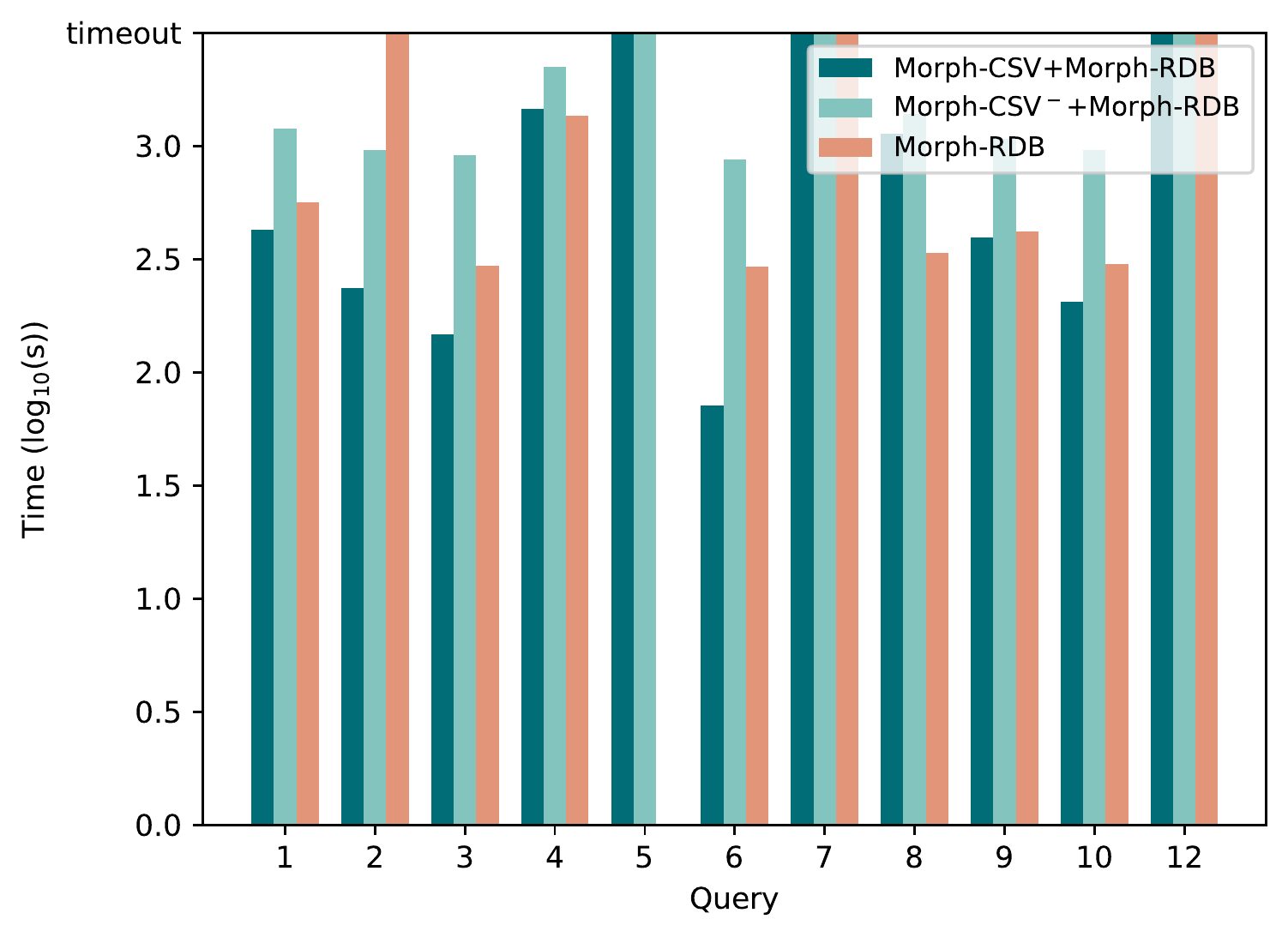}
  \caption{Query execution time for BSBM-360 with Morph-RDB.}
  \label{fig:morphbsbm360}
\end{subfigure}
\caption{\textbf{Query execution Time of Tabular Datasets in BSBM with Morph-RDB.} Execution time in seconds of the tabular datasets from the BSBM benchmark with scale values 45K, 90K, 180K and 360K. The baseline Morph-RDB approach (red columns) is compared with the combination of Morph-CSV (dark green) and Morph-CSV$^-$ (light green) together with Morph-RDB.}
\label{fig:morphbsbm}
\end{figure*}

\begin{figure*}[th]

\begin{subfigure}{.48\textwidth}
  \centering
  \includegraphics[width=1\linewidth]{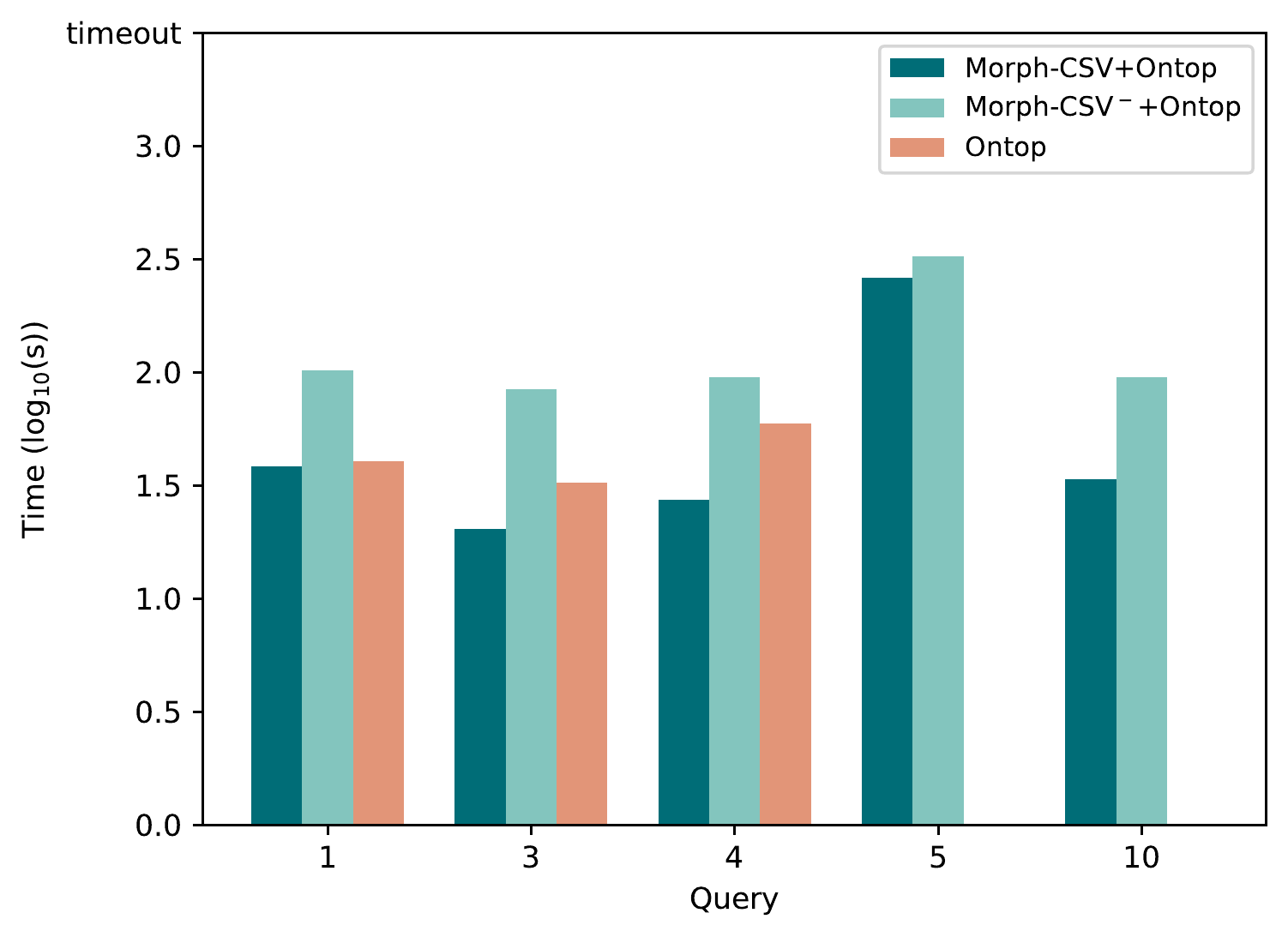}  
  \caption{Query execution time for BSBM-45 with Ontop.}
  \label{fig:ontopbsbm45}
\end{subfigure}
\begin{subfigure}{.48\textwidth}
  \centering
  \includegraphics[width=1\linewidth]{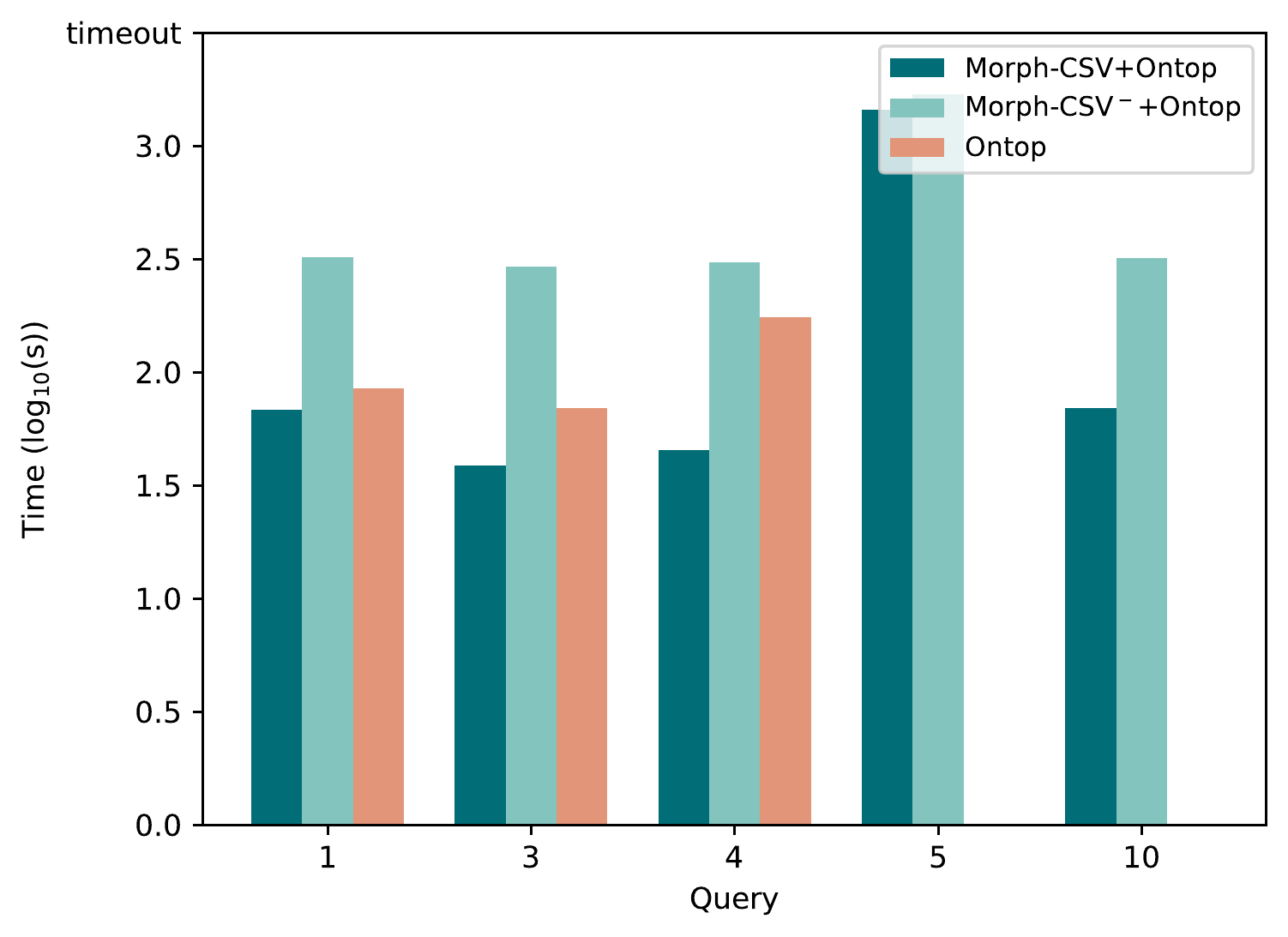}  
  \caption{Query execution time for BSBM-90 with Ontop.}
  \label{fig:ontopbsbm90}
\end{subfigure}
\begin{subfigure}{.48\textwidth}
  \centering
\includegraphics[width=1\linewidth]{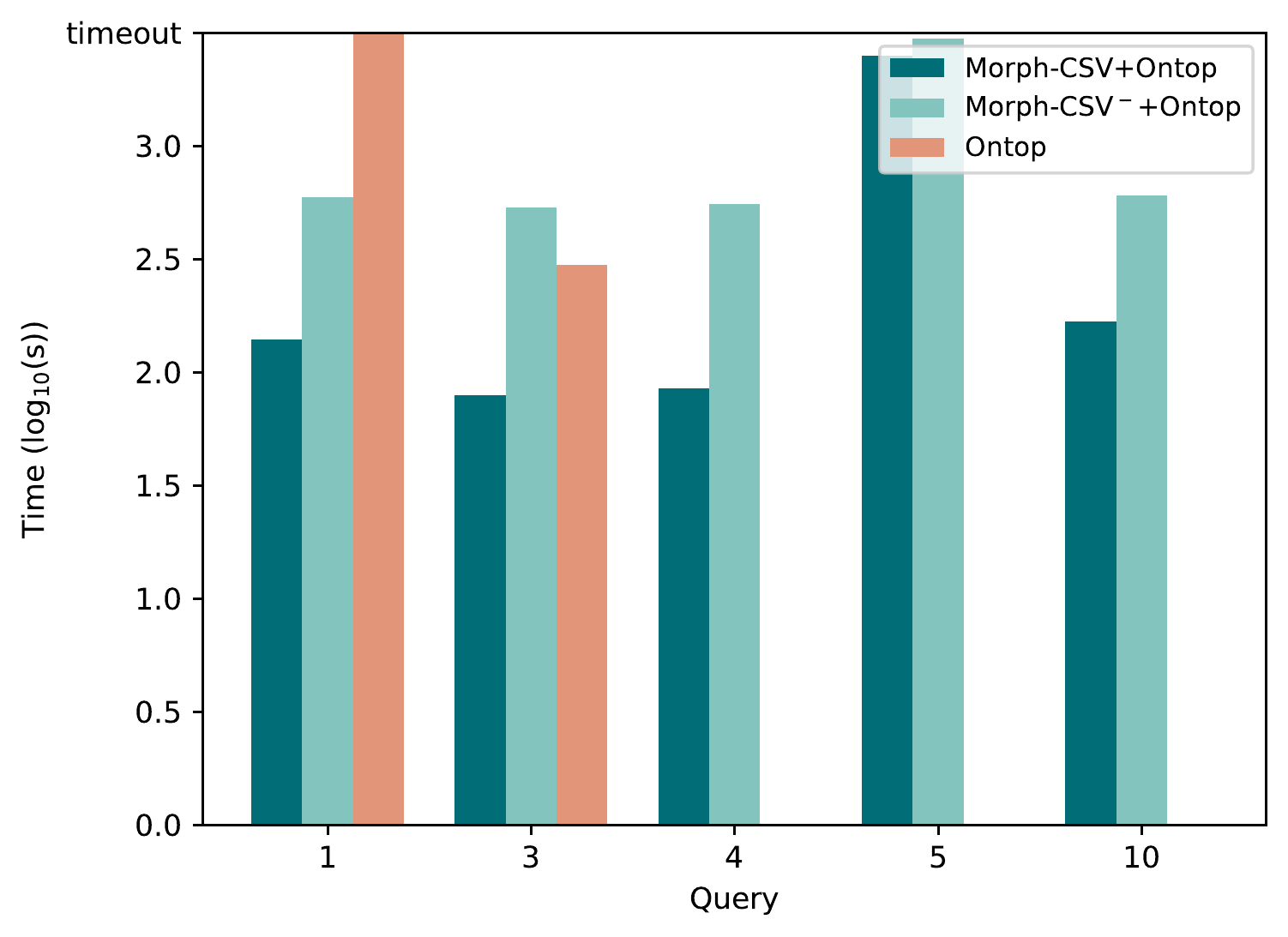}  
  \caption{Query execution time for BSBM-180 with Ontop.}
  \label{fig:ontopbsbm180}
\end{subfigure}
\begin{subfigure}{.48\textwidth}
  \centering
  \includegraphics[width=1\linewidth]{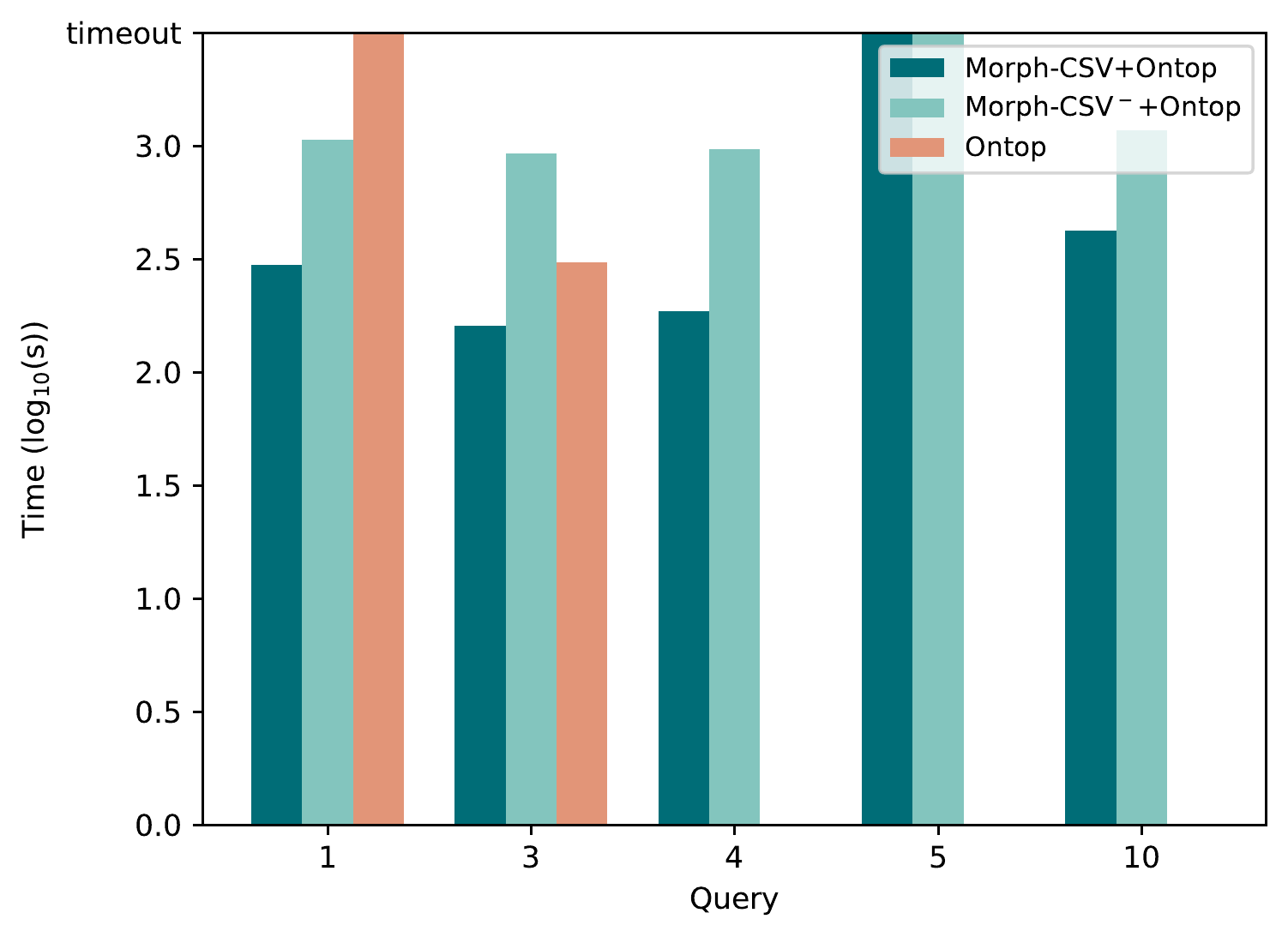}
  \caption{Query execution time for BSBM-360 with Ontop.}
  \label{fig:ontopbsbm360}
\end{subfigure}
\caption{\textbf{Query execution Time of Tabular Datasets in BSBM with Ontop.} Execution time in seconds of the tabular datasets from the BSBM benchmark with scale values 45K, 90K, 180K and 360K. The baseline Ontop approach (red columns) is compared with the combination of Morph-CSV (dark green) and Morph-CSV$^-$ (light green) together with Ontop.}
\label{fig:ontopbsbm}
\end{figure*}

\noindent\textbf{Datasets, annotations and queries.} In order to test our proposal we decided to adapt BSBM, extracting the tabular data sources in CSV format from the SQL generated instances. Additionally, we create the corresponding mapping rules in RML and the metadata following the CSVW specification. We measure the loading time of the two proposals (baseline and Morph-CSV) for each query in the benchmark. Since the focus of Morph-CSV is not the improvement of the support of SPARQL features in the query translation process, we only select the queries of the benchmark that include the supported features  by each engine. This means that Morph-RDB will be evaluated over the queries  Q1, Q2, Q3, Q4, Q5, Q6, Q7, Q8, Q9, Q10 and Q12  and Ontop will be evaluated over Q1, Q3, Q4, Q5 and Q10, both of them using the corresponding R2RML mapping document. For the baseline approach we manually create the RDB schema without  constraints.

\subsubsection{BSBM Results}

\noindent\paragraph*{\textbf{Loading Time}.}
The results of the load time for each query and dataset size are shown in Figure \ref{fig:bsbmload}. The main difference between baseline and Morph-CSV$^-$ in comparison with Morph-CSV is that while the loading time for the first two methods is constant for each size, Morph-CSV loading time depends on several input parameters such as the query and the number and type of constraints. In the case of Morph-CSV, it could be understandable that the application of a set of constraints over the raw data in order to improve query performance and completeness, would have a negative impact in the loading time. This happens in queries Q8 and Q11, where the number of sources and the application of the constraints (mainly integrity constraints),  impact negatively on the loading time of the data in the RDB instance in comparison with the baseline approach. However, in the rest of the queries, the Morph-CSV steps focus on the selection of constraints, sources and columns, and on exploiting the information in query and mapping rules, improving the loading time for each query in comparison with the baseline loading time. This means that, although the engine is including a set of additional steps during the starting phase of an OBDA system, the application of these steps only over the data that is required to answer the query, has a positive impact in the total query execution time. Additionally, we can observe that Morph-CSV is able to process, apply the different constraints, and generate the corresponding instance of the RDB for any query. In the case of Morph-CSV$^-$, applying all the constraints defined for the whole dataset each time a query has to be answered, has a negative impact in the loading time, obtaining the worst results in the loading phase.

\noindent\paragraph*{\textbf{Evaluation Time with Morph-RDB}.} The query execution time using Morph-RDB as the back-end OBDA engine is shown in Figure \ref{fig:morphbsbm}. The first remarkable observation can be seen in query Q5. Although this query contains features supported by Morph-RDB, the engine reports an error when evaluating the query over the database generated by the baseline approach, because it is not able to evaluate the arithmetic expressions in the FILTER clauses. On the contrary, the datatype of each column in the database generated by Morph-CSV (and also Morph-CSV$^-$) is properly defined, making it possible for Morph-RDB to evaluate the query without any problem and obtaining the expected results. Another remarkable difference is in query Q2, which contains a large number of joins, Morph-RDB reports a timeout error for 180K and 360K with the database generated by the baseline approach. However, it is still able to evaluate this query in reasonable time over the databases generated by Morph-CSV and Morph-CSV$^-$. The effect of the application of integrity constraints in the generation of the RDB instance can also be seen in most of the queries (i.e., Q1, Q2, Q3, Q6, Q9, Q10) reducing considerably the query execution time in the database generated by Morph-CSV in comparison with the baseline approach. There are cases (i.e., Q4, Q7, Q12) where the amount of data to retrieve is large, minimizing the effect of the optimizations. Finally, there are cases where optimizations over the indexes cannot be applied (e.g. querying all the properties of a class). We observe this behavior in Q8, in which the difference between the Morph-CSV+Morph-RDB and Morph-RDB approaches is minimal and this behavior is consistent in all size of datasets. In general, Morph-CSV$^-$ obtains worse results than Morph-CSV+Morph-RDB and Morph-RDB alone. The results are understandable as this configuration has to invest time in preparing the full RDB instance for each query, executing many unnecessary steps in comparison with Morph-CSV. However, in some cases the evaluation time is better than the one obtained over the Morph-RDB configuration, where clearly the creation of indexes and integrity constraints play a key role in the performance of the query execution (see Q2).

\noindent\paragraph*{\textbf{Evaluation Time with Ontop}.}
The query execution time using Ontop as the back-end OBDA engine is shown in Figure \ref{fig:ontopbsbm}. Like Morph-RDB, Ontop needs the Morph-CSV generated databases to be able to evaluate Q5 due to the arithmetic expressions of its FILTER operators. Additionally, it also fails in Q10 because it cannot process a FILTER with a date value. In the rest of the queries (Q1, Q3, Q4) we can see that the query evaluation time in Ontop with Morph-CSV is lower than the query evaluation time over the baseline database. Note that in larger databases (180K and 360K), Q1 and Q4 can only be evaluated over the databases generated by Morph-CSV. The Morph-CSV$^-$ configuration is also able to answer the queries just as the Morph-CSV standard configuration, but in comparison with this configuration, the performance is being affected due the inclusion of the additional and unnecessary steps.

As mentioned in the Ontop repository page\footnote{\url{https://github.com/ontop/ontop/wiki/MappingDesignTips}}, integrity constraints are essential for the correct behavior of the engine. Although it is out of the scope of this paper, we observe in our experiments that the main reason why Ontop is only able to answer half of the queries in this benchmark, is related to some issues about maintaining the desirable properties~\cite{corcho2020towards} when translating R2RML mapping rules to its own mappings, called OBDA. The engine also fails to evaluate queries with OPTIONAL clauses when there are NULL values in the answers, as they acknowledged, it is possible that this support has not been implemented in the engine~\cite{xiao2018efficient}.

\noindent\paragraph*{\textbf{Query completeness.}} In Table \ref{tab:completeness-bsbm} we show the query completeness obtained with the BSBM benchmark. It is important to remark that our intention to use this benchmark is for testing performance capabilities of our proposals, the input sources are extracted from the BSBM relational model, which is a well formed and normalized RDB instance. However, there are still some cases where we identify the need of applying constraints over the relational database, which are Q5 in the evaluation over Morph-RDB and Q5 and Q10 over Ontop. In these cases, the baseline configurations of the engines are not able to answer those queries, not because they do not support a feature of the SPARQL query or  cannot do it on time, but because they cannot perform the correct comparison among different datatypes in the relational database instance. We demonstrate  with the application of Morph-CSV that queries can be answered and  the correct number of results can be obtained. Additionally, thanks to the application of indexes and integrity constraints there are some queries such as Q1 and Q2 that can be answered by Morph-CSV configuration but not by the baseline, which means that thanks to these steps we are ensuring the effectiveness of the optimizations provided by Ontop and Morph-RDB in the SPARQL-to-SQL translation process. 

\begin{figure*}[th]

\begin{subfigure}{.48\textwidth}
  \centering
  \includegraphics[width=1\linewidth]{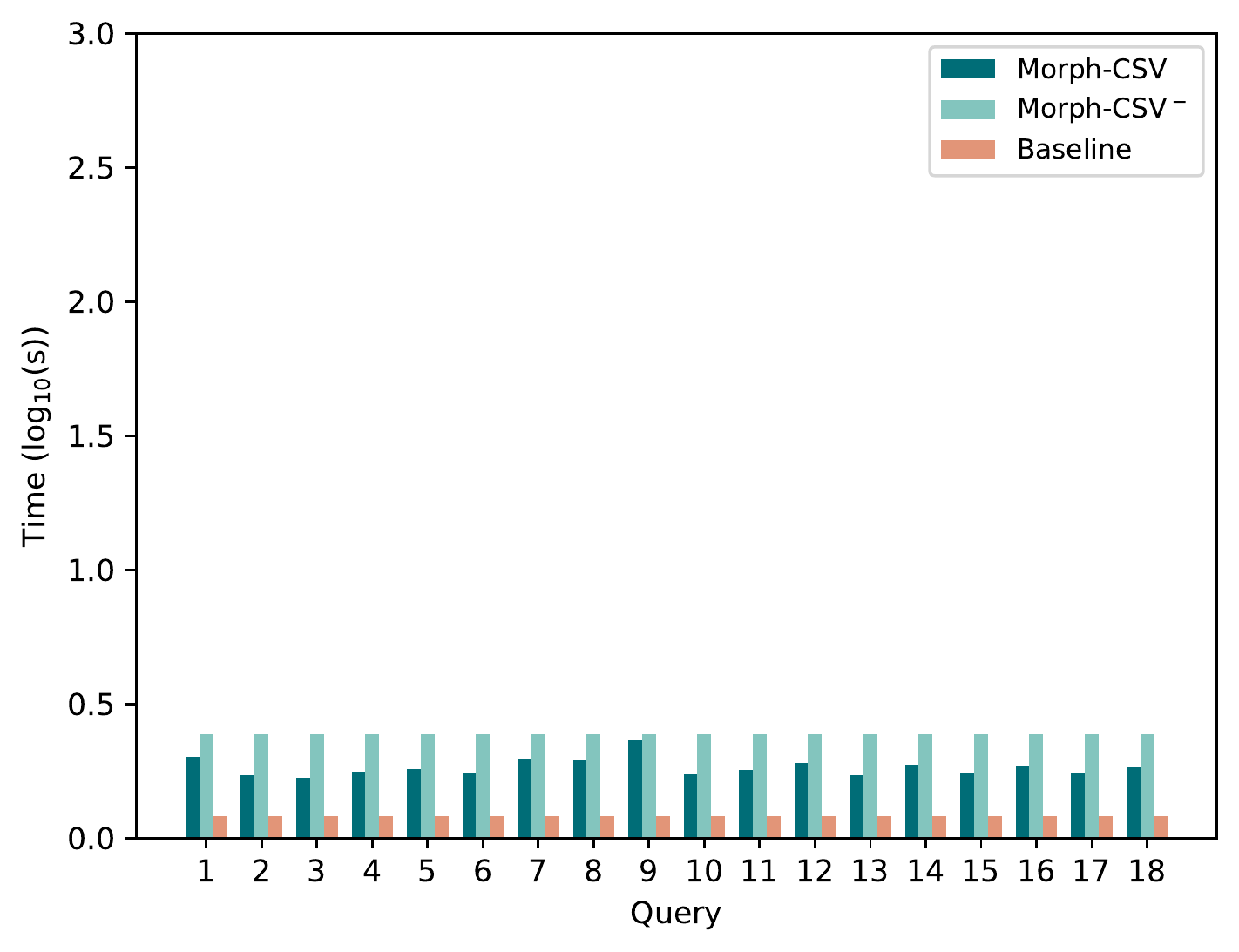}  
  \caption{Loading time for GTFS-1.}
  \label{fig:gtfsload1}
\end{subfigure}
\begin{subfigure}{.48\textwidth}
  \centering
  \includegraphics[width=1\linewidth]{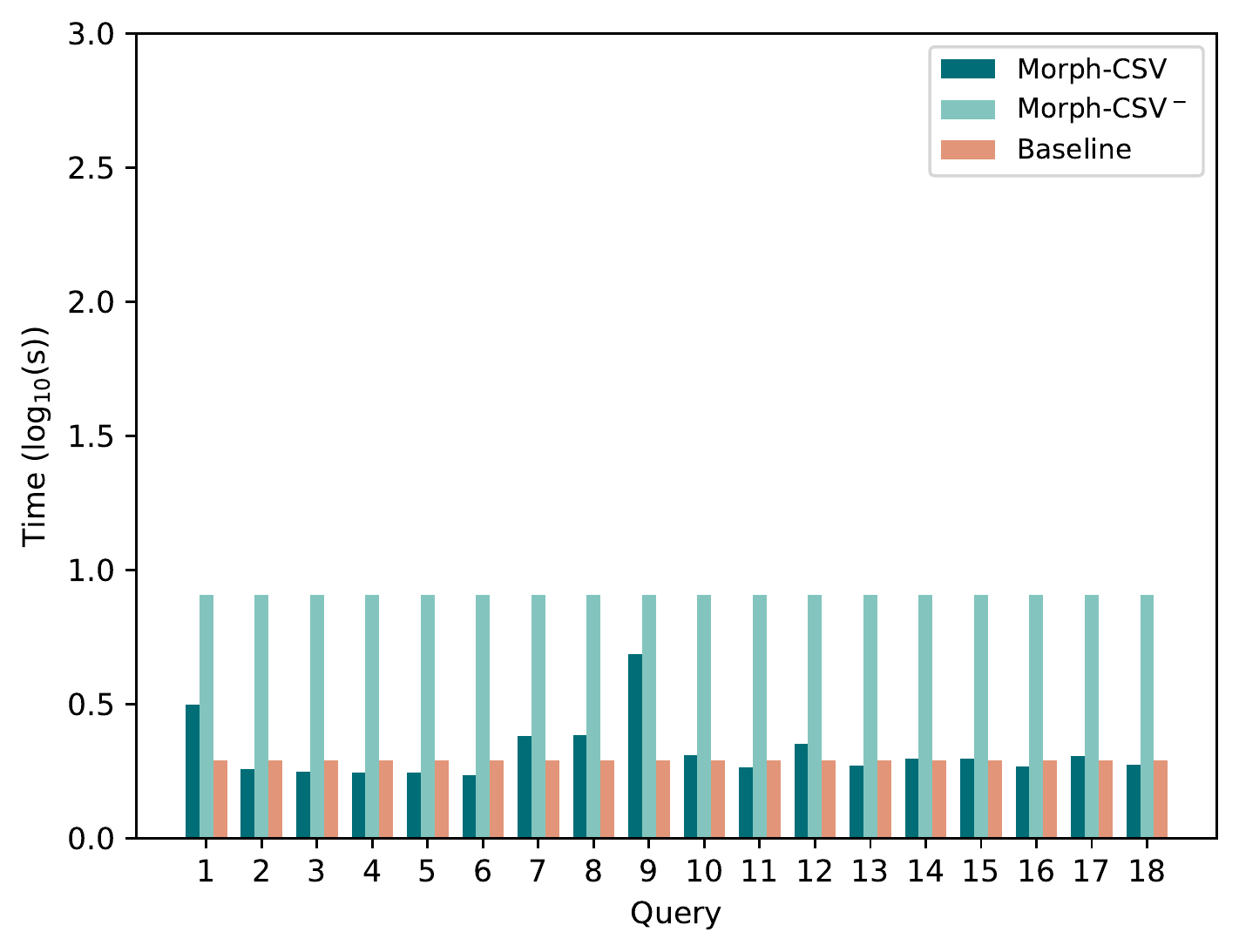}  
  \caption{Loading time for GTFS-10.}
  \label{fig:gtfsload10}
\end{subfigure}
\begin{subfigure}{.48\textwidth}
  \centering
  \includegraphics[width=1\linewidth]{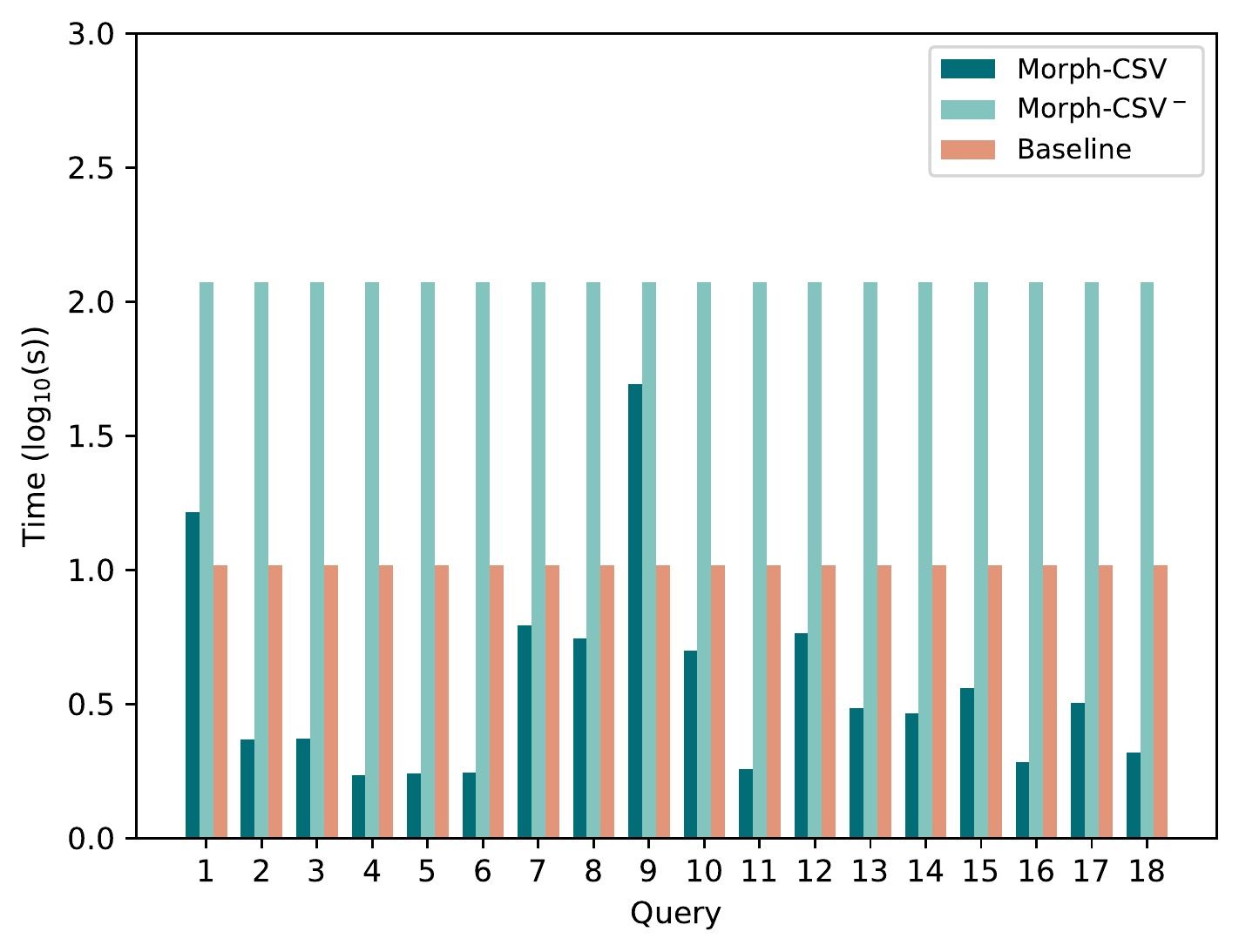}  
  \caption{Loading time for GTFS-100.}
  \label{fig:gtfsload100}
\end{subfigure}
\begin{subfigure}{.48\textwidth}
  \centering
  \includegraphics[width=1\linewidth]{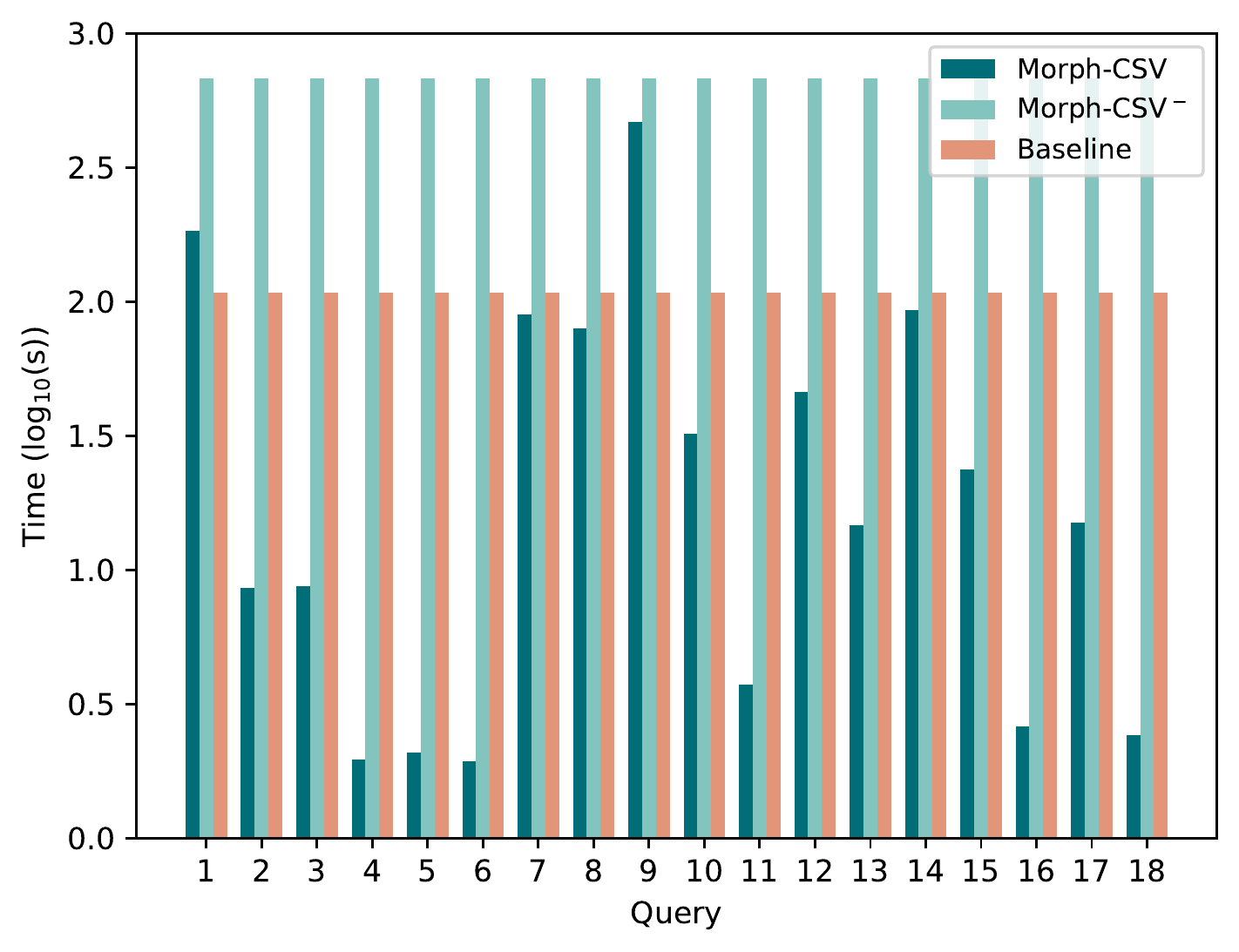}
  \caption{Loading time for GTFS-1000.}
  \label{fig:gtfsload1000}
\end{subfigure}
\caption{\textbf{Loading Time of Tabular Datasets in GTFS.} Loading time in seconds of the tabular datasets from the Madrid-GTFS-Bench with scale values 1, 10, 100 and 1000. The baseline approach (red columns) and Morph-CSV$^-$ (light green) are constant for each dataset and query, while Morph-CSV (dark green) depends on the query and number of constraints to be applied over the selected sources.}
\label{fig:gtfsload}
\end{figure*}

\begin{figure*}[th]

\begin{subfigure}{.48\textwidth}
  \centering
  \includegraphics[width=1\linewidth]{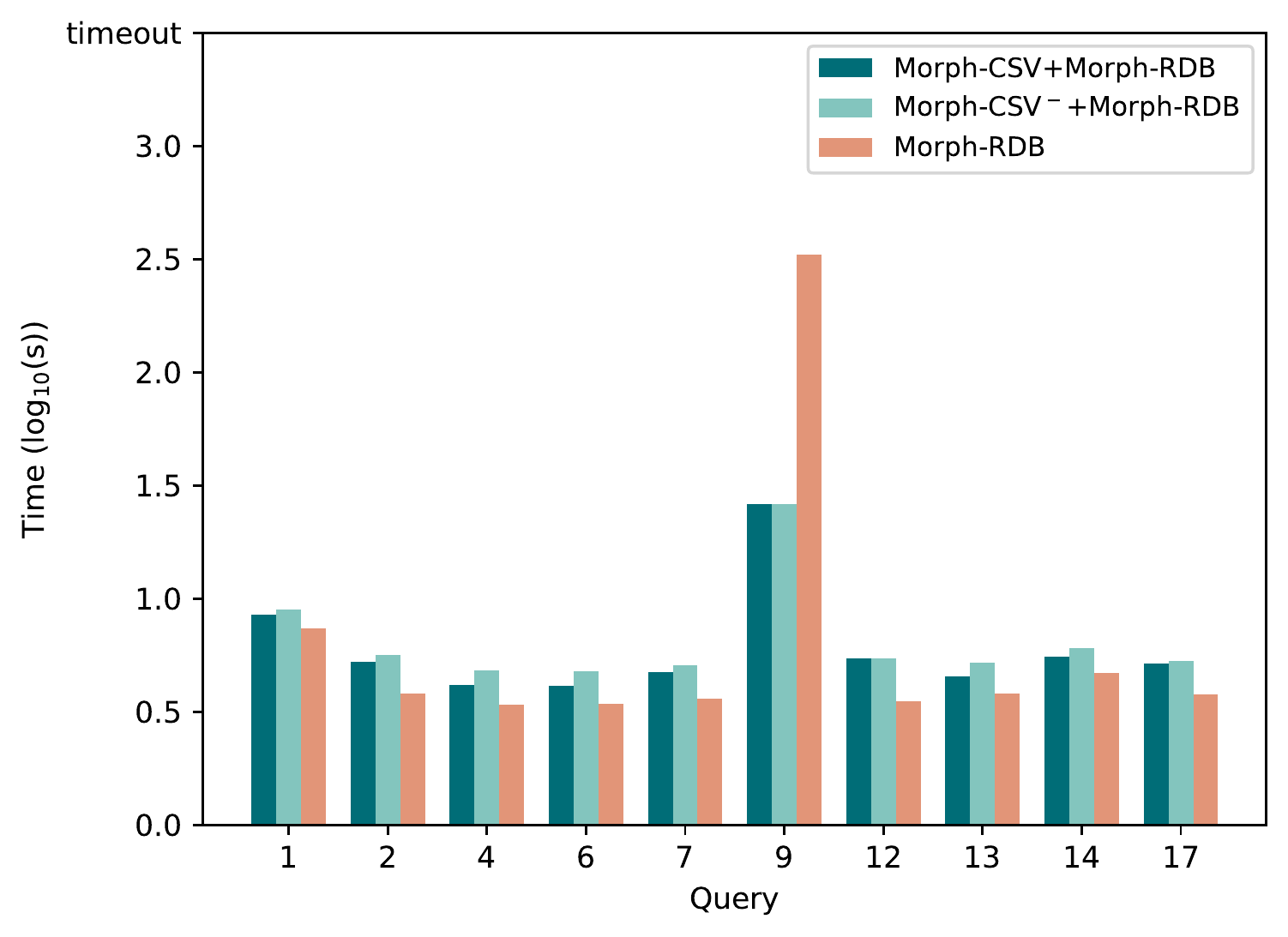}  
  \caption{Query execution time for GTFS-1 with Morph-RDB.}
  \label{fig:morphgtfs1}
\end{subfigure}
\begin{subfigure}{.46\textwidth}
  \centering
  \includegraphics[width=1\linewidth]{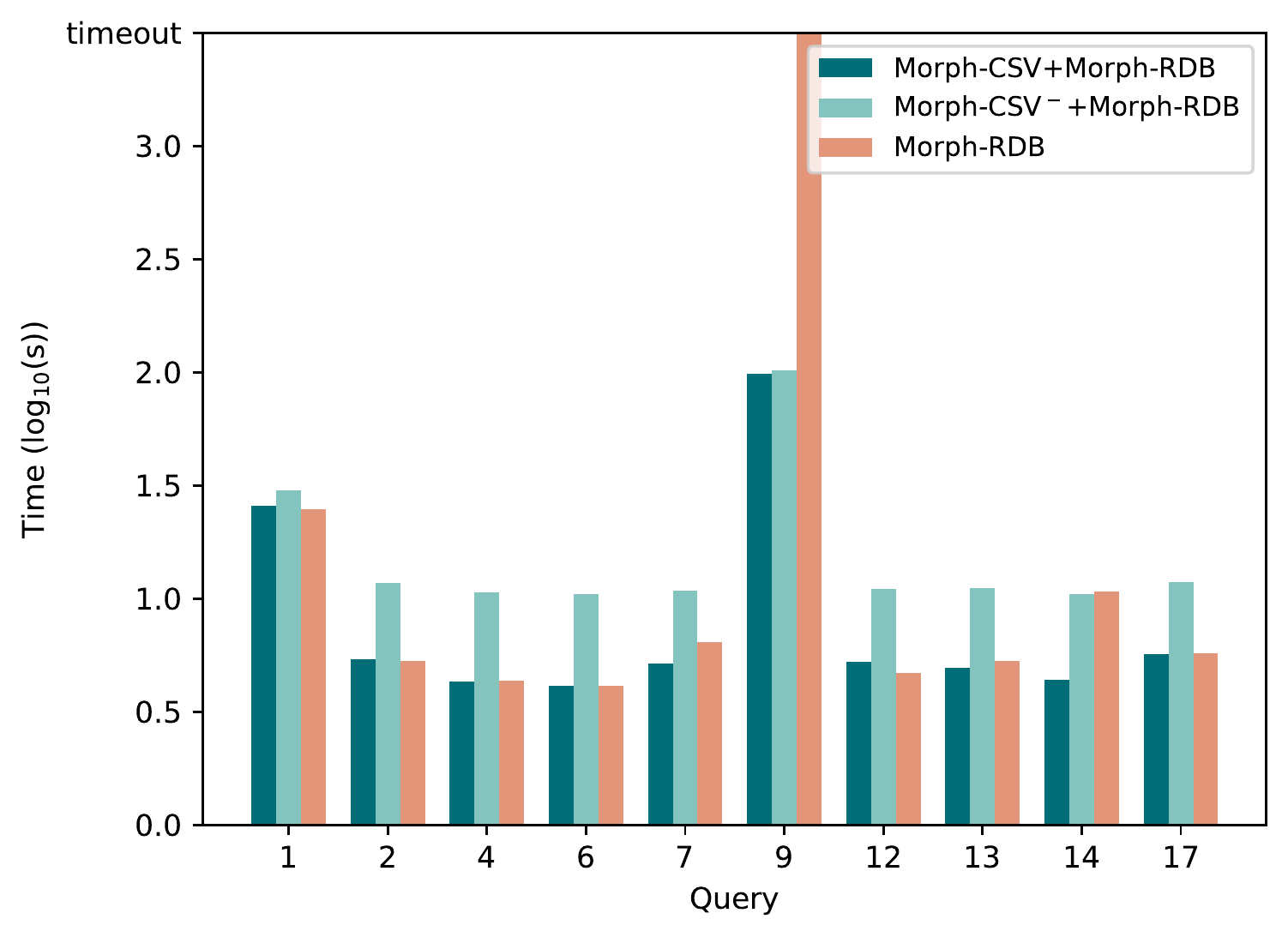}  
  \caption{Query execution time for GTFS-10 with Morph-RDB.}
  \label{fig:morphgtfs10}
\end{subfigure}
\begin{subfigure}{.48\textwidth}
  \centering
  \includegraphics[width=1\linewidth]{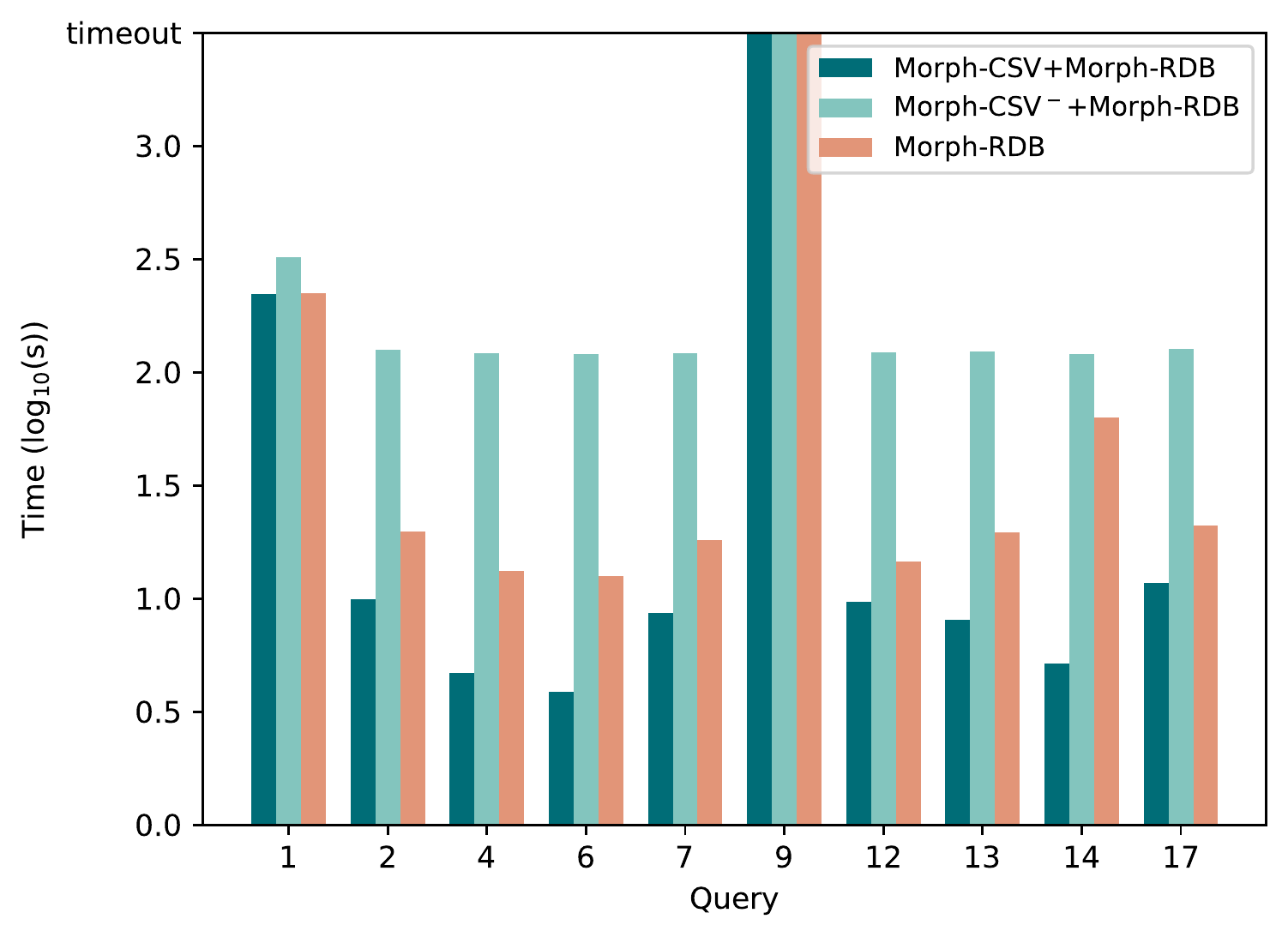}  
  \caption{Query execution time for GTFS-100 with Morph-RDB.}
  \label{fig:morphgtfs100}
\end{subfigure}
\begin{subfigure}{.48\textwidth}
  \centering
  \includegraphics[width=1\linewidth]{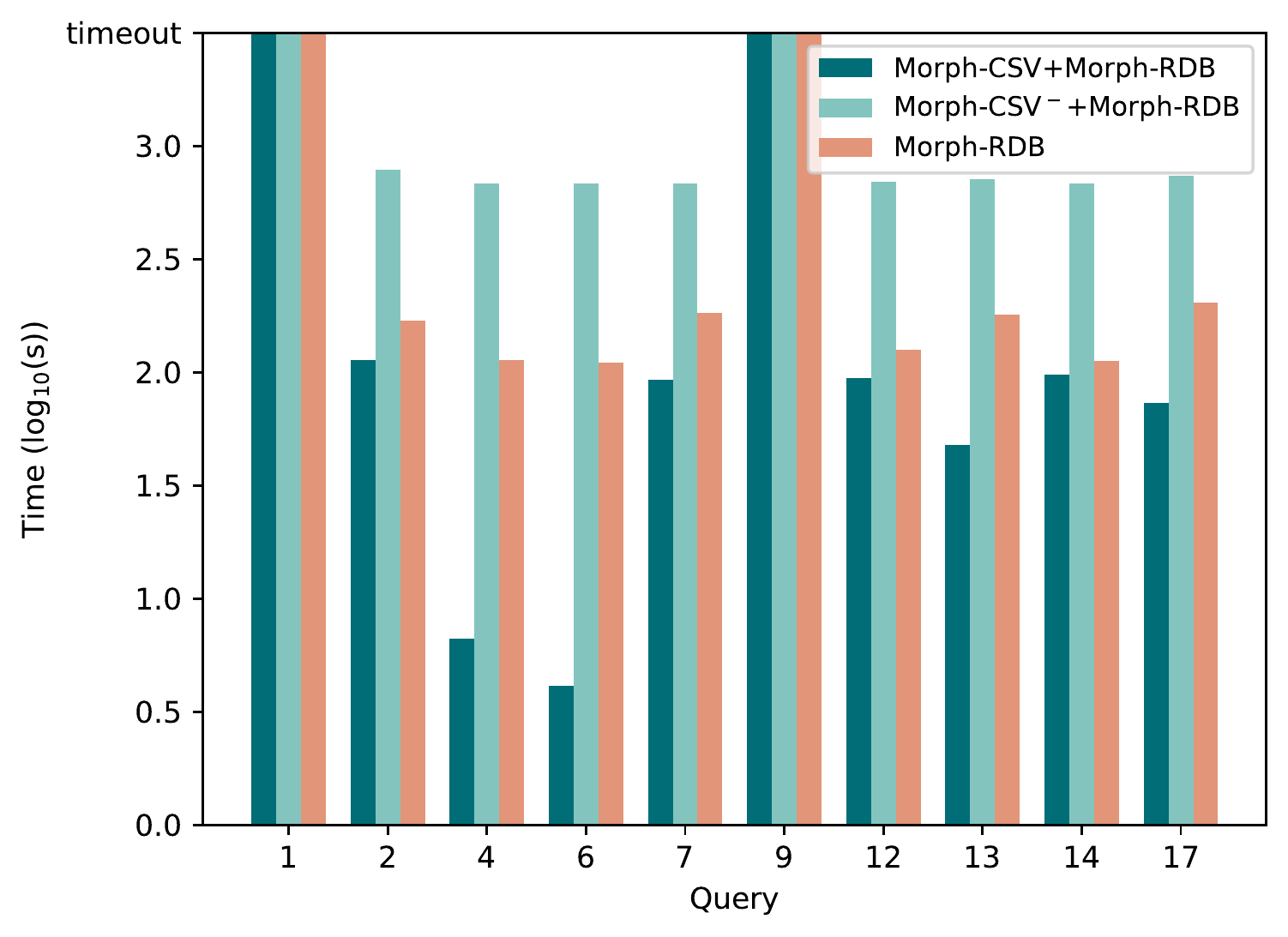}
  \caption{Query execution time for GTFS-1000 with Morph-RDB.}
  \label{fig:morphgtfs1000}
\end{subfigure}
\caption{\textbf{Query execution Time of Tabular Datasets in GTFS with Morph-RDB.} Execution time in seconds of the tabular datasets from the Madrid-GTFS-Bench with scale values 1, 10, 100 and 1000.  The baseline Morph-RDB approach (red columns) is compared with the combination of Morph-CSV (dark green) and Morph-CSV$^-$ (light green) together with Morph-RDB.}
\label{fig:morphgtfs}
\end{figure*}

\begin{figure*}[th]

\begin{subfigure}{.48\textwidth}
  \centering
  \includegraphics[width=1\linewidth]{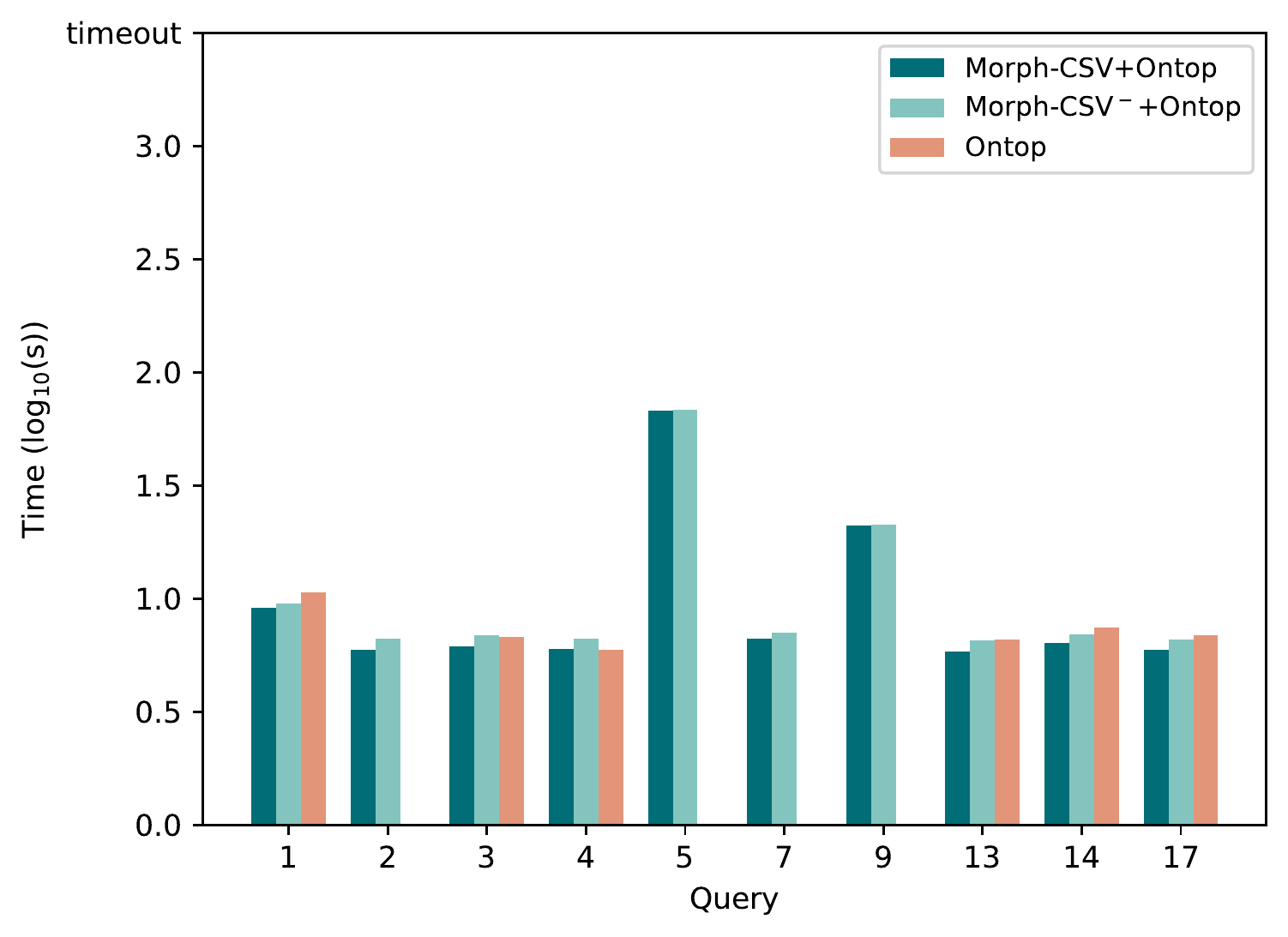}  
  \caption{Total query execution time for GTFS-1 with Ontop.}
  \label{fig:ontop1}
\end{subfigure}
\begin{subfigure}{.46\textwidth}
  \centering
  \includegraphics[width=1\linewidth]{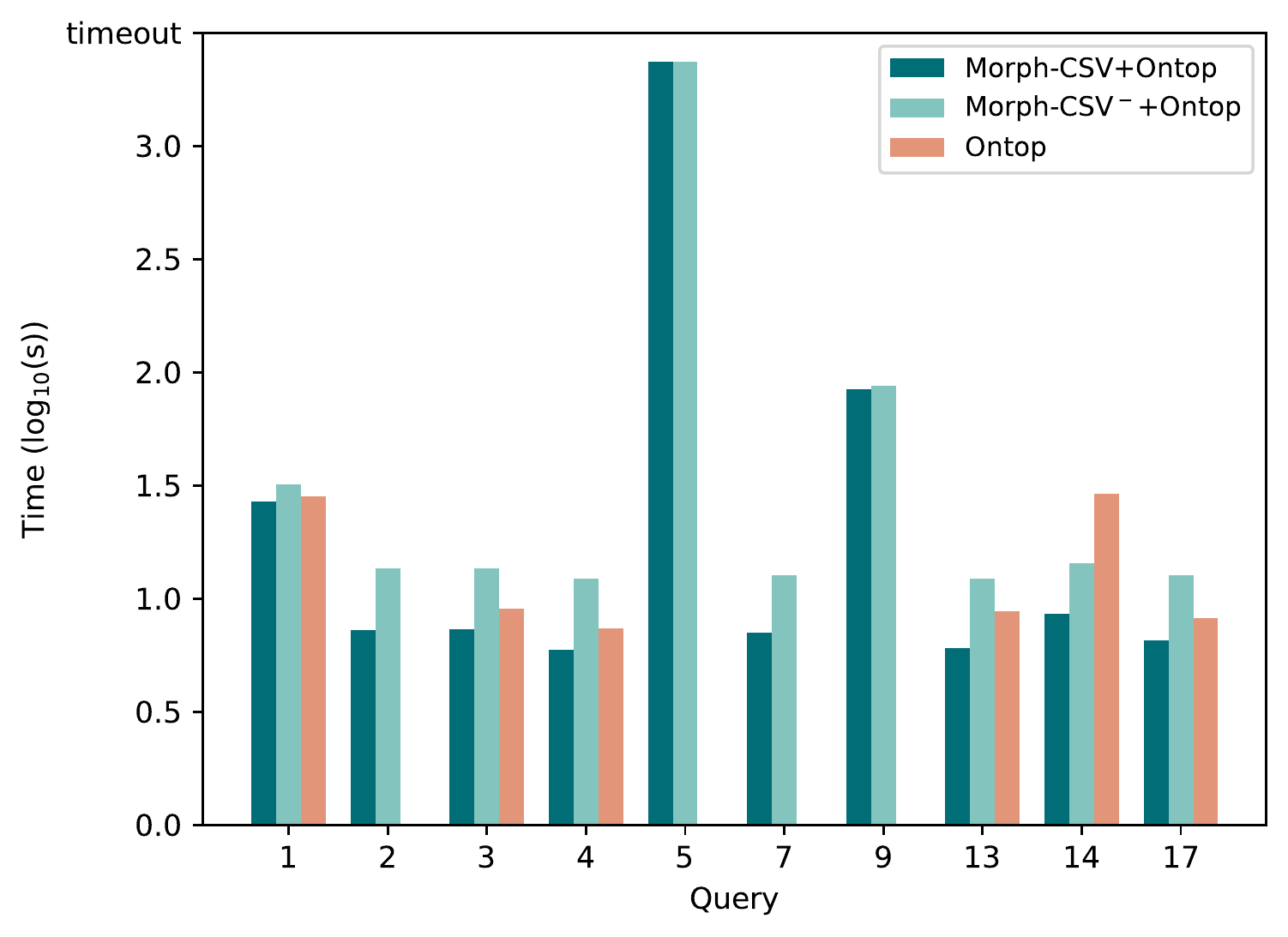}  
  \caption{Total query execution time for GTFS-10 with Ontop.}
  \label{fig:ontop10}
\end{subfigure}
\begin{subfigure}{.48\textwidth}
  \centering
  \includegraphics[width=1\linewidth]{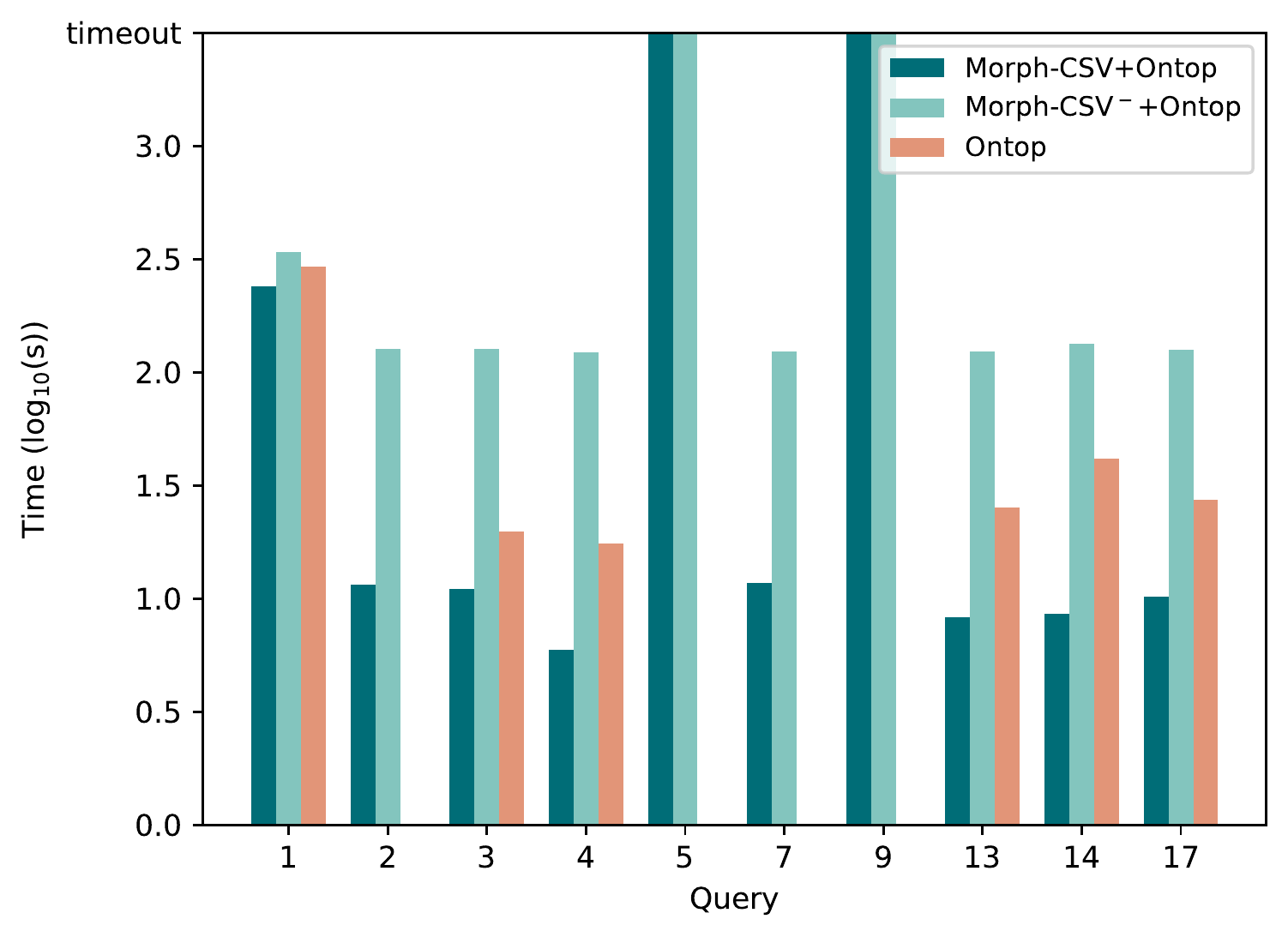}  
  \caption{Total query execution time for GTFS-100 with Ontop.}
  \label{fig:ontop100}
\end{subfigure}
\begin{subfigure}{.48\textwidth}
  \centering
  \includegraphics[width=1\linewidth]{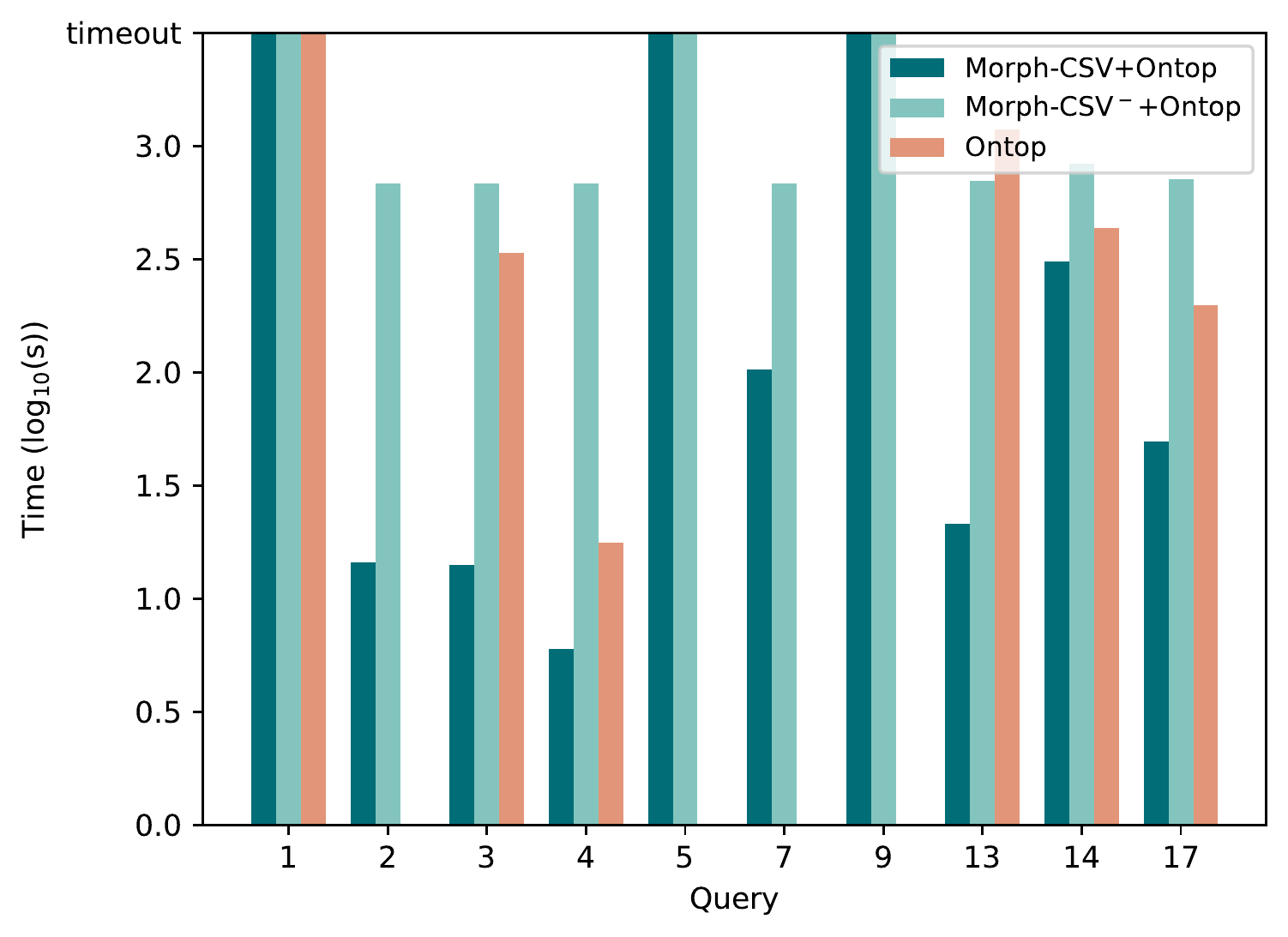}
  \caption{Total query execution time for GTFS-1000 with Ontop.}
  \label{fig:ontop1000}
\end{subfigure}
\caption{\textbf{Query execution Time of Tabular Datasets in GTFS with Ontop.} Execution time in seconds of the tabular datasets from the Madrid-GTFS-Bench with scale values 1, 10, 100 and 1000. The baseline Ontop approach (red columns) is compared with the combination of Morph-CSV (dark green) and Morph-CSV$^-$ (light green) together with Ontop.}
\label{fig:ontopgtfs}
\end{figure*}

\subsection{GTFS-Madrid-Bench}
The GTFS-Madrid Benchmark~\cite{chaves2020gtfs} consists of an ontology, an initial dataset of the metro system of Madrid following the GTFS model, a set of mappings in several specifications, a set of queries according to the ontology that cover relevant features of the SPARQL query language, and a data scaler based on a state of the art proposal~\cite{lanti2014npd}. 

\noindent\textbf{Datasets, annotations and queries.} We select the tabular sources of this benchmark (i.e., the CSV files) and we scale up the original data in several instances (scale factors 10, 100 and 1000). Each generated dataset is denoted as GTFS-$S$ where S is the scale factor. The resources of the benchmark already include the necessary mapping rules and tabular metadata. Like our previous evaluation with BSBM benchmark, we only select the queries with features that are supported by each engine: Morph-RDB will be evaluated using  queries Q1, Q2, Q4, Q6, Q7, Q9, Q12, Q13, Q14, Q17 and Ontop will be evaluated using  queries Q1, Q2, Q3, Q4, Q5, Q7, Q9, Q13, Q14, Q17. The description and features of each query are also available online\footnote{\url{https://github.com/oeg-upm/gtfs-bench/}}.

\subsubsection{Madrid-GTFS-Bench Results}

\noindent\paragraph*{\textbf{Loading Time}.}
The loading time of the GTFS-Madrid-Bench queries is shown in Figure \ref{fig:gtfsload}. For GTFS-1 the baseline approach clearly has better performance than Morph-CSV. However, when the size of the datasets increases, the positive effects of applying constraints become more apparent. For most of the queries, the loading time needed by Morph-CSV is lower in comparison to the loading time in the baseline approach. Additionally, similarly to BSBM, there are a set of queries where the application of integrity constraints has a negative impact on the loading time (queries Q1 and Q9). The impact of the application of all of the constraints for answering each query, presented by the configuration Morph-CSV$^-$, clearly impacts over the performance in the loading time. 

\noindent\paragraph*{\textbf{Evaluation Time with Morph-RDB}.}
The query execution time with Morph-RDB as the back-end OBDA engine is shown in Figure \ref{fig:morphgtfs}. Analyzing the results, we generally observe that the incorporation of Morph-CSV in the workflow of OBDA enhances query performance. With respect to the results of each query, we can observe that on the one hand the behavior of the engine over simple queries (Q1, Q2, Q7, Q12 and Q17) is similar. This is understandable as the selected data sources needed to answer the query do not include the application of several constraints (e.g. there are no joins in the query). On the other hand, in the case of complex queries such as Q4, Q6, Q9, Q13 and Q14, where several tabular sources are needed to answer the queries, the application of constraints has a better impact in comparison to the the baseline approach. Similar behavior is shown over Morph-CSV$^-$, where the complexity of the GTFS data model, with many sources, columns and relations among them, has a clear impact on the total execution time of each query, obtaining worse performance than the baseline in most of the cases. However, for example, in the case of query Q9, Morph-RDB is not able to evaluate the query over the 10th scale database generated by the baseline approach, while in the case of the database generated by Morph-CSV and Morph-CSV$^-$, the query can be answered in reasonable time. In general, due the complexity of GTFS model, we can observe that for small datasets (GTFS-1), the cost of applying the proposed steps of Morph-CSV impacts total execution time. However, when the size of the dataset increases, the baseline approach is impacted due to the fact that it has to load all of the input data sources in the RDB before executing the query, low performance is reported for GTFS-100 and GTFS-1000, including timeout in some queries of the latter. Thanks to the application of the constraints and to the source selection step, for Morph-CSV together with Morph-RDB, the return of the results of the queries has a high performance most of the time. In the cases where Morph-CSV reports a timeout (e.g., Q1 in GTFS-1000); it is because the extremely high number of obtained results cannot be handle by Morph-RDB.

\noindent\paragraph*{\textbf{Evaluation Time with Ontop}.}
The experimental evaluation of the query execution in Ontop as the back-end OBDA engine is shown in Figure \ref{fig:ontopgtfs}. This engine is more strict with datatypes in the RDB in comparison with Morph-RDB, and it is why Q2, Q5, Q7 and Q9 produce a failure in the execution over the databases generated by the baseline approach. All these queries have a FILTER clause on a specific datatype (e.g., date, integer, etc) and Ontop proceeds to check the domain constraints before executing the queries. Morph-CSV solves this problem by exploiting the annotations from the metadata and defines the correct datatypes of each column before evaluating the query. For the queries that can be answered by both approaches (Q1, Q3, Q4, Q13, Q14, Q17), the absence of integrity constraints has a negative impact in Ontop, resulting in lower execution time over the databases generated by Morph-CSV. However, similar to the evaluation over Morph-RDB, the complexity of the GTFS data model with a larque quantity of domain and integrity constraints to be applied over the whole dataset, makes that the behavior observed over Morph-CSV$^-$ is being impacted, hence, obtaining worse results that Morph-CSV configuration and the baseline in most of the cases. Finally, in the case where Ontop is not able to evaluate the query under the defined threshold, we report it as a timeout.

\noindent\paragraph*{\textbf{Query completeness.}} In the same manner as BSBM benchmark, the focus of the GTFS-Madrid-Bench is on testing the performance and scalability issues of virtual OBDA and OBDI engines. The input dataset is also well formed and normalized. The completeness results of the evaluation are shown in Table \ref{tab:complete-gtfs}, where as we describe before, Morph-RDB has a mechanism to infer the datatypes of the database using the \textit{rr:dataType} annotation from R2RML, which allows  the engine to answer the queries of this benchmark without the need of applying  datatype constraints over the RDB instance. However, Ontop does not include such a mechanism and it needs the declaration of the correct datatypes over the RDB instance, which has a negative impact in the execution of many queries of the benchmark, that cannot be answered using the baseline database but they retrieve the correct results including Morph-CSV (or Morph-CSV$^-$) in the pipeline.

\subsection{Use Case: The Bio2RDF project}
Bio2RDF is one of the most popular projects that integrates and publishes biomedical datasets as Linked Data~\cite{belleau2008bio2rdf}. Its community has actively contributed to the generation of those datasets using ad-hoc programming scripts, such as PHP. In our previous work~\cite{iglesias2019enhancing} we proposed an alternative way of generating the datasets using a set of declarative mapping rules to improve the maintainability, readability and understanding of the procedure. In comparison with the other benchmarks where the focus of the evaluation was the improvement of the query evaluation time, this real use case contains multiple heterogeneity challenges that, for example, enforce the application of ad-hoc transformation functions (i.e., mappings in the form of RML+FnO). Thus, with this use case we want to demonstrate the benefits of exploiting declarative annotations (metadata and mappings) over the raw data in order to improve query completeness and the need of incorporating the proposed steps for executing queries over real world data sources.

\noindent\textbf{Dataset, annotations, and queries.} Tabular datasets in CSV or Excel formats cover over 35\% of the total datasets in the Bio2RDF project~\cite{iglesias2019enhancing}. In order to test the capabilities of Morph-CSV, we select a subset of the tabular datasets guaranteeing that they cover all of the identified challenges. Additionally, as far as we are aware, there is no standard benchmark over the Bio2RDF project; we also propose a set of SPARQL queries in order to exploit the selected data. Their main features are shown in Appendix \ref{apppendix:queries}).

\subsubsection{Bio2RDF Results}
\begin{figure}[t]
    \centering
    \includegraphics[width=1\linewidth]{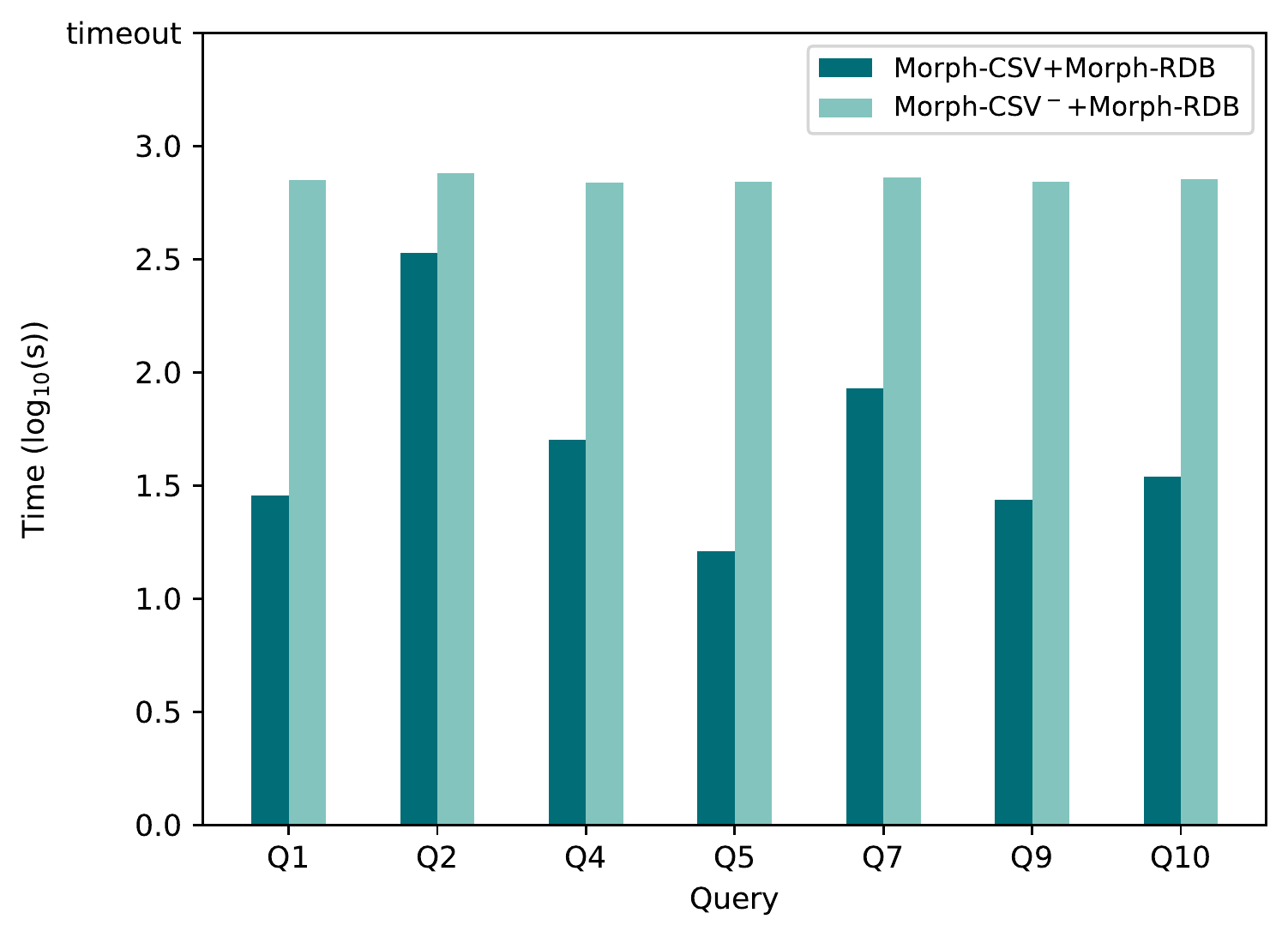}
    \caption{\textbf{Query execution Time of Tabular Datasets in Bio2RDF with Morph-RDB.} Execution time in seconds of the tabular datasets from  Bio2RDF of Morph-CSV and Morph-CSV$^-$ using Morph-RDB as back-end engine. The baseline is not reported as the loading over the RDB instance reports an error.}
    \label{fig:bio2rdfmoprh}
\end{figure}

\begin{figure}[t]
    \centering
    \includegraphics[width=1\linewidth]{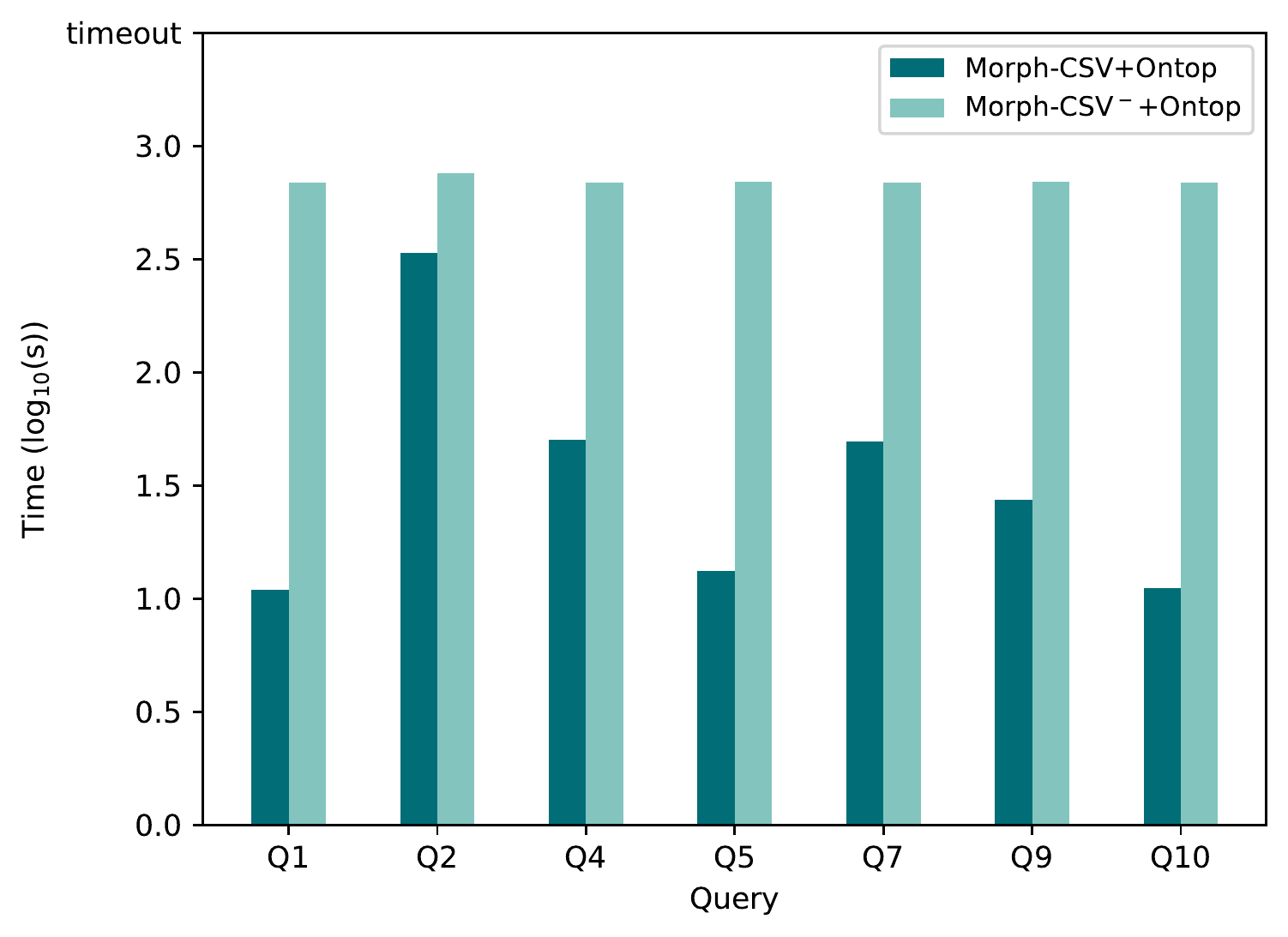}
    \caption{\textbf{Query execution Time of Tabular Datasets in Bio2RDF with Ontop.} Execution time in seconds of the tabular datasets from  Bio2RDF of Morph-CSV and Morph-CSV$^-$ using Ontop as back-end engine. The baseline is not reported as the loading over the RDB instance reports an error. }
    \label{fig:bio2rdfontop}
\end{figure}

The results obtained for query evaluation in Bio2RDF are shown in Figure \ref{fig:bio2rdfmoprh} with Morph-RDB as back-end engine and in Figure \ref{fig:bio2rdfontop} with Ontop. The detailed results obtained by Morph-CSV and Morph-CSV$^-$ are shown in Table \ref{tab:bio2rdfdeatiledresults} and the completeness in Table \ref{tab:completness-bio2rdf}. Analyzing the obtained results, we can observe that there are no results for the baseline approach, this means it was not possible to create an RDB schema and load the input data manually. The main reasons are the heterogeneity problems of a real use case that do not exist in the previous evaluations. GTFS and BSBM have well formed and standard source data models. Problems such as the absence of column names, multiple formats of same datatype in different files (numbers, dates) and the use of delimiters inside the column data, make it impossible to generate the baseline approach without a manual and ad-hoc pre-processing step. However, exploiting declarative annotations, Morph-CSV is able to apply the proposed workflow to this dataset, and successfully answer the proposed queries with both back-end OBDA engines. Similar to the previous benchmarks, loading the complete dataset for answering the input query (Morph-CSV$^-$ configuration) has an negative impact on the total execution time. We can observe that for the proposed queries, most of the total evaluation time of each query is spent in the loading process, as the total execution time in Morph-CSV$^-$ is pretty similar for all the queries. Contrary, query execution is benefited by this previous step obtaining the results in reasonable time for all of the queries.

\section{Discussion of Experimental Results}
\label{sec:discussion}
We have run an experimental evaluation to analyze what are the effects on the use of declarative annotations to extract and apply constraints to enhance virtual OBDA approaches. We have tested our approach over three different cases: (i) a well known benchmark (BSBM) from the e-commerce domain; (ii) a benchmark focused on a virtual OBDA approach for the transport domain; and (iii) a real use case from the biological domain. We describe the main conclusions and findings based on the results obtained:
\begin{itemize}
    \item \textbf{Query complexity:} Clear benefits are obtained from being able to analyze and take advantage of the information provided by the input query, before translating and running it. It allows to only select sources and constraints that are going to be useful for answering the query, avoiding carrying out additional and unnecessary functions over the raw data. Together with the mapping rules, the queries are essential to make relevant decisions during the on-the-fly physical design of the RDB instance (e.g., integrity constraints). Approaches such as the Morph-CSV$^-$ configuration can be valuable when the freshness of the results is not a main requirement, for example to perform a materialization process, which will ensure high quality RDF files where the domain constraints have been applied.
    \item \textbf{Data size:} The total query evaluation time is being impact from how the engine manages the input dataset and the application of constraints. The delegation of these operations to the RDBMS system after loading the full dataset may not be efficient enough. Morph-CSV pushes down the source selection and the application of domain constraints over the raw data. Although it incorporates a set of additional steps in comparison with the baseline, the benefits in the query execution time by the SPARQL-to-SQL engine are already demonstrated, enhancing the total execution time of the queries in most of the cases.
    \item \textbf{Declarative annotations:} The use of declarative and standard mapping rules and metadata makes it possible the generalization of the proposal, avoiding ad-hoc and manual steps. It also incorporates a set of important benefits for the process such as the improvement of its maintainability, readability, and understandability.
    \item \textbf{Querying raw data in OBDA:} Most of the data shared on the web is currently raw data in well known formats such as CSV, JSON, and XML. Semantic Web and more specifically, OBDA technologies, play a key role in starting to see the web as an integrated  database that can be queried. With this approach, we demonstrate that querying tabular data is: i) neither a trivial nor an easy task that can be delegated to na\"ive querying approaches and ii) optimizations and improvements can still be proposed taking advantage and exploiting current annotation proposals to not only enhance performance but also completeness.
\end{itemize}

\section{Related Work}
\label{sec:related}
In this section, we first refer to previous works in data integration systems that precede the OBDA approach. Then, we refer to the general techniques used in systems that handle raw data. Next, we describe current Ontology Based Data Integration (OBDI) systems that handle tabular data. Finally, we describe existing tabular annotation languages and the use of transformation functions in  mappings.

The most relevant concept that predates the OBDA data integration approach is that of mediator ~\cite{wiederhold1992mediators}, defined in the early 90's by Wiederhold. In the proposed architecture for information systems, mediators form a middle layer that makes  user applications independent of the data resources. The idea is to transform heterogeneous data sources into a common data model, which can then be processed and integrated. Classical examples of systems that implemented the original mediator architecture were TSIMMIS~\cite{tsimmis1994}, Information Manifold~\cite{rajaraman1996querying}, and GARLIC~\cite{roth1997don}. The problem of inconsistent formats is not new, and in general mediators may convert attributes of several sources into a common format. The TSIMMIS~\cite{tsimmis1994} architecture includes a Constraints Manager component which handles integrity constraints across different sources. It supports the definition of the interfaces that a source supports for the constraint (e.g., a trigger), the specification of the desired constraint, and the specification of the strategy for enforcing the constraint or for detecting violations. Information Manifold~\cite{rajaraman1996querying} is an integration system for heterogeneous sources on the Web. It uses source content and capabilities descriptions in order to prune the space of sources that are accessed to answer a query. Garlic~\cite{roth1997don} is a system that provides an integrated view over legacy data sources. Each source or repository has its own data model, schema, programming interface, and query capability. Each Garlic object has an interface and may have several implementations, corresponding to different data sources. The system uses these implementations to optimize and execute a query. Both these systems neither handle domain constraints nor constraints across sources.

The work presented in \cite{vidal1999wrappergeneration} provides a toolkit for the generation of wrappers for web-accessible heterogeneous sources (may be represented as HTML tables) through the description of their capabilities. It provides an specification language to define the capability  for each source, and generates a wrapper according to this specification. It also provides a graphical interface for specifying domains of input attributes and  built-in operators to manipulate the data that is extracted. Similarly to this work, the Morph-CSV framework takes into account the specification of domain constraints and transformation functions, but using established standards for tabular annotations and mapping function definitions.

Throughout the years these ideas have evolved from the use of description logics~\cite{Calvanese2002} to the use of ontologies as a common model for data access~\cite{calvanese2017ontop}, what is called Ontology-Based Data Access. Most of the works proposed under this framework are focused on providing access to relational databases~\cite{priyatna2014formalisation,calvanese2017ontop,sequeda2013ultrawrap} and optimizations on the  SPARQL-to-SQL translation process. In this context, the term constraint has been used in~\cite{hovland2016obda}, where the authors defined two new properties extending the concept of OBDA instance. They propose a set of optimizations  during  SPARQL-to-SQL translation with techniques that take into account these constraints. However, the main assumptions made over the OBDA framework (e.g, the data source is an RDB o has an RDB wrapper, or the schema contains a set of constraints) are maintained. There are other works such as~\cite{michel2015translation,botoeva2019ontology} that apply the OBDA framework over document-based databases, i.e., MongoDB. Morph-CSV follows an OBDA approach including the exploitation of additional information from mappings, tabular metadata and queries for 
tabular datasets.

Related to our work are those approaches that allow querying directly information stored in flat files~\cite{idreos2011here}, Drill\footnote{\url{https://drill.apache.org/}}, NoDB~\cite{alagiannis2012nodb}. These systems provide a layer where ``raw'' data is queried, the data is adaptively loaded and stored, and then the query is executed using an assortment of strategies. Although these systems evaluate queries on raw tabular data and may exploit information encoded in the query, they do not make use of annotations or any sort of description of the data as Morph-CSV does.

Current OBDI open source systems that take tabular data as input are Ontario \cite{endris2019ontario} and Squerall \cite{mami2019squerall}. Ontario is a federated query processing approach for heterogeneous data sources. In its source selection step, Ontario uses source descriptions named RDF Molecule Templates~\cite{endris2018querying} which keep information on the sources. The system handles tabular data among other formats, and implements a virtualization approach of query answering techniques for efficient execution. Similarly, Squerall is also an OBDI system that takes as its inputs data and mappings and uses a middleware to aggregate the intermediate results in a distributed manner. Although the aforementioned systems evaluate queries against raw tabular data, they do not exploit the constraints declared in annotations or mapping rules. 

CSV on the Web (CSVW)\footnote{https://www.w3.org/TR/tabular-data-primer/} is a W3C proposal for the definition of metadata on CSV files such as datatypes, valid values, data transformations, and primary and foreign key constraints. A related W3C proposal\footnote{https://www.w3.org/TR/csv2rdf/} defines a procedure and rules for the generation of RDF from tabular data and a few implementations that refer to this proposal are already available.  The CSV2RDF tool is presented in~\cite{Mahmud2018}, the authors define algorithms to transform CSV data into RDF using CSVW metadata annotations, and their experimental study uses datasets from the CSVW Implementation Report \footnote{https://w3c.github.io/csvw/tests/reports/index.html}. Another tool, COW: Converter for CSV on the Web\footnote{https://csvw-converter.readthedocs.io/en/latest/} allows the conversion of datasets in CSV format and uses a JSON schema expressed in an extended version of the CSVW standard. Both are focused on RDF materialization. To the best of our knowledge, no existing tool exploits information in CSVW annotations for querying tabular data in an OBDA context.

Another area related to our work is  the definition and application of data transformation functions. An approach independent of a specific implementation context is described in~\cite{demeester2019implementation}. It enables the description, publication and exploration of functions and  instantiation of associated implementations. The proposed model is the Function Ontology and the publication method follows the Linked Data principles 
Previous works related to this topic focus on developing ad-hoc and programmed functions. For example, R2RML-F~\cite{debruyne2016r2rml} allows using functions in the value of the \texttt{rr:objectMap} property, so as to modify the value of the table columns from a relational database. KR2RML~\cite{slepicka2015kr2rml}, used in Karma, extends R2RML by adding transformation functions in order to deal with nested values. OpenRefine enables such transformations with the usage of GREL functions, which can be used in its RDF extension. Morph-CSV uses the extension of RML together with the Function Ontology~\cite{de2017declarative} that allows to incorporate ad-hoc transformation functions over the data sources in a declarative manner.

\section{Conclusions and Future Work}
\label{sec:conclusions}
In this paper, we have presented an extension of the common OBDA specification to address the problem of query translation over tabular data. We describe and evaluate Morph-CSV, a framework that exploits the information of mapping rules and metadata OBDA annotations to extract and apply a set of relevant constraints. It pushes down the application of these elements directly over the raw data in order to improve query evaluation and query completeness. 
One of the main contributions of this proposal is that it can be used together with any OBDA framework. From the set of experiments that we have performed with two existing state-of-the-art OBDA engines (Morph-RDB and Ontop), we can see that the use of those engines inside the Morph-CSV framework brings several positive impacts: more queries can be answered and less time is needed to answer most queries.

The definition, application and optimization of new functions and constraints to address other challenges for querying tabular data is one of the main lines for future work~\cite{iglesias2019enhancing}. We also want to study the performance of the proposed workflow over OBDA distributed query systems such as the ones proposed in~\cite{endris2019ontario,mami2019squerall}. More in detail, we want to analyze if the outcomes of the proposed steps by Morph-CSV can help in distributed environments where physical design of knowledge graphs are being proposed~\cite{rohde2020optimizing,upm63647} to enhance query performance (i.e., deciding which input sources have to be transformed to RDF and which ones have to be maintained in their original format). Additionally, one of the possible future work lines is the comparison of the proposed approached, that exploits semantic web technologies and annotations, against non-semantic web solutions that provide support to deal with the identified challenges for querying tabular data (e.g., Apache Drill, Presto, Spark, etc.). The results obtained can also be useful to machine learning approaches that identify when the application of the integrity constraints is needed or not, as we observe that there are special cases where it can have a negative impact. We will also study the challenges for querying other data formats (e.g., XML, JSON) in an OBDA context and extend our approach to incorporate them. We also want to remark the importance of having standard and shared methods and vocabularies to publish metadata of raw data on the web, available for tabular data but not for tree data formats such as XML and JSON. Finally, we will adapt this proposal for a materialization process and study its effects comparing it with previous proposals. We also want to study how the materialization process can be improved when historical versions of the same RDF-based knowledge graph are needed, for example, analyzing which input sources have been changed or not, to decide which parts of that knowledge graph have to be generated again.
\begin{acks}
We are very thankful to Anastasia Dimou, Ben de Meester and Pieter Heyvaert (the RML team), who helped us in the initial discussions about the main contributions of our approach and in the creation of (YARR)RML mappings. We are also very thankful to the developers of Morph-CSV: Jhon Toledo and Luis Pozo. The work presented in this paper is supported by the Spanish Ministerio de Econom\'ia, Industria y Competitividad and EU FEDER funds under the DATOS 4.0: RETOS Y SOLUCIONES - UPM Spanish national project (TIN2016-78011-C4-4-R) and by an FPI grant (BES-2017-082511).
\end{acks}

\nocite{*} 
\bibliographystyle{ios1}           
\bibliography{bibliography}        
\onecolumn
\begin{appendix}
\section{Morph-CSV algorithm}
\label{apppendix:algorithm}
The Morph-CSV algorithm exploiting the mapping rules and metadata to enhance virtual ontology based data access for tabular datasets.
\begin{algorithm}[h]
\SetAlgoLined
\KwResult{SPARQL query result set}
 $M \longleftarrow mapping\_rules$\;
 $MD \longleftarrow metadata$\;
 $Q \longleftarrow query$\;
 $D \longleftarrow tabular\_dataset$\;
 $SSG \longleftarrow \emptyset$\;
 \For{$tp\leftarrow 0$ \KwTo $Q.getTP().size()$}{
    $SSG.add(tp)$\;
 }
 \For{$i\leftarrow 0$ \KwTo $SSG.size()$}{
    $p \leftarrow SSG.getPredicates(i)$\;
    \For{$j\leftarrow 0$ \KwTo $M.getTM().size()$}{
        \If{$p.isContainedIn(M.getTM(j)$}{
            $M^{'} \leftarrow M.getTM(j)$\;
            $MD^{'} \leftarrow getMD(M.getTM(j))$\;
        }   
    }
    $M \leftarrow M^{'}$\;
    $MD \leftarrow MD^{'}$\;
 }
 \For{$i\leftarrow 0$ \KwTo $M.getTM().size()$}{
    $path \leftarrow M.getTM(i).getSource()$\;
    $ref \leftarrow M.getTM(i).getReferences()$\;
    $ts \leftarrow D.get(path)$\;
    $D{'}.add(TS.project(ref))$)\;
 }
 $D \leftarrow D^{'}$\;
 \For{$i\leftarrow 0$ \KwTo $D.size()$}{
    $path \leftarrow D[i].getPath()$\; 
    $norm\_2NF(D[i],MD.getMetadata(path))$\;
    $norm\_3NF(D[i],M)$\;
    $duplicates(D[i])$\;
    $substitute(D[i],MD.getMetadata(path))$\;
    $create(D[i],M.getDeclarativeFunctionFragment(path))$\;
 }
 $S \leftarrow schema(D,M,MD)$\;
 $D^{'} \leftarrow load(D,S)$\;
 $M^{'} \leftarrow translate(M)$\;
 $PI = (O,M^{'},S,D^{'})$\;
 \Return{$run\_query(Q,PI)$}\;
\caption{Morph-CSV algorithm}
\end{algorithm}

\newpage
\onecolumn
\section{Query Features}
\label{apppendix:queries}
\begin{table}[h]
\centering
\caption{\textbf{Query features of the evaluation of Morph-CSV.} Domain constraints are described based on the function performed by Morph-CSV and reflect the number of the columns where that functions has been applied. Improvement functions (duplicates, source selection) are always applied.}
\label{tab:queries-gtfs}
\resizebox{0.728\textwidth}{!}{%
\begin{tabular}{c|c|c|c|c}
\hline
\multirow{2}{*}{\textbf{Query}} & \multirow{2}{*}{\textbf{Query characteristics}} & \multicolumn{2}{c|}{\textbf{Constraints}} & \multirow{2}{*}{\textbf{\# Sources}} \\ \cline{3-4}
 &  & \textbf{Integrity} & \textbf{Domain} &  \\ \hline
\multicolumn{5}{c}{\textbf{Madrid-GTFS-Bench}} \\ \hline
Q1 & 4 TP & - & 3 DataType, 4 Sub & 1 \\ \hline
Q2 & 5 TP, 2 OPT, 1 Filter & 1 INDEX & 3 DataType, 5 Sub & 1 \\ \hline
Q3 & 5 TP, 3 OPT, 1 Filter & 1 INDEX & 4 DataType, 5 Sub & 1 \\ \hline
Q4 & 9 TP, 1 Join, 4 OPT & 2 PK, 1 FK & 7 Sub & 2 \\ \hline
Q5 & 5 TP, 2 Join, 1 Filter & 2 PK & 2 DataType, 2 Sub & 2 \\ \hline
Q6 & 3 TP, 1 Join, 1 Filter & 2 PK, 1 FK & - & 2 \\ \hline
Q7 & 15 TP, 5 Join, 5 OPT, 1 Filter & 6 PK, 5 FK & 3 DataType, 8 Sub & 6 \\ \hline
Q8 & 14 TP, 4 Join, 3 OPT & 6 PK, 5 FK & 3 DataType, 8 Sub & 6 \\ \hline
Q9 & 7 TP, 5 Join, 1 OPT, 1 Filter & 5 PK, 3 FK & 2 DataType, 3 Sub & 5 \\ \hline
Q10 & 4 TP, 1 Join, 1 Filter & 2 PK, 1 FK & 2 Sub & 2 \\ \hline
Q11 & 10 TP, 3 Join, 3 Filter (1 not exists) & 3 PK, 2 FK & 2 DataType, 2 Sub & 3 \\ \hline
Q12 & 10 TP, 3 Joins & 4 PK, 3 FK & 1 DataType, 4 Sub & 4 \\ \hline
Q13 & 6 TP, 1 Join, 1 OPT & 1 PK, 1 FK & 1 DataType, 3 Sub & 1 \\ \hline
Q14 & 8 TP, 3 Join, 1 OPT & 4 PK, 3 FK & 1 DataType, 3 Sub & 3 \\ \hline
Q15 & 3 TP, 1 Filter & 1 PK, 1 FK & 4 DataType, 11 Sub & 1 \\ \hline
Q16 & 8 TP, 3 Join, 2 Filter & 4 PK, 2 FK & 2 DataType, 2 Sub & 3 \\ \hline
Q17 & 9 TP, 2 Join & 3 PK, 2 FK & 1 DataType, 4 Sub & 3 \\ \hline
Q18 & 8 TP, 1Union, 3 Join & 4 PK, 3 FK & 1 DataType, 3 Sub & 4 \\ \hline
\multicolumn{5}{c}{\textbf{Bio2RDF}} \\ \hline
Q1 & 4 TP & - & 3 Sub & 1 \\ \hline
Q2 & 4 TP, 1 Join, 1 Filter & 1 PK, 1 INDEX & 7 Sub & 2 \\ \hline
Q3 & 4 TP, 1 Join & 1 PK, 3 INDEX & 5 Sub & 3 \\ \hline
Q4 & 4 TP, 1 Join & 1 PK, 1 INDEX & 7 Sub & 2 \\ \hline
Q5 & 5 TP, 1 Join & 1 PK, 2 INDEX & 6 Sub & 2 \\ \hline
Q6 & 4 TP & - & 2 Sub & 1 \\ \hline
Q7 & 6 TP, 1 Join, 2 Filter & 1 PK & 1 DataType, 4 Sub, 1 Create & 1 \\ \hline
\multicolumn{5}{c}{\textbf{BSBM}} \\ \hline
Q1 & 5 TP, 3 Join, 1 Filter & 3 PK, 2FK & 7 DataType, 1 Sub & 3 \\ \hline
Q2 & 15 TP, 3 Join, 3 OPT & 4 PK, 3 FK & 10 DataType, 12 Sub & 4 \\ \hline
Q3 & 7 TP, 3 Join, 2 Filter, 1 OPT & 3 PK, 2FK & 8 DataType, 3 Sub & 3 \\ \hline
Q4 & 12 TP, 1 Union, 6 Join, 2 Filter & 3 PK, 2FK & 2 DataType, 4 Sub & 2 \\ \hline
Q5 & 7 TP, 2 Join, 2 Filter & 2 PK, 1FK & 6 DataType, 3 Sub & 2 \\ \hline
Q6 & 2 TP, 1 Filter & - & 1 Sub & 1 \\ \hline
Q7 & 14 TP, 5 Join, 1 Filter, 2 OPT & 5 PK, 4 FK & 11 DataType, 2 Sub & 5 \\ \hline
Q8 & 10 TP, 2 Join, 4 OPT & 3 PK, 2 FK & 8 DatatType, 8 Sub & 3 \\ \hline
Q9 & DESCRIBE, 1 TP & - & - & 1 \\ \hline
Q10 & 7 TP, 3 Join, 2 Filter & 3 PK, 3 FK & 7 DatatType, 2 Sub & 3 \\ \hline
Q11 & 2 TP, 1 Union & 11 PK, 11 FK & 29 DataType, 53 Sub & 11 \\ \hline
Q12 & CONSTRUCT, 9 TP, 2 Join & 3 PK, 2 FK & 6 DataType, 7 Sub & 3 \\ \hline
\end{tabular}%
}
\end{table}
\clearpage
\section{Query Completeness}
\label{appendix:completeness}
\begin{table}[h]
\centering
\caption{Query completeness over multiple sizes of a GTFS dataset (the number indicates the scale factor: 1, 10, 100 and 100). The absence of a value means that the
OBDA engine does not support the features of the SPARQL query.}
\label{tab:complete-gtfs}
\resizebox{0.9\textwidth}{!}{%
\begin{tabular}{c|c|c|c|c|c|c|c|c|c|c|c|c|c}
\hline
\textbf{Engines/Queries} & \textbf{Q1} & \textbf{Q2} & \textbf{Q3} & \textbf{Q4} & \textbf{Q5} & \textbf{Q6} & \textbf{Q7} & \textbf{Q9} & \textbf{Q12} & \textbf{Q13} & \textbf{Q14} & \textbf{Q17} & \textbf{Total} \\ \hline
\multicolumn{14}{c}{\textbf{GTFS-1}} \\ \hline
Virtuoso & 58540 & 765 & 765 & 13 & 28 & 1 & 2 & 151439 & 6 & 734 & 2364 & 855 & 156972 \\ \hline
Morph-RDB & 58540 & 765 & - & 13 & - & 1 & 2 & 151439 & 6 & 734 & 2364 & 855 & 156179 \\ \hline
\begin{tabular}[c]{@{}c@{}}Morph-CSV \&\\ Morph-RDB\end{tabular} & 58540 & 765 & - & 13 & - & 1 & 2 & 151439 & 6 & 734 & 2364 & 855 & 156179 \\ \hline
Ontop & 58540 & 0 & 765 & 13 & 0 & - & 0 & 0 & - & 734 & 2364 & 855 & 4731 \\ \hline
\begin{tabular}[c]{@{}c@{}}Morph-CSV \&\\Ontop\end{tabular} & 58540 & 765 & 765 & 13 & 28 & - & 2 & 151439 & - & 734 & 2364 & 855 & 156965 \\ \hline
\multicolumn{14}{c}{\textbf{GTFS-10}} \\ \hline
Virtuoso & 353660 & 6312 & 4207 & 130 & 350 & 1 & 67 & 718317 & 130 & 2650 & 23640 & 8550 & 764354 \\ \hline
Morph-RDB & 353660 & 6312 & - & 130 & - & 1 & 67 & \textit{timeout} & 130 & 2650 & 23640 & 8550 & 41480 \\ \hline
\begin{tabular}[c]{@{}c@{}}Morph-CSV \&\\ Morph-RDB\end{tabular} & 353660 & 6312 & - & 130 & - & 1 & 67 & 718317 & 130 & 2650 & 23640 & 8550 & 759797 \\ \hline
Ontop & 353660 & 0 & 4207 & 130 & 0 & - & 0 & 0 & - & 2650 & 23640 & 8550 & 39177 \\ \hline
\begin{tabular}[c]{@{}c@{}}Morph-CSV \&\\Ontop\end{tabular} & 353660 & 6312 & 4207 & 130 & 350 & - & 67 & 718317 & - & 2650 & 23640 & 8550 & 764223 \\ \hline
\multicolumn{14}{c}{\textbf{GTFS-100}} \\ \hline
Virtuoso & 3536600 & 63100 & 42067 & 1300 & 3500 & 1 & 67 & 7183874 & 1300 & 26500 & 236400 & 85500 & 7643609 \\ \hline
Morph-RDB & 3536600 & 63100 & - & 1300 & - & 1 & 67 & \textit{timeout} & 1300 & 26500 & 236400 & 85500 & 414168 \\ \hline
\begin{tabular}[c]{@{}c@{}}Morph-CSV \&\\ Morph-RDB\end{tabular} & 3536600 & 63100 & - & 1300 & - & 1 & 67 & \textit{timeout} & 1300 & 26500 & 236400 & 85500 & 414168 \\ \hline
Ontop & 3536600 & 0 & 42067 & 1300 & 0 & - & 0 & 0 & - & 26500 & 236400 & 85500 & 391767 \\ \hline
\begin{tabular}[c]{@{}c@{}}Morph-CSV \&\\Ontop\end{tabular} & 3536600 & 63100 & 42067 & 1300 & \textit{timeout} & - & 67 & \textit{timeout} & - & 26500 & 236400 & 85500 & 454934 \\ \hline
\multicolumn{14}{c}{\textbf{GTFS-1000}} \\ \hline
Virtuoso & 35366000 & 1261368 & 420667 & 13000 & 35000 & 1 & 69 & 19077083 & 13000 & 420666 & 2364000 & 855000 & 24459854 \\ \hline
Morph-RDB & \textit{timeout} & 1261368 & - & 13000 & - & 1 & 69 & \textit{timeout} & 13000 & 420666 & 2364000 & 855000 & 4927104 \\ \hline
\begin{tabular}[c]{@{}c@{}}Morph-CSV \&\\ Morph-RDB\end{tabular} & 35366000 & 1261368 & - & 13000 & - & 1 & 69 & \textit{timeout} & 13000 & 420666 & 2364000 & 855000 & 4927104 \\ \hline
Ontop & \textit{timeout} & 0 & 420667 & 13000 & 0 & - & 0 & 0 & - & 420666 & 2364000 & 855000 & 4073333 \\ \hline
\begin{tabular}[c]{@{}c@{}}Morph-CSV \&\\Ontop\end{tabular} & \textit{timeout} & 1261368 & 420667 & 13000 & \textit{timeout} & - & 69 & \textit{timeout} & - & 420666 & 2364000 & 855000 & 5334770 \\ \hline
\end{tabular}%
}
\end{table}

\begin{table}[h]
\centering
\caption{Query completeness over of Bio2RDF tabular dataset.}
\label{tab:completness-bio2rdf}
\resizebox{0.9\textwidth}{!}{%
\begin{tabular}{c|c|c|c|c|c|c|c|c}
\hline
\textbf{Engines/Queries} & \textbf{Q1} & \textbf{Q2} & \textbf{Q3} & \textbf{Q4} & \textbf{Q5} & \textbf{Q6} & \textbf{Q7} & \textbf{Total} \\ \hline
Morph-RDB & 0 & 0 & 0 & 0 & 0 & 0 & 0 & 0 \\ \hline
\begin{tabular}[c]{@{}c@{}}Morph-CSV + \\ Morph-RDB\end{tabular} & 1000 & 1190181 & 10 & 102594 & 200 & 28224 & >10000 & >1422209 \\ \hline
Ontop & 0 & 0 & 0 & 0 & 0 & 0 & 0 & 0 \\ \hline
\begin{tabular}[c]{@{}c@{}}Morph-CSV + \\ Ontop\end{tabular} & 1000 & 1190181 & 10 & 102594 & 200 & 28224 & 13481 & 1335690 \\ \hline
\end{tabular}%
}
\end{table}

\begin{table}[h]
\centering
\caption{Query completeness over multiple sizes of a BSBM dataset. The absence of a value means that the
OBDA engine does not support the features of the SPARQL query.}
\label{tab:completeness-bsbm}
\resizebox{0.9\textwidth}{!}{%
\begin{tabular}{c|c|c|c|c|c|c|c|c|c|c|c|c}
\hline
\textbf{Engines/Queries} & \textbf{Q1} & \textbf{Q2} & \textbf{Q3} & \textbf{Q4} & \textbf{Q5} & \textbf{Q6} & \textbf{Q7} & \textbf{Q8} & \textbf{Q9} & \textbf{Q10} & \textbf{Q12} & \textbf{Total} \\ \hline
\multicolumn{13}{c}{\textbf{45K}} \\ \hline
Virtuoso & 10 & 19672 & 10 & 10 & 5 & 3 & 580691 & 20 & 450000 & 10 & 900000 & 1950431 \\ \hline
Morph-RDB & 10 & 19672 & 10 & 10 & 0 & 3 & 580691 & 20 & 450000 & 10 & 900000 & 1950426 \\ \hline
\begin{tabular}[c]{@{}c@{}}Morph-CSV \&\\ Morph-RDB\end{tabular} & 10 & 19672 & 10 & 10 & 5 & 3 & 580691 & 20 & 450000 & 10 & 900000 & 1950431 \\ \hline
Ontop & 10 & - & 10 & 10 & 0 & - & - & - & - & 0 & - & 30 \\ \hline
\begin{tabular}[c]{@{}c@{}}Morph-CSV \& \\ Ontop\end{tabular} & 10 & - & 10 & 10 & 5 & - & - & - & - & 10 & - & 45 \\ \hline
\multicolumn{13}{c}{\textbf{90K}} \\ \hline
Virtuoso & 10 & 38665 & 10 & 10 & 5 & 5 & 1161448 & 20 & 900000 & 10 & 1800000 & 3900183 \\ \hline
Morph-RDB & 10 & 38665 & 10 & 10 & 0 & 5 & 1161448 & 20 & 900000 & 10 & 1800000 & 3900178 \\ \hline
\begin{tabular}[c]{@{}c@{}}Morph-CSV \&\\ Morph-RDB\end{tabular} & 10 & 38665 & 10 & 10 & 5 & 5 & 1161448 & 20 & 900000 & 10 & 1800000 & 3900183 \\ \hline
Ontop & 10 & - & 10 & 10 & 0 & - & - & - & - & 0 & - & 30 \\ \hline
\begin{tabular}[c]{@{}c@{}}Morph-CSV \& \\ Ontop\end{tabular} & 10 & - & 10 & 10 & 5 & - & - & - & - & 10 & - & 45 \\ \hline
\multicolumn{13}{c}{\textbf{180K}} \\ \hline
Virtuoso & 10 & 69434 & 10 & 10 & 5 & 9 & 2168792 & 20 & 1800000 & 10 & 3600000 & 7638300 \\ \hline
Morph-RDB & 10 & \textit{timeout} & 10 & 10 & 0 & 9 & 2168792 & 20 & 1800000 & 10 & 3600000 & 7568861 \\ \hline
\begin{tabular}[c]{@{}c@{}}Morph-CSV \&\\ Morph-RDB\end{tabular} & 10 & 69434 & 10 & 10 & 5 & 9 & 2168792 & 20 & 1800000 & 10 & 3600000 & 7638295 \\ \hline
Ontop & \textit{timeout} & - & 10 & \textit{timeout} & 0 & - & - & - & - & 0 & - & 10 \\ \hline
\begin{tabular}[c]{@{}c@{}}Morph-CSV \& \\ Ontop\end{tabular} & 10 & - & 10 & 10 & 5 & - & - & - & - & 10 & - & 45 \\ \hline
\multicolumn{13}{c}{\textbf{360K}} \\ \hline
Virtuoso & 10 & 137359 & 10 & 10 & 5 & 18 & 4337584 & 20 & 3600000 & 10 & 7200000 & 15275026 \\ \hline
Morph-RDB & 10 & \textit{timeout} & 10 & 10 & 0 & 18 & \textit{timeout} & 20 & 3600000 & 10 & \textit{timeout} & 3600078 \\ \hline
\begin{tabular}[c]{@{}c@{}}Morph-CSV \&\\ Morph-RDB\end{tabular} & 10 & 137359 & 10 & 10 & \textit{timeout} & 18 & \textit{timeout} & 20 & 3600000 & 10 & \textit{timeout} & 3737437 \\ \hline
Ontop & \textit{timeout} & - & 10 & \textit{timeout} & 0 & - & - & - & - & 0 & - & 10 \\ \hline
\begin{tabular}[c]{@{}c@{}}Morph-CSV \& \\ Ontop\end{tabular} & 10 & - & 10 & 10 & \textit{timeout} & - & - & - & - & 10 & - & 40 \\ \hline
\end{tabular}%
}
\end{table}
\clearpage

\section{Detailed Loading Times for Morph-CSV}
\label{appendix:loadingtime}
\begin{table}[h]
\centering
\caption{Detailed results of Morph-CSV over GTFS-Madrid-Bench. As the input sources of this benchmark are extracted from a well-formed data model, the normalization step is not performed.}
\label{tab:gtfsdeatiledresults}
\resizebox{\textwidth}{!}{%
\begin{tabular}{l|c|c|c|c|c|c|c|c|c|c|c|c|c|c|c|c|c|c|c}
\hline
\multicolumn{1}{c|}{\textbf{Step/Query}} & \textbf{Q1} & \textbf{Q2} & \textbf{Q3} & \textbf{Q4} & \textbf{Q5} & \textbf{Q6} & \textbf{Q7} & \textbf{Q8} & \textbf{Q9} & \textbf{Q10} & \textbf{Q11} & \textbf{Q12} & \textbf{Q13} & \textbf{Q14} & \textbf{Q15} & \textbf{Q16} & \textbf{Q17} & \textbf{Q18} & \textbf{Morph-CSV$^-$} \\ \hline
\multicolumn{20}{c}{\textbf{GTFS-1}} \\ \hline
Selection & 0.370 & 0.387 & 0.374 & 0.376 & 0.381 & 0.368 & 0.381 & 0.386 & 0.390 & 0.378 & 0.377 & 0.373 & 0.430 & 0.396 & 0.363 & 0.370 & 0.389 & 0.379 & 0.410 \\ \hline
Normalization & - & - & - & - & - & - & - & - & - & - & - & - & - & - & - & - & - & - & - \\ \hline
Preparation & 0.286 & 0.065 & 0.062 & 0.113 & 0.114 & 0.106 & 0.337 & 0.359 & 0.464 & 0.126 & 0.169 & 0.226 & 0.073 & 0.246 & 0.069 & 0.205 & 0.169 & 0.226 & 0.772 \\ \hline
Creation \& Load & 0.345 & 0.075 & 0.074 & 0.063 & 0.059 & 0.051 & 0.176 & 0.182 & 0.381 & 0.090 & 0.084 & 0.130 & 0.064 & 0.141 & 0.055 & 0.087 & 0.091 & 0.101 & 0.612 \\ \hline
M. Translation & 0.506 & 0.521 & 0.532 & 0.500 & 0.525 & 0.524 & 0.536 & 0.545 & 0.532 & 0.535 & 0.523 & 0.509 & 0.540 & 0.581 & 0.498 & 0.530 & 0.543 & 0.519 & 0.635 \\ \hline
Total & 1.507 & 1.049 & 1.041 & 1.052 & 1.080 & 1.049 & 1.430 & 1.472 & 1.767 & 1.129 & 1.153 & 1.237 & 1.107 & 1.365 & 0.985 & 1.192 & 1.193 & 1.224 & 2.430 \\ \hline
\multicolumn{20}{c}{\textbf{GTFS-10}} \\ \hline
Selection & 0.998 & 1.005 & 1.033 & 1.059 & 1.010 & 1.012 & 1.023 & 1.041 & 1.004 & 1.030 & 1.021 & 0.994 & 1.006 & 1.019 & 1.009 & 1.019 & 1.013 & 1.028 & 1.0717 \\ \hline
Normalization & - & - & - & - & - & - & - & - & - & - & - & - & - & - & - & - & - & - & - \\ \hline
Preparation & 1.201 & 0.139 & 0.147 & 0.123 & 0.130 & 0.125 & 0.524 & 0.504 & 1.378 & 0.247 & 0.193 & 0.401 & 0.143 & 0.408 & 0.176 & 0.234 & 0.220 & 0.239 & 2.1550 \\ \hline
Creation \& Load & 1.296 & 0.139 & 0.137 & 0.067 & 0.067 & 0.060 & 0.475 & 0.467 & 2.214 & 0.323 & 0.116 & 0.460 & 0.160 & 0.453 & 0.221 & 0.117 & 0.198 & 0.117 & 4.2042 \\ \hline
M. Translation & 0.509 & 0.536 & 0.525 & 0.531 & 0.524 & 0.507 & 0.522 & 0.530 & 0.522 & 0.516 & 0.536 & 0.538 & 0.503 & 0.577 & 0.513 & 0.522 & 0.536 & 0.542 & 0.6442 \\ \hline
Total & 4.004 & 1.820 & 1.842 & 1.780 & 1.732 & 1.704 & 2.545 & 2.542 & 5.119 & 2.116 & 1.866 & 2.393 & 1.811 & 2.458 & 1.920 & 1.892 & 1.967 & 1.926 & 8.0750 \\ \hline
\multicolumn{20}{c}{\textbf{GTFS-100}} \\ \hline
Selection & 7.181 & 7.249 & 7.257 & 7.294 & 7.195 & 7.254 & 7.209 & 7.305 & 7.566 & 7.581 & 7.333 & 7.274 & 7.314 & 7.242 & 7.328 & 7.373 & 7.241 & 7.276 & 8.156 \\ \hline
Normalization & - & - & - & - & - & - & - & - & - & - & - & - & - & - & - & - & - & - & - \\ \hline
Preparation & 11.434 & 0.789 & 0.941 & 0.259 & 0.316 & 0.252 & 1.946 & 1.955 & 11.446 & 1.201 & 0.280 & 1.858 & 0.690 & 1.899 & 1.108 & 0.441 & 0.666 & 0.459 & 16.411 \\ \hline
Creation \& Load & 11.812 & 0.751 & 0.679 & 0.075 & 0.120 & 0.085 & 3.369 & 3.811 & 35.058 & 2.435 & 0.285 & 3.839 & 0.981 & 3.038 & 1.244 & 0.346 & 1.093 & 0.296 & 92.785 \\ \hline
M. Translation & 0.531 & 0.519 & 0.507 & 0.507 & 0.520 & 0.532 & 0.526 & 0.571 & 0.540 & 0.538 & 0.556 & 0.534 & 0.524 & 0.519 & 0.534 & 0.538 & 0.533 & 0.578 & 0.761 \\ \hline
Total & 30.959 & 9.308 & 9.384 & 8.135 & 8.151 & 8.123 & 13.050 & 13.642 & 54.609 & 11.755 & 8.454 & 13.504 & 9.509 & 12.698 & 10.213 & 8.698 & 9.533 & 8.611 & 118.113 \\ \hline
\multicolumn{20}{c}{\textbf{GTFS-1000}} \\ \hline
Selection & 76.815 & 73.390 & 71.395 & 71.686 & 71.770 & 72.521 & 72.749 & 73.408 & 78.764 & 73.982 & 72.248 & 73.084 & 71.511 & 73.874 & 73.003 & 71.692 & 72.449 & 71.849 & 72.920 \\ \hline
Normalization & - & - & - & - & - & - & - & - & - & - & - & - & - & - & - & - & - & - & - \\ \hline
Data Preparation & 140.784 & 8 & 6.826 & 0.657 & 0.737 & 0.441 & 19.167 & 18.768 & 126.915 & 12.364 & 1.318 & 17.717 & 6.718 & 18.356 & 10.506 & 1.505 & 4.200 & 1.552 & 184.215 \\ \hline
Creation \& Load & 123.770 & 6.843 & 7.123 & 0.239 & 0.665 & 0.271 & 52.349 & 52.294 & 3121.927 & 66.614 & 1.620 & 69.820 & 9.169 & 48.674 & 13.434 & 2.028 & 10.732 & 1.905 & 420.123 \\ \hline
M. Translation & 0.546 & 0.521 & 0.528 & 0.541 & 0.533 & 0.532 & 0.541 & 0.550 & 0.557 & 0.524 & 0.535 & 0.532 & 0.511 & 0.532 & 0.557 & 0.551 & 0.528 & 0.541 & 0.607 \\ \hline
Total & 341.915 & 88.795 & 85.871 & 73.123 & 73.705 & 73.766 & 144.808 & 145.021 & 3328.163 & 153.485 & 75.722 & 161.153 & 87.909 & 141.437 & 97.500 & 75.776 & 87.909 & 75.847 & 677.865 \\ \hline
\end{tabular}%
}
\end{table}

\begin{table}[t]
\centering
\caption{Detailed results of Morph-CSV over BSBM. As the input sources of this benchmark are extracted from a well-formed relational database, the normalization step is not performed. } 
\label{tab:detailedresultsbsbm}
\resizebox{\textwidth}{!}{%
\begin{tabular}{l|c|c|c|c|c|c|c|c|c|c|c|c|c}
\hline
\multicolumn{1}{c|}{\textbf{Step/Query}} & \textbf{Q1} & \textbf{Q2} & \textbf{Q3} & \textbf{Q4} & \textbf{Q5} & \textbf{Q6} & \textbf{Q7} & \textbf{Q8} & \textbf{Q9} & \textbf{Q10} & \textbf{Q11} & \textbf{Q12} & \textbf{Morph-CSV$^-$} \\ \hline
\multicolumn{14}{c}{\textbf{BSBM-45K}} \\ \hline
Selection & 0.004 & 0.004 & 0.003 & 0.004 & 0.004 & 0.004 & 0.004 & 0.004 & 0.004 & 0.004 & 0.004 & 0.004 & 0.007 \\ \hline
Normalization & - & - & - & - & - & - & - & - & - & - & - & - & - \\ \hline
Preparation & 3.291 & 5.699 & 3.424 & 3.507 & 3.425 & 2.318 & 20.750 & 28.657 & 13.397 & 7.433 & 43.732 & 8.443 & 43.452 \\ \hline
Creation \& Load & 3.044 & 4.355 & 2.707 & 3.049 & 2.648 & 0.102 & 5.819 & 14.168 & 1.223 & 3.894 & 27.034 & 5.151 & 28.036 \\ \hline
M. Translation & 0.514 & 0.564 & 0.498 & 0.522 & 0.505 & 0.524 & 0.564 & 0.551 & 0.519 & 0.569 & 0.572 & 0.541 & 0.547 \\ \hline
Total & 6.853 & 10.622 & 6.633 & 7.082 & 6.581 & 2.949 & 27.137 & 43.380 & 15.142 & 11.900 & 71.342 & 14.139 & 72.043 \\ \hline
\multicolumn{14}{c}{\textbf{BSBM-90K}} \\ \hline
Selection & 0.004 & 0.004 & 0.004 & 0.004 & 0.004 & 0.003 & 0.004 & 0.004 & 0.004 & 0.004 & 0.005 & 0.004 & 0.005 \\ \hline
Normalization & - & - & - & - & - & - & - & - & - & - & - & - & - \\ \hline
Preparation & 6.882 & 10.022 & 6.334 & 6.378 & 7.167 & 3.149 & 42.191 & 61.591 & 24.135 & 13.907 & 85.432 & 16.059 & 208.798 \\ \hline
Creation \& Load & 6.118 & 8.667 & 5.529 & 5.711 & 6.067 & 0.168 & 12.668 & 30.614 & 2.227 & 8.003 & 56.776 & 10.638 & 58.119 \\ \hline
M, Translation & 0.540 & 0.525 & 0.516 & 0.509 & 0.527 & 0.509 & 0.546 & 0.551 & 0.512 & 0.519 & 0.569 & 0.546 & 0.574 \\ \hline
Total & 13.544 & 19.219 & 12.384 & 12.602 & 13.764 & 3.830 & 55.409 & 92.761 & 26.877 & 22.434 & 142.783 & 27.247 & 267.496 \\ \hline
\multicolumn{14}{c}{\textbf{BSBM-180K}} \\ \hline
Selection & 0.004 & 0.004 & 0.004 & 0.004 & 0.004 & 0.004 & 0.004 & 0.004 & 0.003 & 0.004 & 0.005 & 0.004 & 0.005 \\ \hline
Normalization & - & - & - & - & - & - & - & - & - & - & - & - & - \\ \hline
Preparation & 12.675 & 19.946 & 11.978 & 12.459 & 11.969 & 5.255 & 83.486 & 122.173 & 47.833 & 29.420 & 185.650 & 34.682 & 339.254 \\ \hline
Creation \& Load & 10.740 & 15.848 & 11.134 & 12.542 & 11.450 & 0.268 & 25.693 & 67.677 & 5.243 & 15.268 & 137.522 & 21.411 & 141.737 \\ \hline
M. Translation & 0.534 & 0.508 & 0.545 & 0.513 & 0.532 & 0.514 & 0.584 & 0.554 & 0.553 & 0.574 & 0.607 & 0.599 & 0.606 \\ \hline
Total & 23.953 & 36.307 & 23.661 & 25.518 & 23.955 & 6.041 & 109.767 & 190.408 & 53.634 & 45.266 & 323.784 & 56.695 & 481.602 \\ \hline
\multicolumn{14}{c}{\textbf{BSBM-360K}} \\ \hline
Selection & 0.004 & 0.004 & 0.004 & 0.004 & 0.004 & 0.004 & 0.004 & 0.004 & 0.004 & 0.004 & 0.005 & 0.004 & 0.005 \\ \hline
Normalization & - & - & - & - & - & - & - & - & - & - & - & - & - \\ \hline
Preparation & 23.846 & 43.880 & 24.970 & 24.597 & 23.670 & 10.401 & 198.975 & 293.087 & 110.852 & 57.878 & 415.052 & 66.759 & 578.798 \\ \hline
Creation \& Load & 26.804 & 44.031 & 24.667 & 30.709 & 23.089 & 0.435 & 55.623 & 136.090 & 10.037 & 32.036 & 262.529 & 44.716 & 260.139 \\ \hline
M. Translation & 0.545 & 0.571 & 0.536 & 0.533 & 0.540 & 0.494 & 0.580 & 0.583 & 0.503 & 0.563 & 0.632 & 0.540 & 0.578 \\ \hline
Total & 51.199 & 88.486 & 50.176 & 55.842 & 47.302 & 11.333 & 255.183 & 429.765 & 121.396 & 90.481 & 678.218 & 112.019 & 839.521 \\ \hline
\end{tabular}%
}
\end{table}

\begin{table}[t]
\centering
\caption{Detailed results of Morph-CSV over Bio2RDF.}
\label{tab:bio2rdfdeatiledresults}
\resizebox{\textwidth}{!}{%
\begin{tabular}{l|c|c|c|c|c|c|c|c|c|c|c}
\hline
\multicolumn{1}{c|}{\textbf{Step/Query}} & \textbf{Q1} & \textbf{Q2} & \textbf{Q3} & \textbf{Q4} & \textbf{Q5} & \textbf{Q6} & \textbf{Q7} & \textbf{Q8} & \textbf{Q9} & \textbf{Q10} & \textbf{Morph-CSV$^-$} \\ \hline
Selection & 3.705 & 3.749 & 3.777 & 3.714 & 3.732 & 3.787 & 3.719 & 3.775 & 3.712 & 3.632 & 48.088\\ \hline
Normalization & - & - & 0.194 & - & - & 0.212 & - & 0.301 & - & - &  38.577\\ \hline
Preparation & 0.903 & 123.798 & 0.580 & 7.414 & 0.628 & 126.852 & 8.457 & 2.660 & 5.531 & 0.555 & 253.461 \\ \hline
Creation \& Load & 0.318 & 131.912 & 0.113 & 34.569 & 0.994 & 147.659 & 32.203 & 0.901 & 8.968 & 0.790 &  265.301 \\ \hline
M. Translation & 0.541 & 0.542 & 0.546 & 0.534 & 0.548 & 0.560 & 0.535 & 0.539 & 0.543 & 0.552 & 0.693 \\ \hline
Total & 5.467 & 260.000 & 5.211 & 46.231 & 5.901 & 279.071 & 44.915 & 8.176 & 18.756 & 5.529 & 606.121 \\ \hline
\end{tabular}%
}
\end{table}

\end{appendix}

\end{document}